\documentclass[epsf]{emulateapj}
\shorttitle{Phase-space substructures in M87}
\shortauthors{Romanowsky et al.}
\slugcomment{Accepted by ApJ, 16 December 2011 }

\def\ngc{N_{\rm GC}}
\def\sigc{\sigma_{\rm c}}
\def\sgc{\sigma_{\rm GC}}
\def\kms{\,km~s$^{-1}$}

\def\gsim{\;\rlap{\lower 2.5pt
 \hbox{$\sim$}}\raise 1.5pt\hbox{$>$}\;}
\def\lsim{\;\rlap{\lower 2.5pt
   \hbox{$\sim$}}\raise 1.5pt\hbox{$<$}\;}

\usepackage{subfigure}

\begin{document}

\title{The ongoing assembly of a central cluster galaxy: \\ Phase-space substructures in the halo of M87}

\author{
Aaron J. Romanowsky\altaffilmark{1}, Jay Strader\altaffilmark{2}, Jean P. Brodie\altaffilmark{1}, J. Christopher Mihos\altaffilmark{3},\\ 
Lee R. Spitler\altaffilmark{4}, Duncan A. Forbes\altaffilmark{4}, Caroline Foster\altaffilmark{4,5}, Jacob A. Arnold\altaffilmark{1}
}

\affil{
\altaffilmark{1}University of California Observatories, Santa Cruz, CA 95064, USA\\
\altaffilmark{2}Harvard-Smithsonian Center for Astrophysics, Cambridge, MA 02138, USA\\
\altaffilmark{3}Department of Astronomy, Case Western Reserve University, 10900 Euclid Ave., Cleveland, OH 44106, USA\\
\altaffilmark{4}Centre for Astrophysics \& Supercomputing, Swinburne University, Hawthorn, VIC 3122, Australia\\
\altaffilmark{5}European Southern Observatory, Alonso de C\'ordova 3107, Vitacura, Casilla 19001, Santiago 19, Chile\\
}

\begin{abstract}
The halos of galaxies preserve unique records of their formation histories.
We carry out the first combined observational and theoretical study
of phase-space halo substructure in an early-type galaxy:
M87, the central galaxy in the Virgo cluster.
We analyze an unprecedented wide-field, high-precision photometric and spectroscopic
data set for 488 globular clusters (GCs),
which includes new, large-radius Subaru/Suprime-Cam and Keck/DEIMOS observations.
We find signatures of two substructures in position-velocity phase-space.
One is a small, cold stream associated with a known stellar filament in the outer halo;
the other is a large shell-like pattern in the inner halo that
implies a massive, hitherto unrecognized accretion event.
We perform extensive statistical tests and independent metallicity analyses
to verify the presence and characterize the properties of these features,
and to provide more general methodologies for future extragalactic studies
of phase-space substructure.
The cold outer stream is consistent with a dwarf galaxy accretion event,
while for the inner shell there is tension between a low progenitor mass
implied by the cold velocity dispersion, and a high mass from the large number of GCs,
which might be resolved by a $\sim0.5\,L^*$ E/S0 progenitor.
We also carry out proof-of-principle numerical simulations of the
accretion of smaller galaxies in an M87-like gravitational potential.
These produce analogous features to the observed substructures,
which should have observable lifetimes of $\sim$~1~Gyr.
The shell and stream GCs together support a scenario where the extended stellar
envelope of M87 has been built up by a steady rain of
material that continues until the present day.
This phase-space method demonstrates unique potential for detailed
tests of galaxy formation beyond the Local Group.
\end{abstract}


\section{Introduction}\label{sec:intro}

The ultimate record of the formational histories of galaxies
is found in the six-dimensional phase space of stellar positions and
velocities, supplemented by additional information about
ages and elemental abundances (primarily iron-based metallicities).
The halos of galaxies are natural hunting grounds for assembly clues in phase-space
because of the preservative effects of the long dynamical times,
and since any material entering a galaxy naturally has to traverse its halo.

Indeed, it is a fundamental prediction of the current cosmological paradigm that the
outer regions of galaxies should be constantly bombarded with smaller, infalling systems
that persist as significant substructures while gradually becoming disrupted.
This idea has been borne out qualitatively in recent years by observations
of halo streams and shells in the nearby Universe
(e.g., \citealt{1999Natur.402...53H,2001Natur.412...49I,2006ApJ...642L.137B,2009AJ....138.1417T,2010AJ....140..962M,2011arXiv1111.2864C}).

Massive elliptical galaxies are touchstones for this issue,
since their spheroidal nature and their residence in the densest areas of the Universe
suggest particularly vigorous accretion histories.
Comparisons of these galaxies' stellar luminosity profiles
at high and low redshifts now provide indirect
evidence that their outer envelopes grow tremendously in size up to the present day
(e.g., \citealt{2008ApJ...687L..61B,2008ApJ...688...48V,2009ApJ...695..101D,2010ApJ...709.1018V,2011arXiv1106.4308C}),
which may be naturally explained by accretion and mergers in a cosmological context
\citep{2006ApJ...648L..21K, 2009ApJ...699L.178N,2010MNRAS.401.1099H,2010ApJ...725.2312O,2011arXiv1106.5490O}.

The most extreme examples of ongoing assembly are expected to be the brightest cluster
galaxies (BCGs), whose typical locations at the centers of clusters should involve
very active merger histories (e.g., \citealt{2009ApJ...696.1094R}).\footnote{Although
the central galaxy is often not technically the brightest object in a cluster,
we will follow the convention of calling this the BCG.}
Current cosmologically-based models find that
around half of the stellar mass growth of a present-day BCG occurred over the past 5~Gyr
via gas-poor mergers
(after an early formation epoch for the core regions;
\citealt{2007MNRAS.375....2D}).

This theoretical picture has received very mixed support from
comparisons of BCG stellar masses and surface brightness distributions 
at different redshifts 
\citep{2008MNRAS.387.1253W,2009MNRAS.395.1491B,2009Natur.458..603C,2010ApJ...721L..19V,2011ApJ...726...69A,2011MNRAS.414..445S}.
Surveys for major mergers in action have found that these contribute significantly
to recent BCG growth
(\citealt{2008MNRAS.388.1537M,2009MNRAS.396.2003L}; see also \citealt{2011MNRAS.414L..80B}), 
but the contributions from minor mergers are unknown and the
overall comparison to theory unclear.
Part of the difficulty is that even if accretion and mergers are important
growth processes, the frequency of observing an ongoing event can be fairly low.
Thus it would be invaluable to go beyond purely photometric observations and
delve into the phase-space of position, velocity, and metallicity
for longer lived signatures (e.g., \citealt{2008ApJ...689..936J,2009MNRAS.394..641Z}).

Within the Local Group, detailed phase-space studies are being undertaken using individual stars 
(e.g., \citealt{2008ApJ...689..958K,2009ApJ...705.1275G,2009ApJ...698..567S,2011ApJ...738...79X}), 
but are currently impossible in more distant galaxies.
Instead, bright stellar proxies may be used, such as planetary nebulae (PNe;
\citealt{2003ApJ...582..170D,2003MNRAS.346L..62M,2010ApJ...725L..97S,2011MNRAS.414..642C})
and globular clusters (GCs).
GCs are thought to form along with field stars, and as collections of $\sim$~$10^6$ stars each,
they are visible at much greater distances, allowing their individual positions, 
line-of-sight velocities, and chemical properties to be measured
as far away as $\sim$~50~Mpc \citep{2011A&A...531A.119R}.

The ``chemodynamics'' of GCs provided the watershed evidence 
for the accretion origin of the Milky Way's outer halo \citep{1978ApJ...225..357S}, 
where it is now thought that disrupting satellite galaxies have deposited
many accompanying GCs (e.g., \citealt{2003AJ....125..188B,2007ChJAA...7..111G,2007PASP..119..939G,2007ApJ...661L..49L,2010MNRAS.404.1203F}).
A similar scenario is now evident in the nearby spiral M31, where some GCs are
associated with halo substructures \citep{2009MNRAS.396.1619C,2010ApJ...717L..11M}.
In BCGs, the presence of GCs in enormous numbers
(up to $5\times10^4$ per system) has long motivated suggestions 
that accretion is an important factor for these galaxies \citep{1982AJ.....87.1465F,1998ApJ...501..554C}, although it is not clear how many of these GCs may have formed
during in situ processes or mergers rather than being accreted.

Various GC and PN-based studies of the halo kinematics of giant ellipticals and BCGs have turned up
evidence for substructures that may imply active accretion events or mergers
\citep{2003ApJ...591..850C,2009AJ....137.4956R,2010A&A...513A..52S,2010A&A...518A..44M,2011AJ....141...27W}.
In other cases, accretion events are suggested by photometric substructures or by
broad kinematical or chemical halo transitions
\citep{2009AJ....138.1417T,2009MNRAS.398...91P,2009MNRAS.394.1249C,2010MNRAS.407L..26C,2010ApJ...715..972J,2011MNRAS.413.2943F,2011ApJ...736L..26A,2011MNRAS.415..993M}.
However, in no case has a halo feature been observed and modeled in enough detail
to determine its origin and the implications for galaxy formation.

To go beyond the previous work on nearby BCGs by carrying out detailed phase-space studies
of their halos,
there are two critical observational requirements
beside sheer instrumental throughput: 
wide-field coverage and spectroscopic resolution.
The former permits blind searches for halo substructures and demands a 
field of view of at least $\sim$~0.5~degree in order to probe galactocentric radii out
to $\sim$~100~kpc or more in galaxies at $\sim$~15~Mpc distances.
The latter allows for a wide mass range of accreted galaxies to be probed;
e.g., if a BCG has a characteristic velocity dispersion of $\sigma \sim$~300~\kms\
and the precision of velocity measurements for halo tracers
is $\Delta v \sim$~50--100~\kms, as in many past
surveys, then the observations are sensitive to mass ratios down 
to $\sim (\Delta v/\sigma)^2 \sim$~1:40--1:10 in dynamical mass.
More minor mergers than these are certainly expected to be more frequent, and may
comprise a dominant mode for halo assembly, so a velocity resolution of
tens of \kms\ ($\Delta v/\sigma \lsim 0.1$) is preferable.

We present here the first wide-field, large-sample spectroscopic survey of a BCG with
velocity resolution of $\Delta v/\sigma \sim 0.04$.
Our subject is M87, the central elliptical in Virgo, the nearest galaxy cluster
at a distance of $\sim$~16.5~Mpc.
The GC system of M87 has seen decades of spectroscopic study
(e.g., \citealt{1987AJ.....93..779H,1997ApJ...486..230C,2001ApJ...559..812H}), but
extending to radii of only $\sim$~40~kpc and with $\Delta v/\sigma \gsim 0.25$.

Until now, M87 could be characterized as a dynamically quiet galaxy, with only
mild signs of interactions such as very faint stellar filaments in its
far outer halo \citep{2005ApJ...631L..41M,2010ApJ...715..972J}.  Our new GC study reveals the
kinematics of one of these faint streams, and unveils an unsuspected,
enormous shell-like substructure in phase-space.

Our paper proceeds as follows.  Section~\ref{sec:obs}
describes the observations and data reduction, and presents a basic overview
of the halo substructures.
Sections~\ref{sec:shell} and \ref{sec:outer}
analyze the characteristics and statistical significance of the large inner shell
and outer stream, respectively.
Theoretical analyses of the substructure origins and dynamics are explored in
Section~\ref{sec:sim}.
Section~\ref{sec:concl} summarizes the findings and outstanding questions.

\section{Observations and basic results}\label{sec:obs}

We have revisited M87 in a new-era high-precision wide-field survey of GCs.
The main component of our data set is presented and discussed in detail in 
\citet[hereafter S+11]{2011arXiv1110.2778S}.  In brief, it is based on photometry
from CFHT/Megacam and Subaru/Suprime-Cam optical images, 
and spectroscopy from MMT/Hectospec, Keck/DEIMOS, and Keck/LRIS.
New line-of-sight velocities were obtained for 451 GCs over a series of
campaigns from 2007 to 2010, extending over a range in galactocentric radius of
$\sim$~1--35~arcmin ($\sim$~5--170~kpc).

The first subset of the new data extended along a narrow track eastwards of M87's center,
in search of any kinematical transition between the BCG and the surrounding
cluster.  A peculiar fluctuation in the velocity dispersion profile was indeed 
found at a radius of $\sim$~9~arcmin ($\sim$~45~kpc), which was the first 
indication of an inner-halo shell-like feature as we will discuss later.
Subsequent data (the large majority of the total sample)
were obtained at generally random position angles around the galaxy.

In a parallel campaign presented here for the first time, 
we observed intensively the region around the outer stellar ``stream A'' 
\citep{2005ApJ...631L..41M,2010ApJ...715..972J,2010ApJ...720..569R,2011ApJ...735...76K},
at radii of $\sim$~35--50~arcmin ($\sim$~170--240~kpc).
GC selection for spectroscopic follow-up was obtained from a variety of images as
they became available: archival Subaru/Suprime-Cam 
\citep{2002PASJ...54..833M} $BVI$ imaging,
the Megacam $gri$ imaging mentioned above \citep{2009ApJ...703..939H},
and our own new Suprime-Cam imaging in $gi$.
Given the incomplete spatial coverage of the first two imaging data-sets, the
third data-set provides our default photometric source for the stream area.
These images were taken on 2009 Apr 21, with intermittent transparency, 
$\sim$~1$^{\prime\prime}$ seeing, and exposure times of 300 sec in each band.

Our spectroscopic follow-up of the stream GCs used Keck/DEIMOS and
observing and analysis techniques established
in our previous work on GC kinematics 
(\citealt{2009AJ....137.4956R,2011ApJ...736L..26A,2011MNRAS.415.3393F}; S+11).
Four masks were observed during the nights 2009 Mar 23, 2010 Mar 11--13, and 2010 Jun 12,
with generally good conditions and exposure times varying from 0.5 to 2 hours.
The spectra cover an approximate spectral range of 6500--9000 
\AA\ at a moderate resolution (1.5 \AA\ FWHM) that allows the derivation of precise line-of-sight velocities. 

The data were reduced in a standard manner, including bias subtraction, flat fielding, wavelength calibration, and 
sky subtraction, followed by optimal extraction. 
Heliocentric line-of-sight velocities were derived through cross-correlation with a range of 
stellar templates, using only the region around the Calcium {\small II} triplet (CaT;
H$\alpha$, if available, was used for confirmation of borderline cases). 
Some background galaxies were identified via examination of the extracted spectra and excluded 
from further analysis. 

The velocity uncertainties were estimated by combining in quadrature the statistical uncertainties from
the cross-correlation with the systematic uncertainties from using different stellar templates.
The uncertainties for the confirmed GCs ranged from 5 to 18~\kms,
with their reliability supported by repeat observations of four GCs and five stars.
The characteristics of the spectroscopically-confirmed GCs (20 of them), stars,
and galaxies in the stream region are provided in Table~\ref{tab:data}.
We also targeted five PN candidates from \citet{2003ApJS..145...65F} but found no
H$\alpha$ emission in their spectra (objects 7-14, 7-18, 7-51, 7-58, 7-65).

\LongTables

\begin{deluxetable}{lccccc}
\tablewidth{0pt}
\tabletypesize{\footnotesize}
\tablecaption{Spectroscopic objects in M87 outer stream region \label{tab:data}}
\tablehead{ID & R.A. & Decl. & $i_0$ & $(g-i)_0$ & $v$ \\
& [J2000] & [J2000] & &  & [\kms]}
\startdata
{\it confirmed:}\\
R8922 & 187.36271 & 12.84842 & 20.07 & 1.01 & $1460\pm6$ \\ 
R10106 & 187.35946 & 12.87679 & 19.52 & 0.99 & $1500\pm5$ \\ 
R10806 & 187.36816 & 12.89351 & 21.36 & 1.08 & $839\pm8$ \\ 
R10814 & 187.35690 & 12.89390 & 21.38 & 0.85 & $1386\pm10$ \\ 
R11987 & 187.36719 & 12.92166 & 22.24 & 0.78 & $1100\pm10$ \\
R13629 & 187.32990 & 12.96460 & 21.65 & 0.80 & $1112\pm7$ \\
R14045 & 187.35667 & 12.97587 & 21.72 & 0.80 & $852\pm8$ \\ 
R14100 & 187.29791 & 12.97751 & 21.77 & 0.97 & $1154\pm11$ \\ 
R15088 & 187.32974 & 13.00221 & 21.09 & 0.76 & $1395\pm10$ \\ 
R15675 & 187.27000 & 13.01773 & 21.98 & 0.84 & $1141\pm18$ \\ 
R16138 & 187.41290 & 13.02971 & 22.07 & 1.13 & $1063\pm12$ \\ 
R16501 & 187.36578 & 13.03980 & 22.33 & 0.71 & $1590\pm12$ \\ 
R16571 & 187.36916 & 13.04153 & 20.44 & 0.79 & $1577\pm9$ \\ 
R17084 & 187.33063 & 13.05539 & 21.18 & 0.83 & $1490\pm8$ \\ 
R17186 & 187.23102 & 13.05824 & 20.41 & 0.84 & $1291\pm6$ \\ 
R17522 & 187.34823 & 13.06668 & 20.89 & 0.78 & $1467\pm5$ \\ 
R18039 & 187.34714 & 13.08054 & 22.08 & 0.76 & $673\pm9$ \\ 
R18045 & 187.31874 & 13.08087 & 21.47 & 0.81 & $1462\pm11$ \\ 
R19909 & 187.33585 & 13.12520 & 21.79 & 0.83 & $1437\pm12$ \\ 
R21956\footnote{Probably bound to NGC~4461.} & 187.23676 & 13.16944 & 19.90 & 0.87 & $1876\pm7$ 
\vspace{0.1cm}
\\
{\it marginal:}\\
R12580 & 187.39820 & 12.93529 & 22.01 & 0.74 & $1085\pm16$ \\ 
R13657 & 187.39522 & 12.96553 & 22.14 & 0.78 & $1222\pm18$ \\ 
R15867 & 187.41155 & 13.02209 & 22.31 & 0.75 & $1244\pm14$ \\ 
R16402 & 187.32270 & 13.03704 & 22.45 & 1.21 & $898\pm17$ 
\vspace{0.1cm}
\\
{\it stars\footnote{A simple velocity boundary of 250~\kms\ has been used to separate
stars and GCs, but some of the higher-velocity stars might in principle be low-velocity GCs.}:}\\
R7403 & 187.37441 & 12.81474 & 17.55 & 0.59 & $-3\pm48$ \\ 
R7754 & 187.37956 & 12.82188 & 19.87 & 1.91 & $14\pm74$ \\ 
R8193 & 187.38146 & 12.83134 & 19.29 & 0.69 & $-148\pm21$ \\ 
R8376 & 187.37963 & 12.83570 & 20.52 & 0.98 & $-134\pm23$ \\ 
R8798 & 187.39239 & 12.84514 & 18.86 & 0.72 & $65\pm21$ \\ 
R9363 & 187.34117 & 12.85885 & 19.34 & 0.88 & $40\pm7$ \\ 
R10146 & 187.34439 & 12.87770 & 20.45 & 0.65 & $58\pm20$ \\ 
R10837 & 187.36675 & 12.89449 & 17.67 & 2.65 & $42\pm14$ \\ 
R11062 & 187.40232 & 12.89997 & 15.38 & 1.74 & $-24\pm5$ \\ 
R11805 & 187.39709 & 12.91762 & 22.47 & 1.33 & $189\pm20$ \\ 
R11872 & 187.36193 & 12.91931 & 19.96 & 1.11 & $45\pm15$ \\ 
R12125 & 187.40527 & 12.92458 & 20.42 & 0.74 & $145\pm6$ \\ 
R12313 & 187.29011 & 12.92963 & 15.75 & 0.49 & $4\pm20$ \\ 
R12657 & 187.34624 & 12.93754 & 17.48 & 1.09 & $10\pm5$ \\ 
R12691 & 187.40031 & 12.93805 & 16.32 & 2.03 & $-20\pm5$ \\ 
R12891 & 187.34345 & 12.94330 & 20.44 & 0.84 & $177\pm51$ \\ 
R13164 & 187.30174 & 12.95084 & 20.11 & 0.69 & $-41\pm25$ \\ 
R13188 & 187.23932 & 12.95184 & 17.91 & 1.33 & $35\pm17$ \\ 
R13506 & 187.46370 & 12.95930 & 15.37 & 0.35 & $-6\pm8$ \\ 
R13992 & 187.33976 & 12.97469 & 19.83 & 1.80 & $4\pm36$ \\ 
R14054 & 187.45815 & 12.97566 & 19.07 & 0.74 & $98\pm5$ \\ 
R14166 & 187.33685 & 12.97877 & 20.88 & 1.03 & $198\pm7$ \\ 
R14317 & 187.34125 & 12.98226 & 15.49 & 0.92 & $30\pm5$ \\ 
R14391 & 187.35850 & 12.98435 & 19.70 & 0.69 & $-87\pm6$ \\ 
R14427 & 187.44737 & 12.98488 & 20.96 & 1.03 & $213\pm14$ \\ 
R14800 & 187.26911 & 12.99484 & 15.70 & 1.67 & $0\pm18$ \\ 
R15119 & 187.40424 & 13.00259 & 17.83 & 2.02 & $9\pm5$ \\ 
R15179 & 187.44657 & 13.00379 & 17.83 & 0.87 & $70\pm5$ \\ 
R15494 & 187.45031 & 13.01196 & 19.09 & 0.82 & $-3\pm5$ \\ 
R15721 & 187.30143 & 13.01873 & 16.86 & 0.59 & $58\pm18$ \\ 
R15838 & 187.42486 & 13.02142 & 21.09 & 0.72 & $-31\pm10$ \\ 
R15983 & 187.36178 & 13.02593 & 21.37 & 1.89 & $55\pm32$ \\ 
R16488 & 187.39562 & 13.03932 & 19.52 & 1.12 & $108\pm5$ \\ 
R17262 & 187.41103 & 13.05953 & 22.24 & 0.64 & $110\pm24$ \\ 
R17841 & 187.35704 & 13.07556 & 22.00 & 0.72 & $185\pm12$ \\ 
R18139 & 187.27810 & 13.08366 & 22.26 & 1.27 & $-84\pm26$ \\ 
R18154 & 187.34945 & 13.08374 & 18.48 & 1.54 & $-57\pm5$ \\ 
R18730 & 187.27131 & 13.09786 & 21.26 & 0.80 & $-214\pm15$ \\ 
R19052 & 187.27947 & 13.10543 & 15.40 & 0.97 & $-35\pm15$ \\ 
R19452 & 187.37246 & 13.11475 & 16.77 & 2.38 & $-1\pm5$ \\ 
R20070 & 187.26784 & 13.12938 & 17.82 & 0.67 & $127\pm14$ \\ 
R20445 & 187.30037 & 13.13758 & 20.15 & 0.76 & $103\pm6$ \\ 
R20496 & 187.38952 & 13.13803 & 21.20 & 0.78 & $32\pm16$ \\ 
R20765 & 187.24172 & 13.14418 & 20.54 & 0.72 & $222\pm18$ \\ 
R21009 & 187.23749 & 13.14917 & 15.18 & 1.85 & $5\pm17$ \\ 
R21082 & 187.29307 & 13.15058 & 17.69 & 1.30 & $-6\pm8$ \\ 
R22195 & 187.33737 & 13.17345 & 20.94 & 0.79 & $73\pm9$ \\ 
R22421 & 187.33604 & 13.17731 & 21.49 & 0.57 & $143\pm16$ \\ 
R22843 & 187.32868 & 13.18426 & 20.83 & 0.71 & $22\pm13$ 
\vspace{0.1cm}
\\
{\it galaxies:}\\
R12858 & 187.23866 & 12.94289 & 22.24 & 0.71 & \nodata\ \\ 
R13471 & 187.25251 & 12.95961 & 21.89 & 1.05 & \nodata\ \\ 
R14229 & 187.22466 & 12.98077 & 22.03 & 0.94 & \nodata\ \\ 
R15594 & 187.25608 & 13.01557 & 22.41 & 0.98 & \nodata\ \\ 
R15685 & 187.23852 & 13.01819 & 22.40 & 1.37 & \nodata\ \\ 
R17134 & 187.23876 & 13.05692 & 22.59 & 0.77 & \nodata\ \\ 
R21550 & 187.21381 & 13.15999 & 22.97 & 0.69 & \nodata\ \\ 
\enddata
\tablecomments{Photometry is from Subaru-Suprime/Cam imaging
using 4-pixel aperture magnitudes, bootstrapped to the Sloan Digital Sky Survey and then
corrected for Galactic extinction following \citet{2010ApJ...719..415P}.
The uncertainties in the stellar velocities are indicative only, and
not as carefully characterized as for the GCs.
The most likely objects belonging to the cold GC stream are
R16501, R16571, R17084, R17522, R18045.
}
\end{deluxetable}

\begin{figure*}
\epsscale{0.56}
\plotone{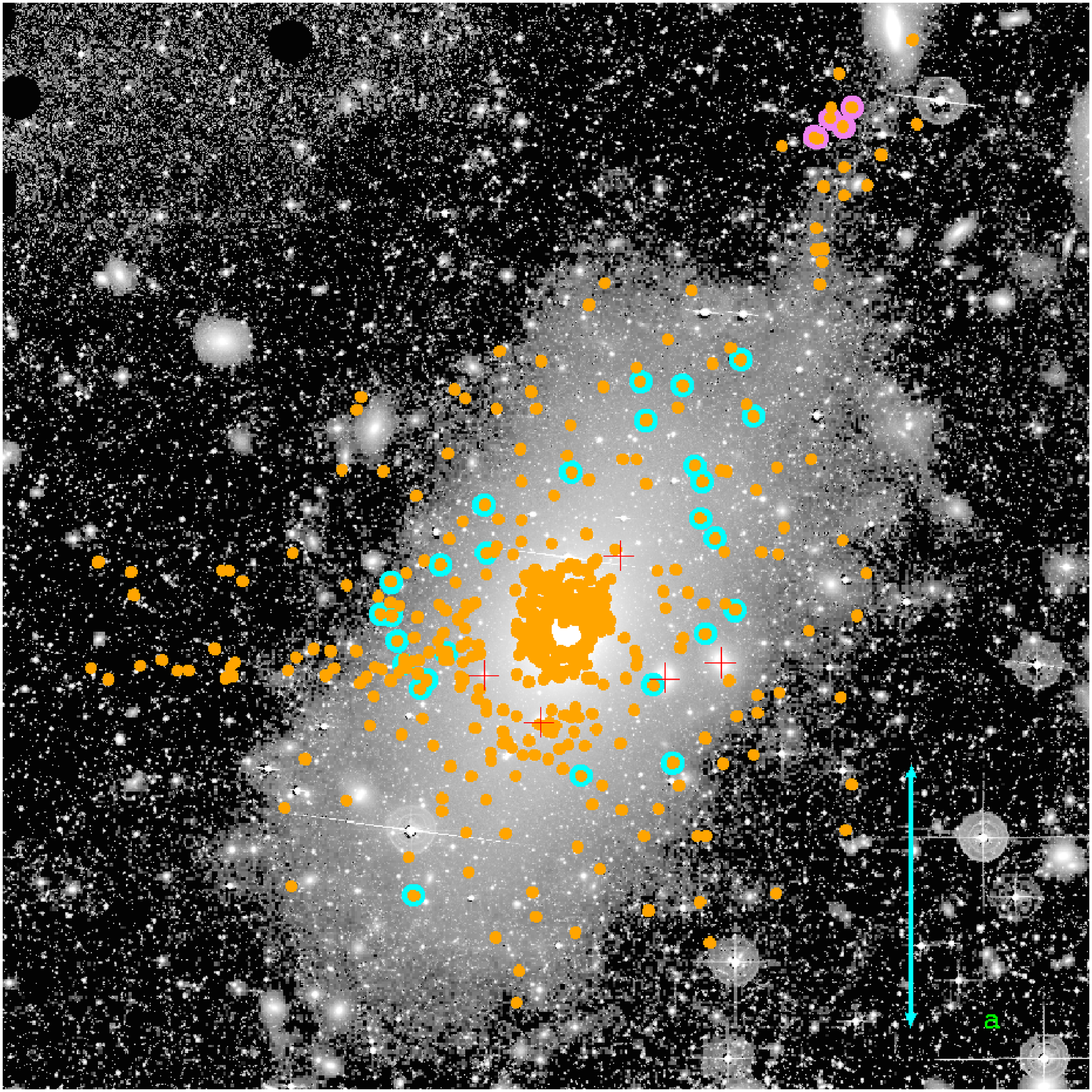}
\hspace{0.1cm}
\plotone{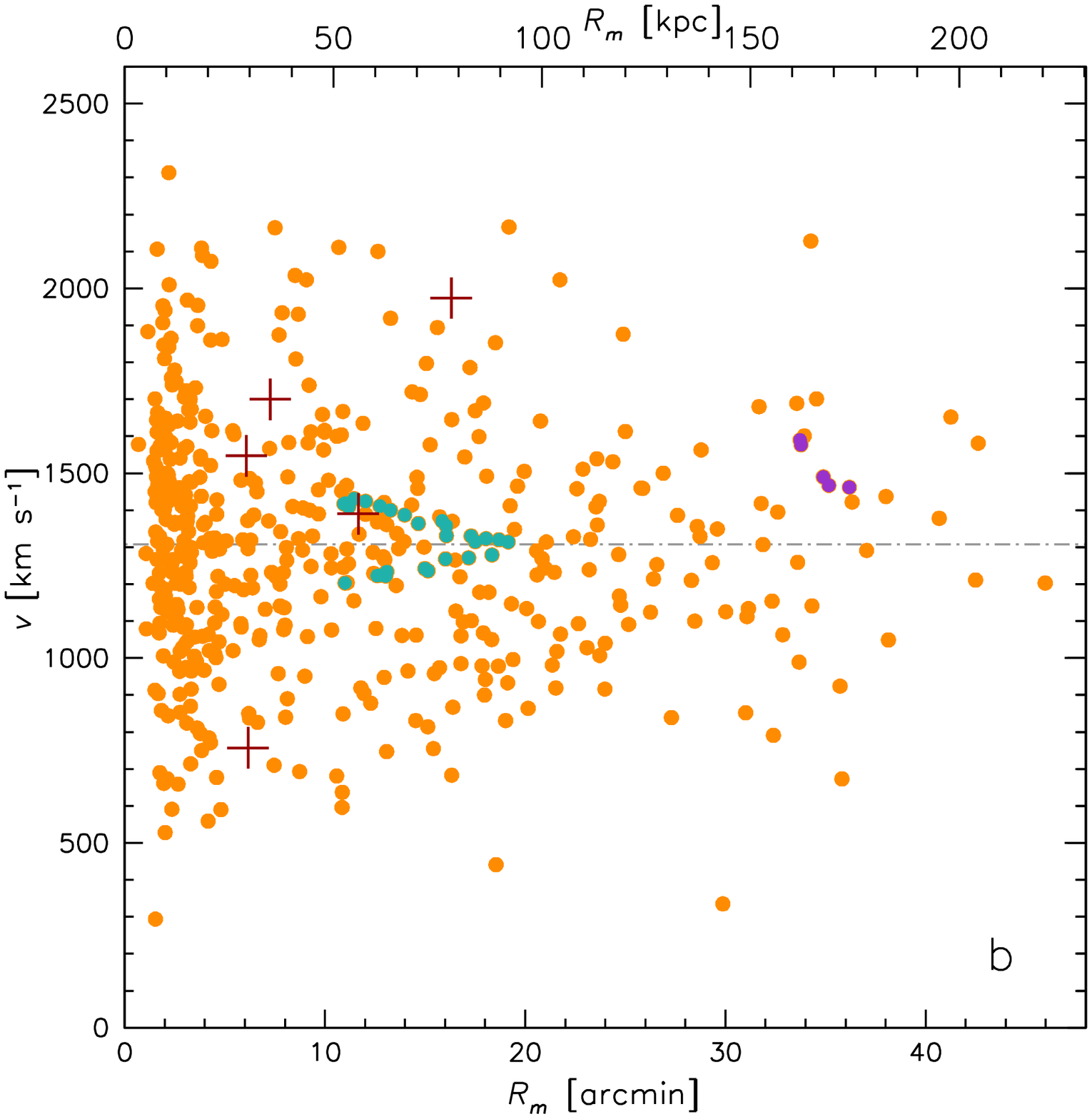}
\figcaption[main]{\label{fig:main}
Overview of spectroscopic GC observations around M87. Orange circles show confirmed GCs,
red crosses show five centrally-located low-luminosity elliptical galaxies, and candidate substructure
members are highlighted with blue and purple.
These are an inner shell (at radii of $\sim$~10--20~arcmin) and an outer stream
(at $\sim$~35~arcmin),
where the most likely members are identified by maximum-likelihood fitting to simple models
(see Section~\ref{sec:max}).
(a): Data locations in positional space, overplotted on a deep optical
image (down to a $V$-band surface brightness of $\mu_V\sim$~28.5~mag~arcsec$^{-2}$;
\citealt{2010ApJ...715..972J}), with a 100 kpc (21 arcmin)
scale illustrated by a bar with arrows. 
(b): The phase space of line-of-sight velocity vs.~\,galactocentric
radius, with the dot-dashed horizontal line representing the systemic velocity of M87
(1307~\kms).
The radius plotted is the intermediate-axis radius (see Section~\ref{sec:rad}).
The velocity
uncertainties are not shown, but are generally smaller than the point sizes in the plot. 
The satellite galaxy velocities are from Hypercat \citep{2003A&A...412...45P},
where we note that the value for NGC~4486A was recently dramatically revised by
\citet{2011A&A...528A.128P}.
}
\end{figure*}

\begin{figure*}
\epsscale{0.56}
\plotone{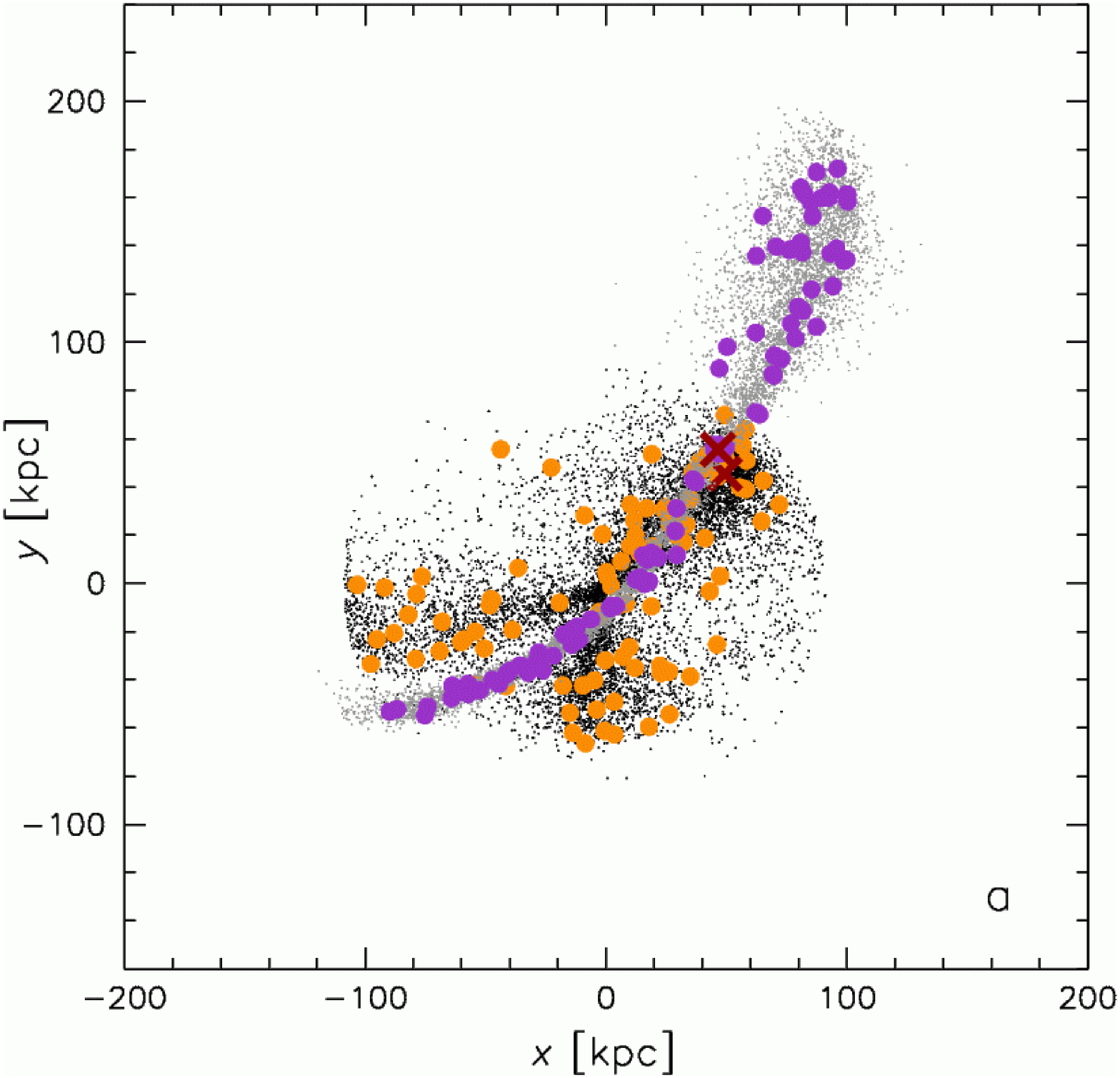}
\hspace{0.1cm}
\plotone{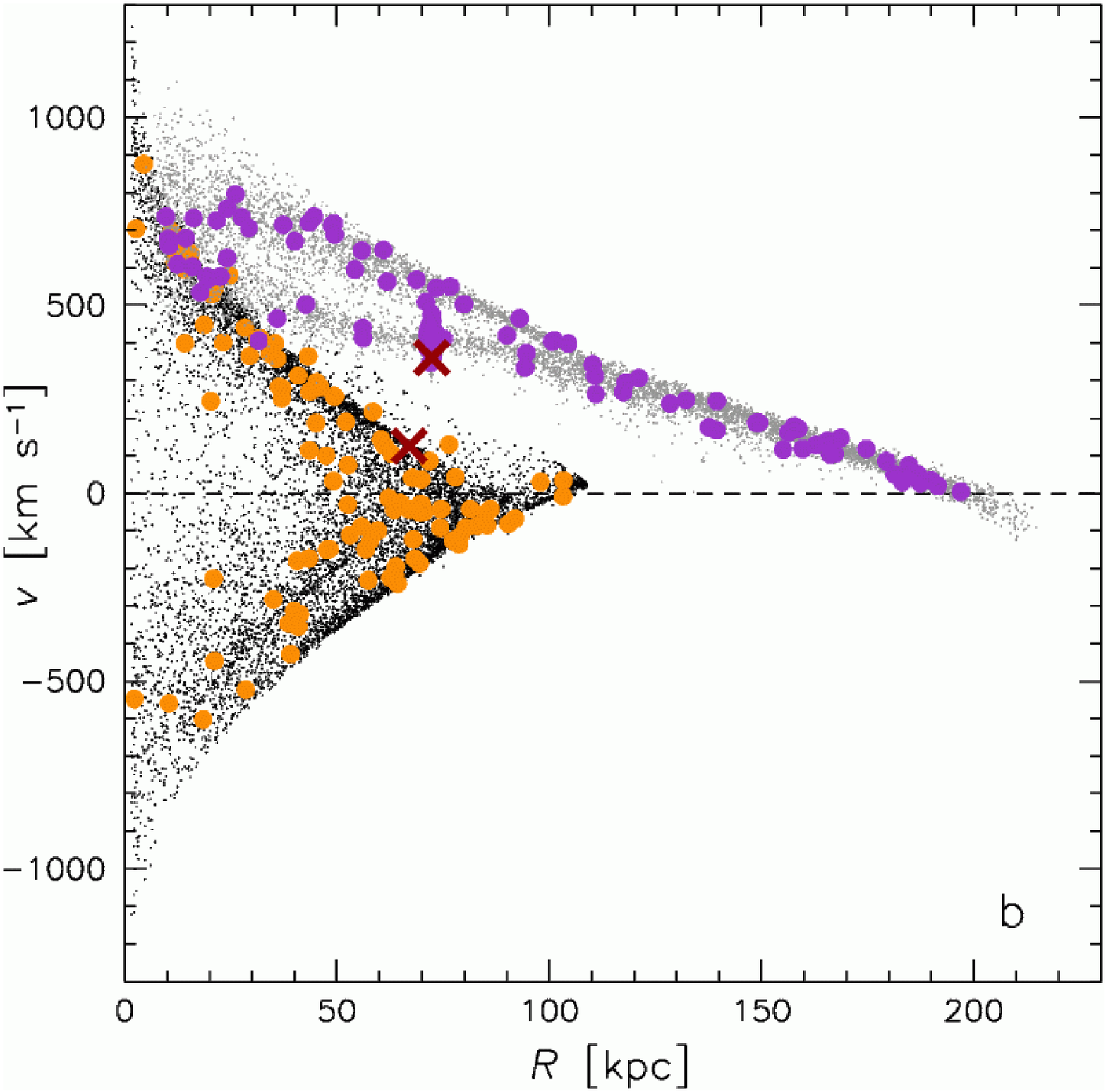}
\figcaption[shell7sshell7r]{\label{fig:sims}
Simulations of galaxies falling into an idealized central cluster potential,
As in Figure~\ref{fig:main}, both positional space (a) and phase space (b) are shown.
In each simulation, subsamples of $100$ and $10^4$ particles
are plotted as large and small dots, respectively.
The red $\times$ symbols show the locations of the progenitor nuclei.
Two independent simulations are superimposed,
representing initial apocentric distances of 90 and 200~kpc
(using small black and large orange dots, and small gray and large purple dots,
respectively).
These two cases are intended to be qualitative analogues to the M87 shell and stream
(Figure~\ref{fig:main}), with the 0.2\% particle subsamples as ``GCs''
(with small measurement errors added).
The snapshots shown correspond to 3.5~Gyr ($\sim$~10--20 dynamical times) after the initial 
apocenters.
In the first case, there are 
sharp features in phase space even when
the tidal debris is well mixed in positional space (which happens more rapidly
for a smaller initial apocenter).
See Section~\ref{sec:sim} for further details.
}
\end{figure*}


Our new spectroscopic data thus provides a total of 468 velocities of M87 GCs
(we exclude one large-radius object probably associated with the Virgo galaxy
NGC~4461, but retain another that may be bound to the small elliptical NGC~4478).
Although a great deal of additional spectroscopic data of M87 GCs are available in the
literature, 
we do {\it not} use most of the previous data, since the
typically large measurement errors impede the study of cold substructures.
Instead, we incorporate only 20 high resolution measurements from the literature
(with median uncertainties of 4~\kms ; \citealt{2005ApJ...627..203H,2007PhDT.........4H,2007AJ....133.1722E}). Note that our treatment of previous data differs slightly from the
convention of S+11, who incorporated all data published since 2003.

\begin{figure*}
\epsscale{0.56}
\plotone{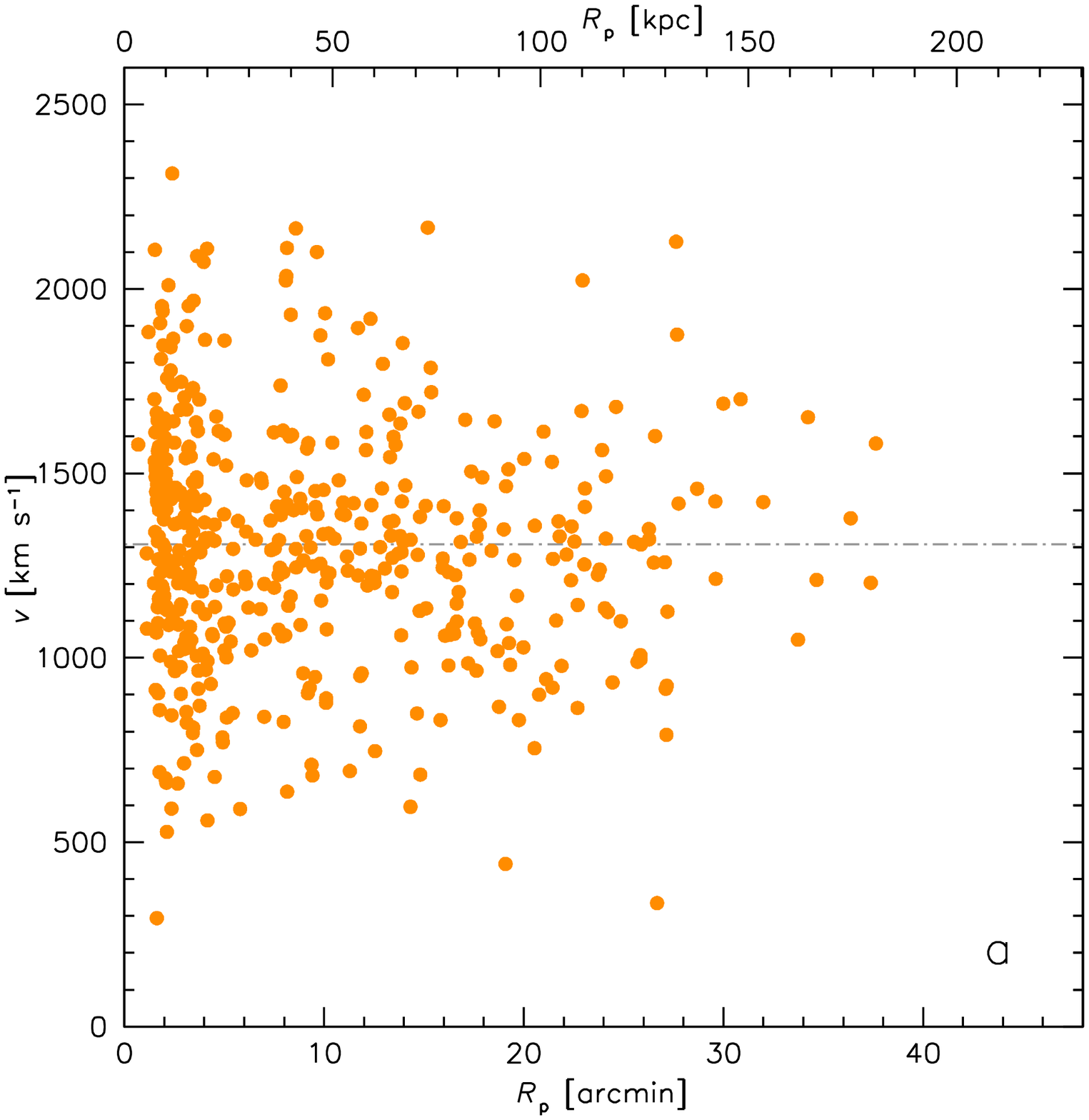}
\hspace{0.1cm}
\plotone{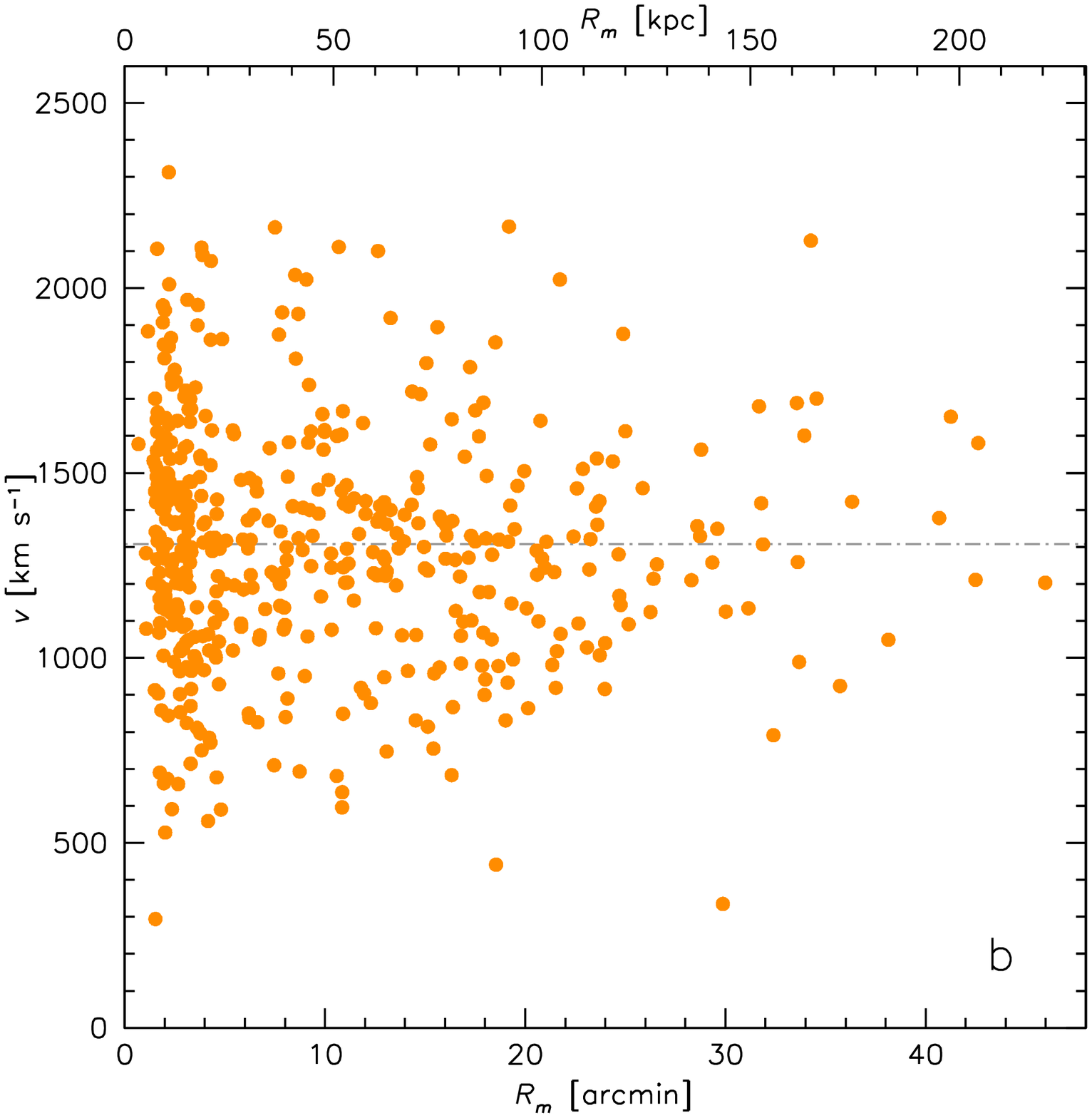}
\figcaption[radii]{\label{fig:radii}
Radius-velocity phase-space of GCs in M87,
omitting the objects in the outer stream region, where
different conventions for the galactocentric radius are used.
(a): simple radius.  (b): ellipse-based radius 
(see Figure~\ref{fig:main}(b), and the main text for details).
The ellipse-based radius causes the shell-like feature to appear more distinct.
}
\end{figure*}

The combined literature and new data provide a total sample of 488 GC line-of-sight velocities
around M87, extending from its central regions out to 200~kpc in its halo,
where the median velocity uncertainties are 14~\kms.
This contrasts with most previous work on GC kinematics beyond the Local Group,
with typical velocity errors of $\sim$~30--100~\kms.
Our GC sample also has 
an extremely low rate of contamination from foreground stars and background 
galaxies.\footnote{Our sample of ``GCs'' includes
some mysterious objects termed ultra-compact dwarfs, which have
sizes of $\sim$~10~pc or more, in contrast to classical GCs with sizes of a few pc. 
Whether or not there are two or more classes of object, or a continuum of GCs
with different properties, is a major scientific question that we cannot resolve here
(see \citealt{2011AJ....142..199B}).
Therefore although we know for a small subset of our spectroscopic sample that the
sizes are unusually large, we retain these in any case as tracer particles in the halo
(excluding only S923 which has a peculiar morphology).
}
This is the largest GC data-set of this quality in any galaxy,
providing an unprecedented opportunity for 
exploring phase space for formational relics.

We present a basic overview of the GC observations in Figure~\ref{fig:main}(a,b),
with a plot of positions in real-space along with a phase-space plot of velocity
vs.~\,galactocentric radius.
It is important to note that most of the {\it spatial} substructure apparent in the GC data
is just a product of inhomogeneous spectroscopic sampling:
e.g., the ``gap'' at a radius of $\sim$~7~arcmin, 
and elongated extensions to the far East and Northwest.
Our analyses will be relatively insensitive to such inhomogeneities
since we will focus on the distribution of velocities as a function of 
galactocentric radius.  Furthermore, the sampling effects will be built in to
our statistical tests for substructure.

Most of the GCs show a broad spread of velocities reflecting a dynamically
relaxed system, but there are two sharper features that stand out in phase-space.
The first is a small elongated ``stream'' of GCs at 150 kpc, associated with the
known stellar filament,
The second is a large chevron-shaped overdensity between 50 and 100 kpc,
which we denote the ``shell''.
It is an enormous structure spread out over $\sim$~$10^4$~kpc$^2$ on the sky, 
and is not identifiable with any previously known photometric feature (Figure~\ref{fig:main}(a)).
The statistical significance of these two substructures will be
demonstrated in Sections~\ref{sec:shell} and \ref{sec:outer},
where we also find independent support from their distinct GC metallicity distributions.

Both features are most naturally explained as the shredded remnants of infalling galaxies,
which we will investigate in Section~\ref{sec:sim} using numerical simulations.
To orient the discussion of the rest of the paper, we present illustrative results
from these simulations in Figure~\ref{fig:sims}, which broadly reproduce some of the
general features of the data.

\section{Inner shell analysis}\label{sec:shell}

Here we characterize the properties of the large shell 
identified in Figure~\ref{fig:main}(b),
and attempt to establish its statistical significance.
We exclude the outer-stream data set from these analyses,
thereby focusing on the
main part of the data set which has relatively unbiased spatial sampling.

In Section~\ref{sec:rad} we compare the use of different conventions for radial distance.
We carry out a group-finding analysis and quantify the entropy in the phase-space data
in Sections~\ref{sec:group} and \ref{sec:ent}, and 
perform maximum-likelihood fitting of simple shell models in Section~\ref{sec:max}.
In Section~\ref{sec:color} we analyze the shell GC color distributions, and in
Section~\ref{sec:prog} consider the implications for the shell progenitor.

\subsection{Radius convention}\label{sec:rad}

One of the greatest challenges in detecting and analyzing substructures in extragalactic
systems is decoding the multi-dimensional information contained in observations.
The dynamical definition of a substructure is that it occupies a distinct region in
some phase-space of physical parameters that are linked to orbital trajectories.
This could be a space defined by the integrals of motion, such as
the energy and angular momentum in a spherical gravitational potential,
or it could be
the six-dimensional observational phase-space of position and velocity.

To pick out a substructure from a background of unrelated structures,
it helps to link the observations to orbital solutions.
This can be tricky enough with 6-D data as available for some objects
within the Milky Way, but in distant galaxies,
only three dimensions are normally observable
(two of position, one of velocity).
This ``projection'' of phase space typically smears out the orbital tracks
and can make substructure inferences degenerate.

If one already knew a priori the orbital characteristics of a substructure,
one could apply an optimal coordinate transformation to the observations that
would maximize the contrast in a ``reduced'' phase-space plot.
One of the most basic reductions to try is converting the two spatial positions
to a galactocentric radius, since in the limit of a spherical potential and
either a radial or a circular orbit, the particles will outline a characteristic
chevron-shaped track in the phase-space of radius and line-of-sight-velocity 
(cf. Figure 1 of \citealt{1997ApJ...488..702R}).

Real galaxies are not perfectly spherical, and in dynamical modeling there
is a standard modification to the above transformation that corrects for
flattening to first-order.  This is to use
an ellipse-based circular-equivalent or ``intermediate-axis'' radius.
If $x$ and $y$ are projected coordinates along a galaxy's major and minor axes,
respectively, then the simple projected radius is:
\begin{equation}
R_{\rm p} = \sqrt{x^2+y^2} .
\end{equation}
The equivalent radius, given an axis ratio $q$, is:
\begin{equation}
R_m = \sqrt{qx^2+\frac{y^2}{q}} .
\end{equation}
This convention allows one to conserve the area (and approximately the
mass) enclosed within an elliptical isophote when transforming to a
circularized one-dimensional model.
It is also equivalent to recovering a constant radius $r$ in three-dimensions
for an inclined thin disk.

The next question is what value for $q$ to use, as both the flattening of the
observable objects and of the unobserved gravitational potential could be important.
Unfortunately,
in the outskirts of M87, neither of the flattenings is well constrained (i.e., for
the GC system or for the potential which is dominated by dark matter).
It was found in S+11 (figure~7) that the GC system has a typical $q\sim0.75$
at $R_m\sim$~2\arcmin--10\arcmin\ ($\sim$~5--50~kpc), 
while possibly decreasing with radius.
At larger radii, the measurement is more difficult, but we have used a
maximum-likelihood technique to estimate a possible range of $q\sim$~0.55--0.75,
at $R_m\sim$~15\arcmin--20\arcmin\ ($\sim$~70--100~kpc).

Our constraints on $q(R_m)$ for the GCs are consistent with the profile
for the {\it stars} derived by \citet{2009ApJS..182..216K}, 
where $q$ declines from $\sim0.9$ at $R_m\sim$~1\arcmin\ to $\sim0.55$ 
outside $R_m\sim$~10\arcmin.
We therefore adopt these stellar results for our default model for the GCs,
for the sake of having smooth, plausible profiles of $q$ and position angle
with radius.\footnote{A more
recent study of M87 with deeper surface photometry found that 
the stellar-light flattening becomes even stronger at large radii, reaching
$q\sim0.4$ at $R_m\sim$~20\arcmin\ \citep{2010ApJ...715..972J}.
This is clearly inconsistent with the flattening of the GC system,
and it would be inappropriate to use this profile for our analyses.
Both the Kormendy and the Janowiecki profiles agree on a position angle
of $\sim 150^\circ$ at large radii, while our GC analysis is more suggestive
of $\sim 105^\circ$--$135^\circ$. We do not ascribe much credence to this
inconsistency, given the difficulties with background contamination
and spatial incompleteness at large radii, and given the much better consistency
between stars and GCs at small radii.}


Figure~\ref{fig:radii} then compares radius-velocity phase-space plots 
for the M87 GCs, using both the simple and the elliptical radius.
The cold shell at $\sim$~50~kpc appears considerably sharper when using the elliptical radius,
and we suspect that it may be easier to find substructures
in other data sets if the $R_m$ convention is used.

We will adopt this $R_m$ transformation as our default
for the analyses used in the rest of this paper,
while issuing some caveats.
One is that it is not entirely clear from a theoretical standpoint
why the transformation should work:
an accretion event generally defines its own 
orbital plane that will not necessarily align with the apparent shape of 
the rest of the galaxy.
It could be that we are seeing an effect
related to the disrupted accretor being highly spread out in the host potential.

The other concern from an observational standpoint is that the $q(R_m)$ profile for
the GCs is not well determined, and fairly small changes in the radius-transformation
(of $\sim$~1\arcmin) would noticeably affect the inferred morphology of the substructure,
causing it to lose some of its apparent sharpness in phase-space.
Such ``blurring'' would be expected if there is a genuine cold feature, and if one deviates
from an optimal coordinate transformation as discussed above
(we have verified this effect using mock data simulations that we will discuss later).
However, we consider it more difficult to artificially
{\it sharpen} phase-space features through a transformation error.

In summary, the choice of the radius in the phase-space diagram
is a source of systematic uncertainty
that is difficult to quantify.  Where relevant in our analyses, we will also
consider the possibility that the substructure is more amorphous than in
our default model (e.g., as in Figure~\ref{fig:radii}(a)).

\subsection{Group finding}\label{sec:group}

\begin{figure}
\epsscale{1.15}
\plotone{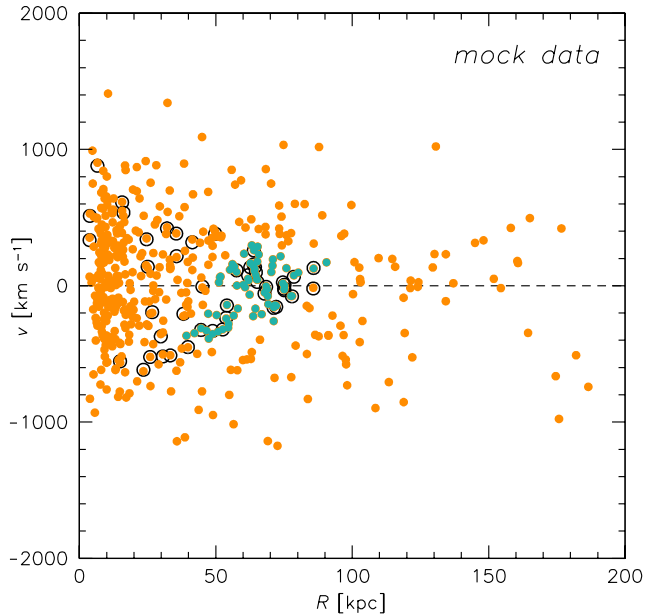}
\figcaption[group2]{\label{fig:group2}
Phase-space group finding in a mock data set
constructed from a simulated accretion event (see Figure~\ref{fig:sims}(b))
superimposed on a randomized population of
GCs. The genuine shell objects are shown as black circles, with velocities that differ slightly from the mock
data set because of simulated observational errors. The general morphology of the substructure is
recovered by the group-finding algorithm (blue dots). 
Compare application to the real data in Figure~\ref{fig:main}(b).
}
\end{figure}

\begin{figure*}
\epsscale{1.15}
\plotone{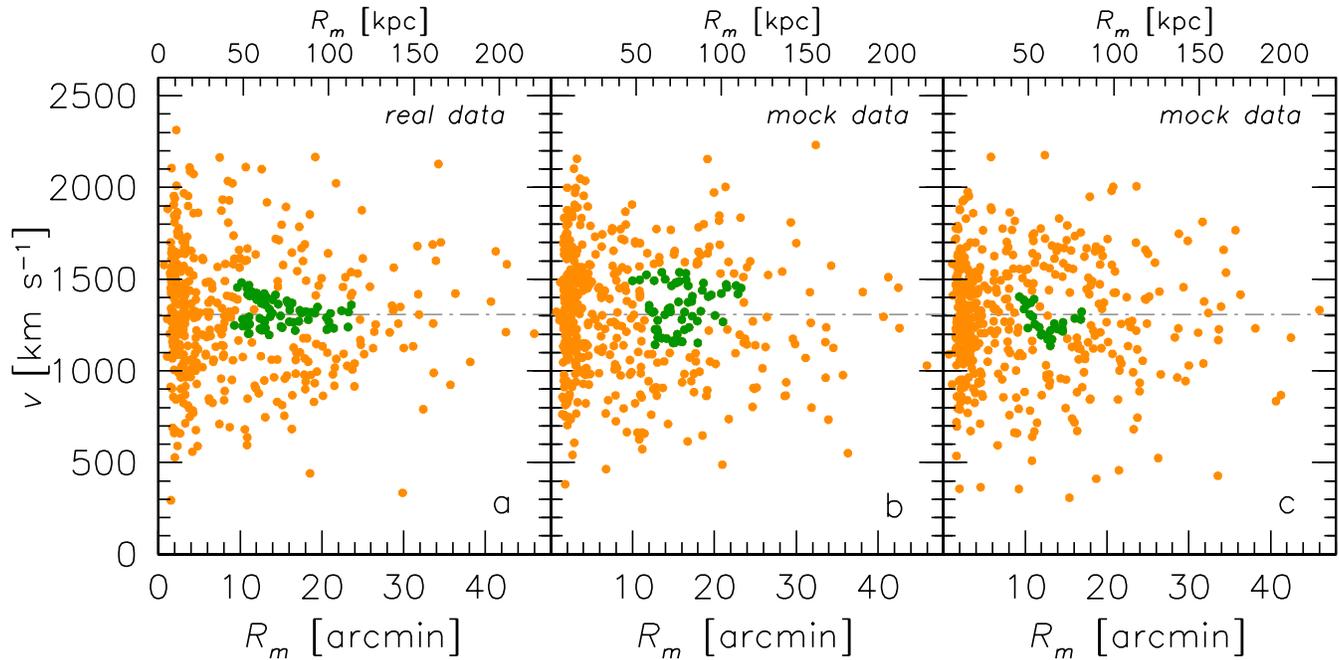}
\figcaption[group]{\label{fig:group}
Group-finding of GCs in phase space.
Orange dots show the parent GC population, while green dots show the
largest identified groups in each data set.
(a): the real M87 data
(where the outer stream region is again omitted).
(b): the mock data set generated from an unstructured model, with the largest ``group'' marked.
(c): a typical mock data set.
The false group detections typically have many fewer objects than in the M87 data,
and lack the clear, fairly symmetric shell-like morphology.
}
\end{figure*}

To ascertain the veracity of the shell, rather than rely only on
the notoriously misled human eye which is good at creating patterns out
of random noise, we examine several different types of models and statistical tests.
This is an aspect that has been underdeveloped in previous work on GC kinematics
around early-type galaxies.

Our first step is to use a group-finding algorithm, based on a standard friends-of-friends
approach (e.g., \citealt{2003ApJ...589..790P,2009ApJ...699.1518R}).
We define a distance between objects of:
\begin{equation}
\Delta s = \sqrt{\left(\frac{\Delta \log_{10}R_m}{w_R}\right)^2 + \left(\frac{\Delta v}{w_v}\right)^2} ,
\end{equation}
where $w_R$ and $w_v$ are radius and velocity scale factors that act inversely as weights.
Note that our adopted two-dimensional phase-space metric differs from the
three-dimensional $(x,y,v)$ metric commonly used, which is suited for
very localized substructures that have had little mixing, but we find that it
does not work well for structures like shells and long streams that have elongated
``diagonal'' morphologies in phase-space.
Using the 3-D metric may have caused large-scale substructures to be missed in the past.

A group is defined as a set of objects separated by less than a linking length
$\Lambda$ in phase space, which is in turn defined relative to
the mean interparticle spacing $\langle\Delta s\rangle$
by a dimensionless parameter $\lambda$.
We then have the criterion:
\begin{equation}
|\Delta s| < \Lambda = \lambda \langle\Delta s\rangle .
\end{equation}

The next step is to choose the free parameters ($w_R,w_v,\lambda$) in the algorithm,
where the objective is to maximize the sensitivity to real features in
the data while minimizing the incidence of false groups and linkages.
As a starting point, we construct mock data sets based on simulations of
an accretion event (which we will discuss in detail later).
To represent the shell, we select $47$ particles, a number that is
chosen to mimic the real data set, in the sense that the simulation has 27 particles
in the same radial region where we
find a shell in the real data (which has an estimated 27 GCs as we will see later).

We then fill out the rest of the mock data set by
adding 422 GCs drawn from a Gaussian velocity distribution with 
$\sigma=$~450~\kms, and with positions corresponding to those in the real data set,
but randomized by 20\% in order to reduce any lingering spatial substructures.
Throughout this paper, we will construct similar, smooth ``background'' GC
populations, which would be unrealistic at some level if the halo were 
rich in substructures of various masses and stages of dissolution.
In the future, one could consider constructing background distributions based on
a range of theoretical models for halo accretion histories, but for the purposes
of this paper, we focus primarily on the null hypothesis of a halo
with {\it no} substructure.
A caveat to then keep in mind is that the large ``shell'' which
we recover could in principle consist of two or more distinct substructures that
overlap in the observable phase-space and conspire to masquerade as a single structure.

Note that the value of $\sigma=$~450~\kms\ is larger than the $\sim$~300~\kms\
of the real data, which is done to be
consistent with the simulations' circular velocity and larger shell velocity width.
We will later rescale the resulting group-finding parameters to suit the real data.
The exact velocities of the mock data set are also convolved with
small errors corresponding to the observational uncertainties in the real data.

\begin{figure*}
\epsscale{0.38}
\plotone{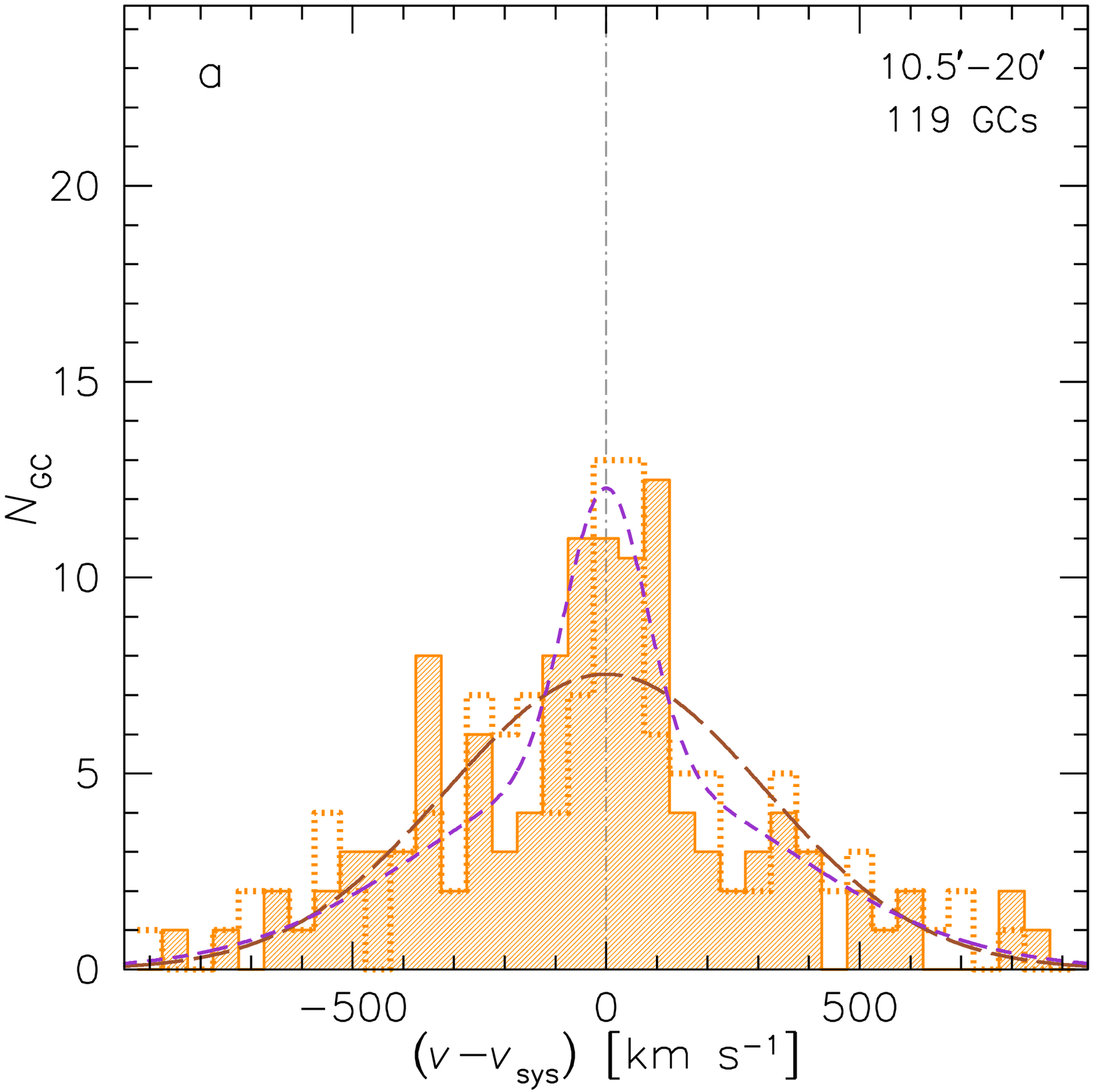}
\plotone{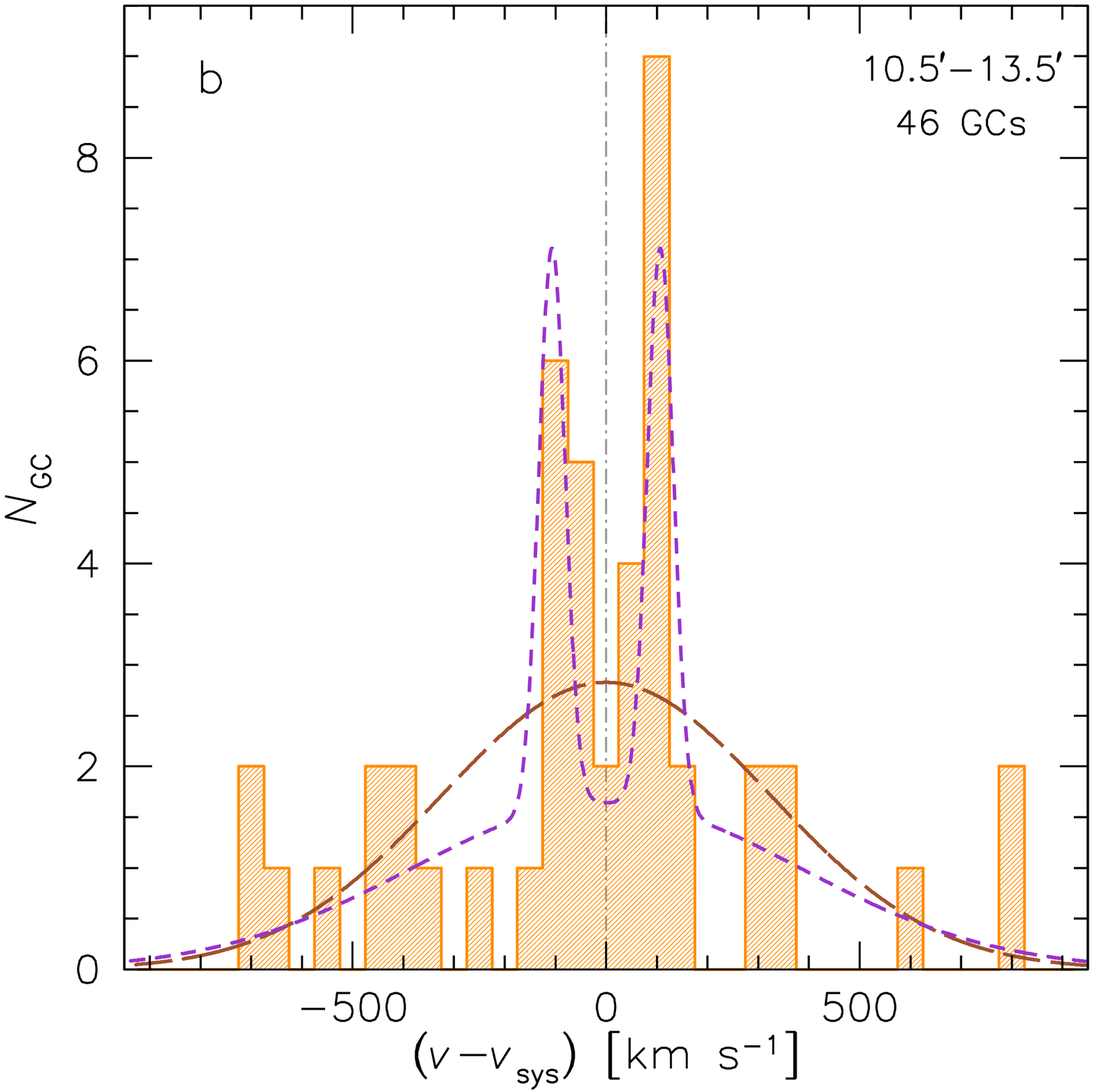}
\plotone{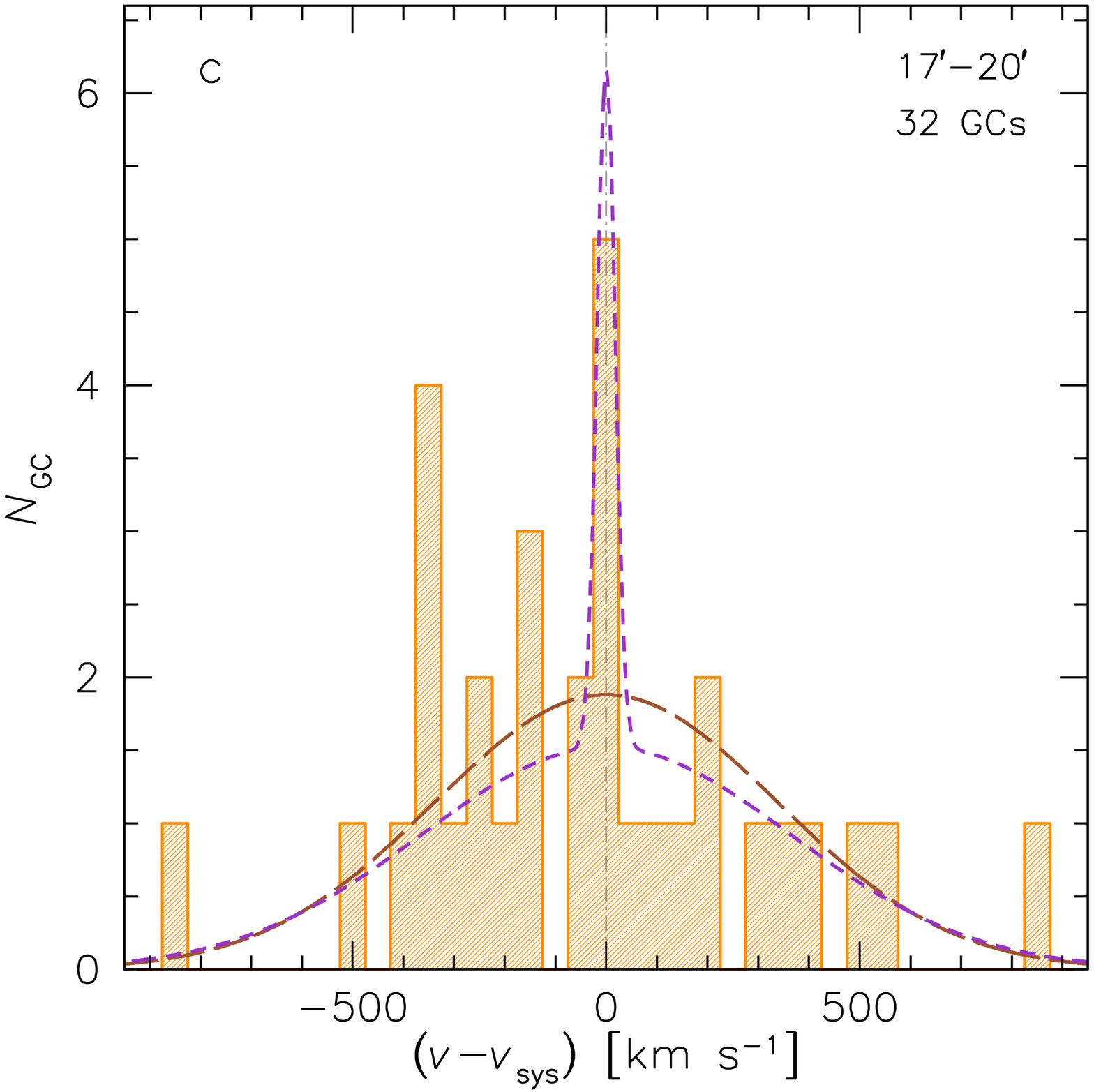}
\figcaption[losvds]{\label{fig:losvds}
Line-of-sight velocity distribution of GCs in the shell region around
M87 ($R_m$=10.5--20~arcmin).
The systemic velocity of 1307~\kms\ has been subtracted from the data.
The filled histograms show the data, while the long- and short-dashed curves
illustrate single and double Gaussian model fits, respectively 
(with additional smoothing to account for measurement errors).
The panels show different radial regions.
(a): entire region.  A mock data set is also shown as a dotted histogram.
(b) and (c): finer radial bins as labeled.
The shell-like substructure is clearly visible as an extra, low-dispersion component 
in the overall velocity distribution.  At smaller radii, it appears as a symmetric
double peak, and at larger radii as a narrow single peak.
}
\end{figure*}

We find optimum parameters for the simulated data set of
$w_R=$~0.035~dex, $w_v=$~50~\kms, $\lambda=0.07$,
with acceptable variations around these values at the $\sim$~20\% level.
The group-finding results are shown in Figure~\ref{fig:group2}:
a group of 70 objects is identified, which reproduces the broad morphology of the
shell's limb (cf. Figure~\ref{fig:sims}(b)) and recovers many of the 
genuine shell members along with many false associations.
This exercise provides some confidence in the simple group-finding
approach for recovering cold phase-space substructures that comprise only $\sim$~10\% of 
a sample.   However, the individual objects assigned to the substructure can include
a high degree of contamination.
This fact, combined with the potential sensitivity of the group-finding parameters
to the details of the data sampling and of the substructure properties,
motivates exploring alternative algorithms (which we do below) as well as the future
development of more efficient phase-space group-finders.

Now after rescaling the foregoing simulation-based parameters, 
we adopt the following for use with the real data:
$w_R=$~0.035~dex, $w_v=$~30~\kms, $\lambda=0.07$.
The group-finding algorithm then recovers a shell-like substructure with 68 members,
around half of which may be false associations,
which we illustrate with Figure~\ref{fig:group}(a).
Given the uncertainty in the radius convention discussed in Section~\ref{sec:rad},
we also carry out the same procedure while using the simple radius $R_{\rm p}$.
We find in this case a group with 80 members, demonstrating that our detection
of a large phase-space substructure in M87 is not strongly sensitive to the radius convention.

To further evaluate the statistical significance of this shell, we carry out 
Monte Carlo simulations of underlying smooth data sets.
We generate 100 mock GC data sets, using in each case the same radial locations
as in the real data, but substituting a random velocity drawn from a 
Gaussian distribution with $\sigma=324$~\kms\
(which is the dispersion found in the real data).

We then run the group-finding algorithm with the same parameters as for the
real data.
The median largest group size found (outside the crowded central regions where
the group-finding parameters are not optimized) has 30 members,
and the largest group of all has 61 members.
Examples of these groups are shown in panels (b) and (c) of Figure~\ref{fig:group},
alongside the real data in panel (a).
We thus exclude finding a 68-member group by chance at the 99\% level.
This test is somewhat conservative in using the radial locations
of the real GCs in the mock data sets, which may already have some clumpiness.

If we were to include the data from the outer stream region, a stream-like phase-space
feature would indeed be picked up by
the group-finding algorithm, but so would be 6 other small ``substructures'' that
(based on the Monte Carlo simulations above) are probably not real.
The outer stream is analyzed separately in a later section, using a slightly
different algorithm because of the much smaller number counts and the importance
of two-dimensional spatial clumping.

\subsection{Entropy}\label{sec:ent}

The next test uses a very general entropy metric $S$ that measures the clumpiness
of data in a set of partitions:
\begin{equation}
S = - \sum_i N_i \ln N_i ,
\end{equation}
where $N_i$ is the number of objects in each partition \citep{1999Natur.402...53H}.
We consider the 4--20~arcmin region of interest and adopt
a gridded partition of (radius by velocity) with
size (2~arcmin $\times$ 80~\kms).
We measure $S$ and compare it to 1000 mock data realizations using
Gaussian velocity distributions.
We find that this region is significantly substructured at the 1~$\sigma$ level.
This is a conservative test because again we have used radii from the real data,
and because the rectangular
partitioning is not optimized to pick up extended diagonal features as the
data appear to show.

\subsection{Maximum likelihood modeling}\label{sec:max}

We next carry out a set of parametric model fits to the data in
the shell region, which both allows us to evaluate the significance
of the substructure and to quantify its properties.
We will in general focus on the $R_m=$~10.5--20~arcmin
(50--95~kpc) region where the shell is most apparent by eye.

\begin{figure*}
\epsscale{0.85}
\plotone{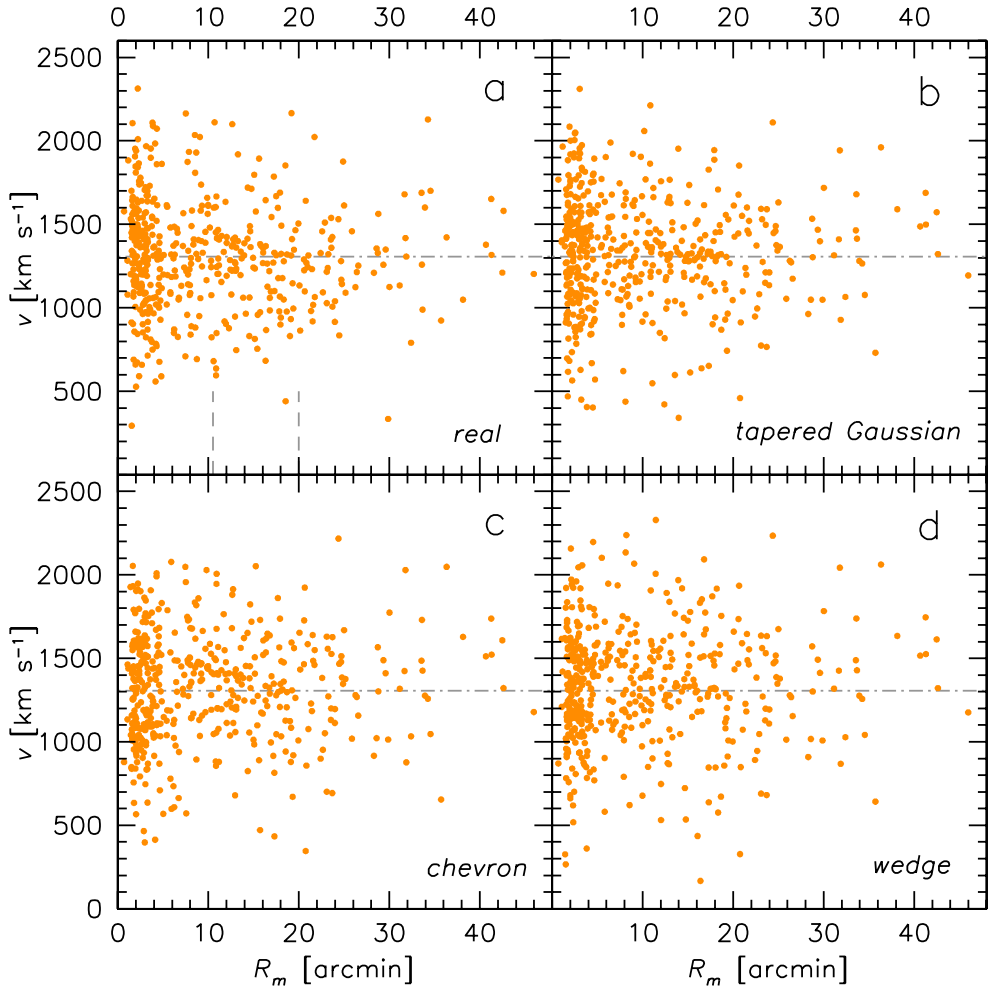}
\figcaption[phase]{\label{fig:phase}
Simple models for GC substructure in phase-space.
(a): the real data.
Vertical dashed lines show the radial range where the models are fitted.
(b): mock data set
using a tapered double-Gaussian model.
(c): mock data set using a Gaussian+chevron model.
(d): mock data set using a Gaussian+wedge model.
Some single-Gaussian models can also be seen in Figure~\ref{fig:group}(b,c) for comparison.
The real data appear to have a phase-space morphology intermediate to these
various two-component models, but closest to the chevron-based model.
}
\end{figure*}

\begin{figure*}
\epsscale{0.85}
\plotone{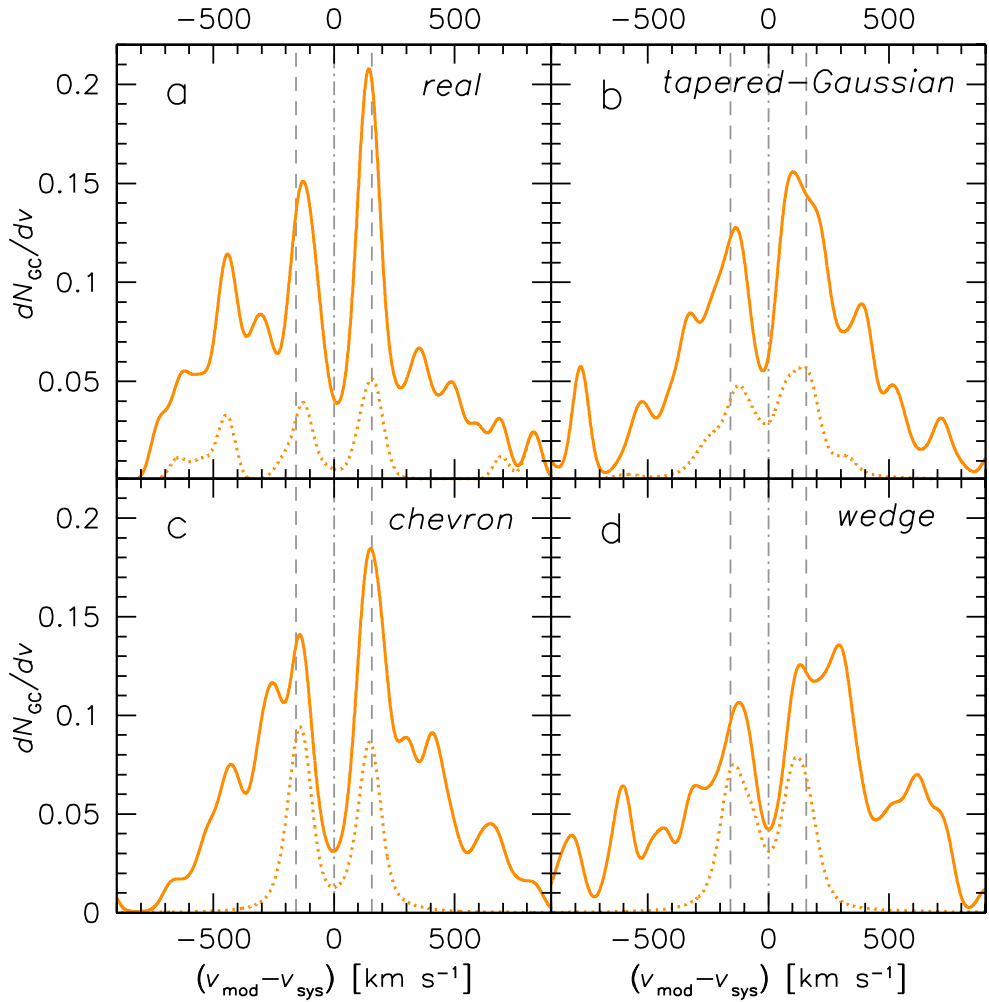}
\figcaption[losvds2]{\label{fig:losvds2}
GC line-of-sight velocity distributions, after projection
with a diagonal velocity model (see main text), smoothed by 30~\kms.
The solid curves shows the overall distribution in each panel, with
sample subpopulations shown as dotted curves.
The vertical dashed lines show $\pm$~160~\kms\ model reference points.
(a): real data around M87 from the $R_m=$~10.5--20~arcmin region, 
with the intermediate-color subpopulation also shown as a dotted curve.
(b): mock data set comprised of a wide Gaussian and a second narrow Gaussian
with a tapered radial dependence (which is shown as the dotted curve).
(c): Gaussian+chevron mock data set, with the chevron subsample also shown.
(d): Gaussian+wedge mock data set, with the wedge subsample also shown.
Only the chevron-based data set has a sharp double-peaked velocity distribution
as in the real data, after projecting with the diagonal model.
}
\end{figure*}

The basic idea is that given a model for the line-of-sight velocity
distribution (LOSVD) $dL/dv$ at a given measurement location $R_i$, we
quantify the likelihood ${\cal L}$ of a velocity measurement $v_i\pm\Delta v_i$:
\begin{equation}
{\cal L}\left(v_i,R_i\right) = \int_{-\infty}^\infty \, \frac{dL}{dv}\left(v,R_i\right) \, e^{-\frac{(v-v_i)^2}{2(\Delta v_i)^2}} \, dv 
\end{equation}
\citep{2001ApJ...553..722R}. 
The likelihood can be considered as equivalent to a $\chi^2$ statistic after the operation
$-2 \ln {\cal L}$.
For a Gaussian model $dL/dv$ with an intrinsic dispersion $\sigma$
centered on a systemic velocity $v_{\rm sys}$, 
the maximum likelihood model
fitting involves minimizing the simple function:
\begin{equation}
-2 \ln {\cal L}= \frac{(v_i-v_{\rm sys})^2}{\sigma^2+(\Delta v_i)^2}+\ln\left[\sigma^2+(\Delta v_i)^2\right] 
\end{equation}
 \citep{1986ApJ...305..600N,2006A&A...448..155B}.

Before constructing detailed models of the system in phase-space, we
consider the simplest possible models, lumping all of the shell-region
velocity data together, with no spatial information.
Using the likelihood formalism above, 
we fit a single Gaussian to these data and find a dispersion of 315~\kms.
The implied LOSVD for this model is shown along with the data in 
Figure~\ref{fig:losvds}(a).

It is apparent that a single Gaussian is not a good representation of the
data, so we next try a double-Gaussian model.
We find a best fit where the hot component has a dispersion of 
$361\pm34$~\kms, and the cold component comprises a fraction
$f_{\rm s}=0.25\pm0.12$ of the GCs, with a dispersion of $86\pm34$~\kms\
(where Monte Carlo simulated data sets were used to estimate these uncertainties
and to correct for slight fitting biases).
We show a mock data set of this solution in Figure~\ref{fig:losvds}(a), 
to illustrate the level at which noise can influence the LOSVD.

By eye, the second Gaussian appears required for adequately fitting the
data, and to quantify this statement we carry out a Monte Carlo analysis
using a series of single-Gaussian mock data sets.
We find that $f_{\rm s}$ values as high as 0.24 are fitted by chance 
in only 15 out of 1000 simulations, while most of these false second components
are hot, with only 1 out of 1000 having a dispersion as low as $\sim$~80~\kms.
We conclude that there is definitely
a secondary cold component, with a dispersion of 120~\kms\ at the most,
comprising 13\%--37\% of the GC data.
It is also important to note that this conclusion does not depend on the
detailed morphology of the substructure, and is insensitive to the adopted
radius convention (Section~\ref{sec:rad}), since we have not used radius information
except to define the overall radial bin, and the results are also very similar if
we define this bin using the circular radius $R_{\rm p}$.

Next, given the radial dependence of the shell that is visible by eye 
in Figure~\ref{fig:radii}(b),
we examine the LOSVD in two radial bins: one at the ``mouth'' of
the substructure and one at the ``apex'' (Figure~\ref{fig:losvds}(a,b)).
We again fit double-Gaussian models, allowing for a split around $v_{\rm sys}$
for the cold component.
In the first bin, we find a dispersion of $23^{+15}_{-14}$~\kms\ with 
$f_{\rm s}=0.30^{+0.13}_{-0.12}$
and a peak separation of $214\pm22$~\kms.
In the second, we find $f_{\rm s}=0.12^{+0.10}_{-0.07}$, and
 a dispersion consistent with zero with a 68\% upper limit of 20~\kms.

These binned results together suggest a cold substructure with $f_{\rm s}\sim0.2$
and a dispersion of $\sim$~15~\kms.
The latter result is similar to our rms velocity measurement errors,
which can be challenging to determine accurately, and consequently a dispersion
anywhere in the range 0--35~\kms\ is plausible.

We now model all of the shell-region data jointly in phase space.
Each model consists of a fraction
$f_{\rm s}$ of substructure, and a Gaussian component 
with fraction $(1-f_{\rm s}$) and dispersion $\sigma_1$.
The substructure LOSVD is parametrized in one of three ways:
as a single Gaussian, a nested double Gaussian, or a step function.
These models are all combined with a linear trend of LOSVD width with radius,
producing a tapered Gaussian, a chevron, or a filled wedge, respectively.
The characteristic width vs.~\,radius is:
\begin{equation}
v_{\rm w} = v_{\rm sys} \pm v^\prime \times (R-R_{\rm x}) ,
\end{equation}
where $v^\prime$ is the slope and $R_{\rm x}$ is the apex radius.
The systemic velocity is set as a fixed parameter $v_{\rm sys}=$~1307~\kms.

We fit the various model alternatives to the shell region and find
$f_{\rm s} \sim$~0.22, 0.23, and 0.26 for the tapered-Gaussian,
chevron, and wedge models, respectively.
Examples of mock data generated from the best model fits are shown in Figure~\ref{fig:phase}.
By eye, the real data suggest a feature that is intermediate between
the different models, but most closely related to the chevron,
owing to the strong ``edge'' seen in the substructure.
More quantitatively, Monte Carlo modeling of likelihood fits is unfortunately not
very informative about which model is the best representation of the substructure.
We find that any of the alternative models can readily 
fit mock data based on the other models.

Instead we turn to a somewhat less parametric representation of the data,
using simply a diagonal model 
(with a characteristic slope $v^\prime=$~17.3~\kms\ arcmin$^{-1}$) 
to project each data point to a velocity at a radius of $R_m=10$~arcmin.
This exercise allows us to collapse all of the position-velocity phase-space data
into single one-dimensional LOSVDs analogous to Figure~\ref{fig:losvds},
comparing real data to simple models.

The results are shown in Figure~\ref{fig:losvds2}, where
a 30~\kms\ Gaussian smoothing kernel has been applied in order to capture the 
shape of the distribution while reducing Poisson noise.
Given an apex in all cases
of $R_{\rm x}\simeq$~19~arcmin,
the model projects to
characteristic velocities of $v_{\rm mod}\sim$~160~\kms\
(relative to $v_{\rm sys}$) at 10~arcmin.
Applying the projection to mock data,
two peaks are always generated but other properties are more model dependent:
the peak widths, locations relative to $v_{\rm mod}$, and dips in between.
In particular, a chevron model when added to a Gaussian produces the most distinct, narrow
peaks close to $v_{\rm mod}$.

We find by inspection of Figure~\ref{fig:losvds2} that the real data most closely resemble
the chevron-based model, with narrow, well-defined peaks near $v_{\rm mod}$.
This plot also indicates a two-sided rather than one-sided chevron
(see also Figure~\ref{fig:losvds}).
There is furthermore an additional narrow feature near $-400$~\kms\
that appears stronger than a random fluctuation and suggests the presence of 
another sharp edge in phase-space terminating at $R_m \sim$~30~arcmin
(see also Figures~\ref{fig:radii}(b) and \ref{fig:losvds}),
i.e., perhaps tracing a second, fainter shell.

Although our overall conclusion from LOSVD analyses 
is that a chevron is the best representation for the shell,
it is difficult to be certain given the available data
and the uncertainties in the coordinate transformations to $R_m$.
The real shell probably has
a morphology somewhat intermediate between the idealized models studied here,
and also shows hints of being asymmetric with respect to $v_{\rm sys}$.
For now, we report the best-fit parameters for the favored chevron model, with
uncertainties based on Monte Carlo simulations:
$f_{\rm s} = 0.23\pm0.06$,
$\sigma_1 = 19\pm10$~\kms,
$v^\prime = 15.2\pm1.5$~\kms\ arcmin$^{-1}$,
$\sigma_2 = 358\pm27$~\kms.
Of the 119 confirmed GCs in the $R_m=$~50--95~kpc region, we thus
estimate that 21--35 of them are in the shell.
We highlight the 27 most likely shell members based on this chevron model
in Figure~\ref{fig:main}(b).

Note that the quoted error bars on these fit parameters are the statistical 
uncertainties {\it for an assumed simple chevron model}.  The systematic 
uncertainties from considering alternative models, and from allowing for
different radius conventions (Section~\ref{sec:rad}), are difficult to quantify.
Above we reported also a maximal-ignorance model that disregards
any spatial tapering of the substructure by fitting a simple double-Gaussian,
and returns a very conservative upper limit of 120~\kms\ on the dispersion.

\subsection{Color and metallicity distributions}\label{sec:color}

One great advantage of using GCs as phase-space tracers
is that they provide additional measurable properties beyond position and velocity.
Because each GC is to a good approximation comprised of a homogeneous
stellar population with a single age and metallicity, the principles of 
stellar-subpopulation tagging used in the Milky Way \citep{2008A&ARv..15..145H} can be extended to
more distant systems.
In particular, GCs are now known to be generally very old objects (ages
of $\sim$~10 Gyr), both in M87 \citep{1998ApJ...496..808C} and in other galaxies \citep{2006ARA&A..44..193B}.
Therefore their optical colors are dominated by metallicity variations
rather than by ages, and we can use {\it color} as a good {\it metallicity proxy}.

Rather than solely relying on literature findings on this point, we consider
the color-metallicity link in our own data set.
From the DEIMOS spectra, there are 47 objects with high enough $S/N$ 
to estimate iron-based metallicities [Fe/H] via the CaT line strengths, 
using techniques that we have developed earlier with the same instrumental set-up
\citep{2010AJ....139.1566F,2011MNRAS.415.3393F}.  
Approximate line-strength uncertainties are estimated from the spectroscopic signal-to-noise
ratio, after calibration from Monte Carlo simulated observations of model spectra.

\begin{figure*}
\epsscale{0.56}
\plotone{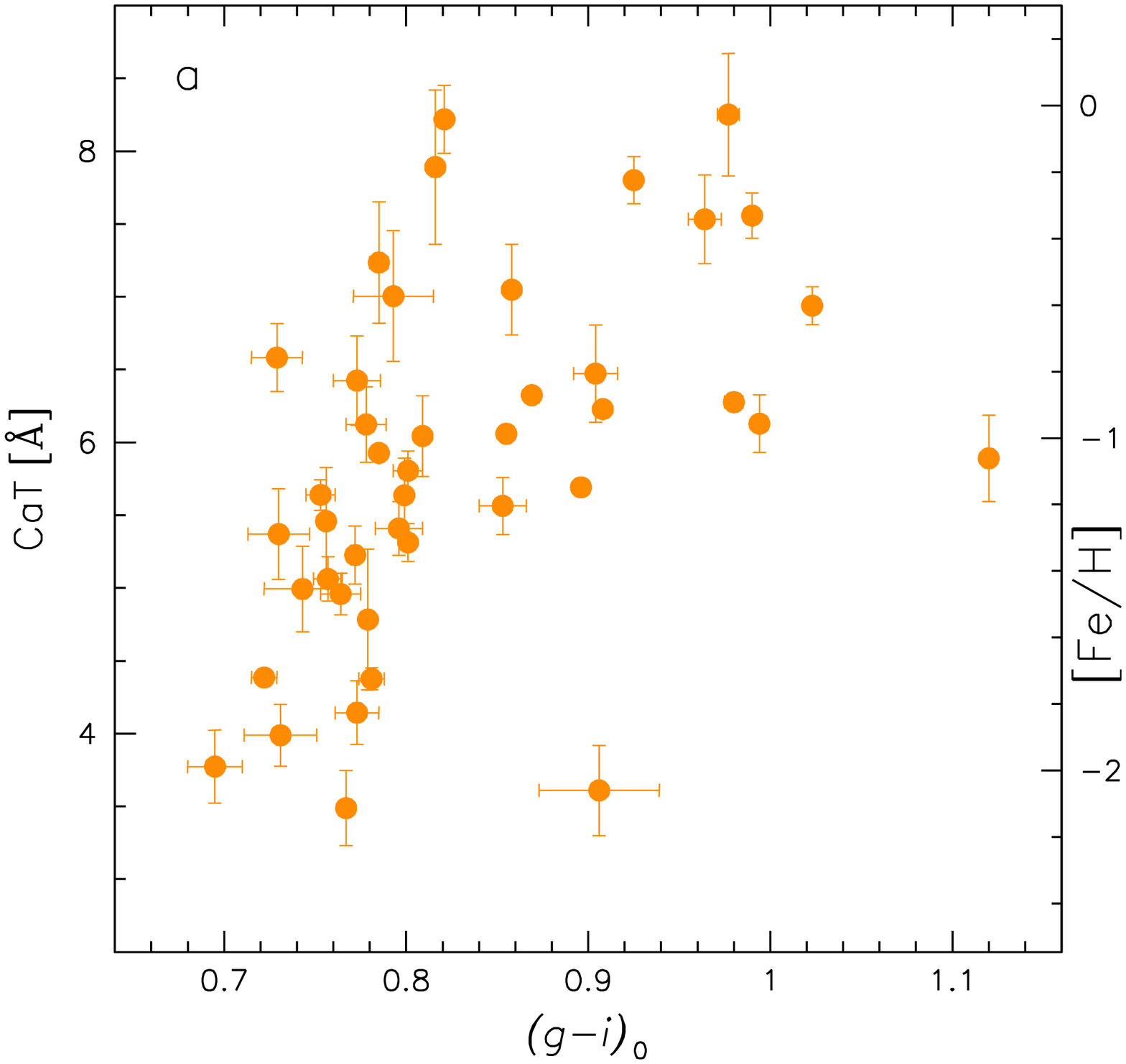}
\hspace{0.1cm}
\plotone{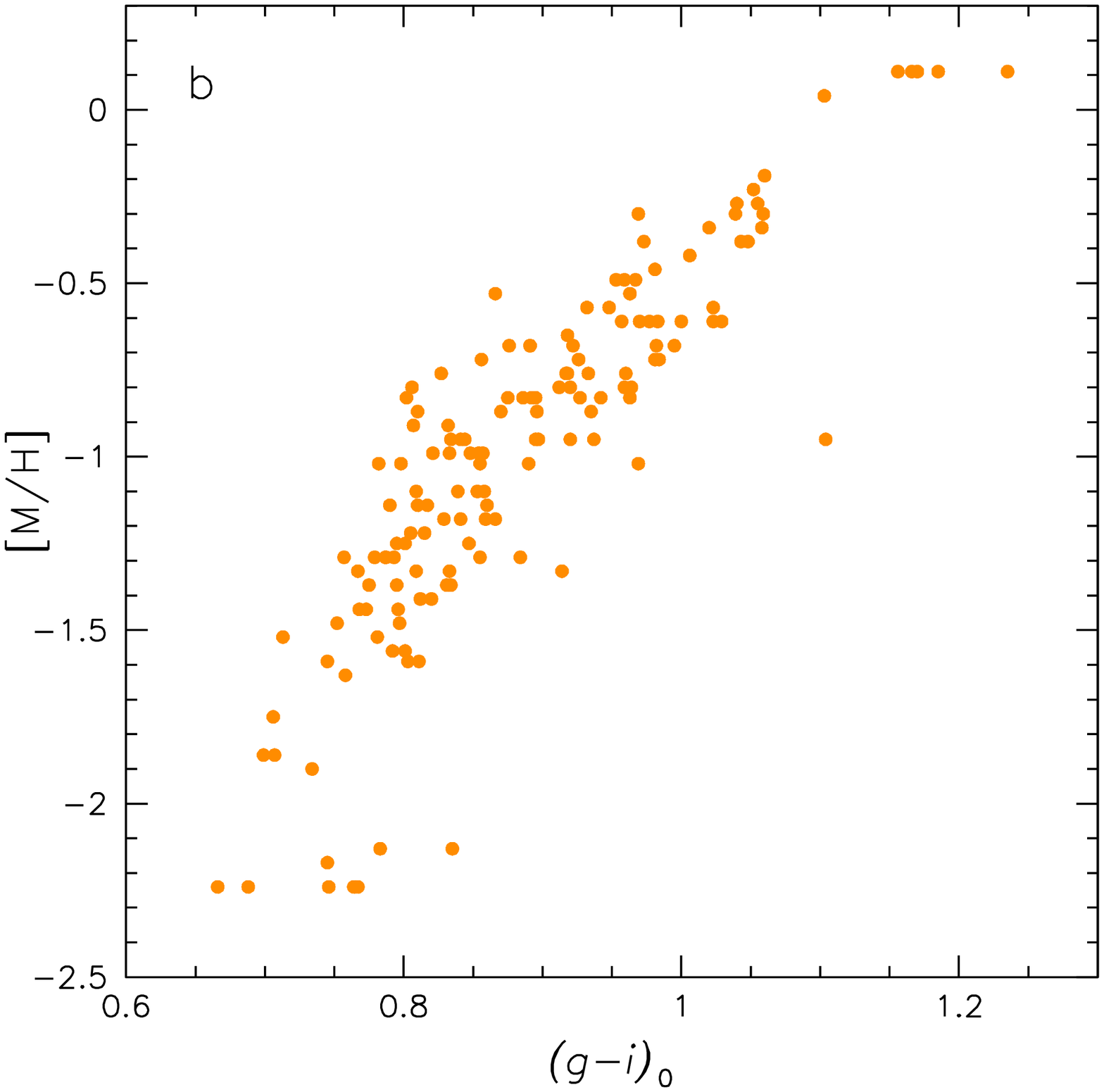}
\figcaption[metals]{\label{fig:metals}
Color vs.~\,metallicity for M87 GCs.
(a): results from CaT analysis of the subset of
our DEIMOS spectra with good enough signal-to-noise
($S/N \gsim 12$) and with $(g-i)_0$ data available (44 objects).
The left axis shows the direct CaT line strengths,
where the error bars indicate 68\% uncertainties.
The right axis shows the equivalent iron metallicity.
The color uncertainties in some cases are smaller than the point sizes.
One additional outlier is off the scale at $(g-i)_0=0.91$, CaT$=$1.9~\AA.
(b): color versus metallicity for 148 GCs from
earlier spectroscopic analysis \citep{1998ApJ...496..808C}.
Note that while in this case the line-index metallicity conversions below [M/H]~$\sim -1.8$
may well be non-linear, and above [M/H]~$\sim 0$ the metallicities are not
well calibrated, the important issue here is not the exact metallicities
but rather the monotonicity of the color-metallicity relation,
which is demonstrated by both data sets.
}
\end{figure*}

We provide these metallicity estimates in Table~\ref{tab:CaT}, and 
in Figure~\ref{fig:metals}(a)
plot them vs.~\,dereddened colors for the same GCs where $(g-i)_0$ data are available.
We find a strong correlation between metallicity and color that 
agrees roughly with model predictions \citep{2003MNRAS.340.1317V}.
Understanding the scatter in the data will require more detailed investigation.

\begin{deluxetable}{lcclcc}
\tablewidth{0pt}
\tabletypesize{\footnotesize}
\tablecaption{CaT measurements of M87 globular clusters\label{tab:CaT}}
\tablehead{ID & R.A. & Decl. & $i_0$ & $(g-i)_0$ & \,\,\,\,\, CaT \,\,\,\,\, \\
& [J2000] & [J2000] & &  & [\AA]}
\startdata
S731   & 187.72452 & 12.28682 & 20.37 & $0.858\pm0.004$ & $7.05\pm0.31$ \\    
S878   & 187.70708 & 12.28149 & 20.62 & $0.816\pm0.004$ & $7.89\pm0.53$ \\    
S952   & 187.69952 & 12.27727 & 20.58 & $0.779\pm0.004$ & $4.78\pm0.48$ \\    
S1007  & 187.69344 & 12.28971 & 19.41 & $1.023\pm0.002$ & $6.94\pm0.13$ \\    
S1074  & 187.68856 & 12.27776 & 20.29 & $1.120\pm0.004$ & $5.89\pm0.30$ \\    
S1265  & 187.67009 & 12.28303 & 19.82 & $0.772\pm0.003$ & $5.23\pm0.20$ \\    
H20493 & 187.87314 & 12.15391 & 20.13 & $0.785\pm0.005$ & $7.24\pm0.42$ \\    
H21863 & 187.87253 & 12.17069 & 20.54 & $0.977\pm0.006$ & $8.25\pm0.42$ \\    
H25785 & 187.86159 & 12.21268 & 19.88 & $0.914\pm0.002$ & $1.95\pm0.22$ \\    
H26690 & 187.72956 & 12.22270 & 19.70 & $0.990\pm0.002$ & $7.56\pm0.16$ \\    
H27496 & 187.80753 & 12.23123 & 19.56 & $0.801\pm0.002$ & $5.31\pm0.13$ \\    
H27916 & 187.71521 & 12.23610 & 20.45 & $0.756\pm0.004$ & $5.46\pm0.37$ \\    
H28411 & 187.70221 & 12.24179 & 20.10 & $0.799\pm0.002$ & $5.64\pm0.26$ \\    
H28415 & 187.78343 & 12.24165 & 19.71 & $0.925\pm0.002$ & $7.80\pm0.16$ \\    
H28866 & 187.75563 & 12.24666 & 19.96 & $0.994\pm0.004$ & $6.13\pm0.20$ \\    
H30757 & 187.71996 & 12.26691 & 20.20 & $0.809\pm0.002$ & $6.04\pm0.28$ \\    
H30772 & 187.74191 & 12.26728 & 19.84 & $0.906\pm0.033$ & $3.61\pm0.31$ \\    
H31134 & 187.72735 & 12.27091 & 20.04 & $0.821\pm0.003$ & $8.22\pm0.23$ \\    
T13456 & 187.92093 & 12.46766 & 21.14 & $0.757\pm0.008$ & $5.06\pm0.15$ \\    
T13500 & 187.93052 & 12.42406 & 20.55 & $0.767\pm0.004$ & $3.49\pm0.26$ \\    
T13559 & 187.93331 & 12.37800 & 21.87 & $0.904\pm0.012$ & $6.47\pm0.33$ \\    
T13569 & 187.98853 & 12.36469 & 21.93 & $0.695\pm0.015$ & $3.77\pm0.25$ \\    
T13571 & 187.89063 & 12.36288 & 21.42 & $0.801\pm0.008$ & $5.81\pm0.14$ \\   
T13589 & 187.92350 & 12.35406 & 21.71 & $0.773\pm0.012$ & $4.14\pm0.22$ \\    
T13593 & 187.90345 & 12.35247 & 21.81 & $0.778\pm0.011$ & $6.12\pm0.26$ \\    
T13609 & 187.96307 & 12.34319 & 21.95 & $0.729\pm0.014$ & $6.58\pm0.23$ \\    
T13611 & 187.92010 & 12.34146 & 22.44 & $0.793\pm0.022$ & $7.00\pm0.45$ \\    
T13623 & 187.90570 & 12.33556 & 21.49 & $0.764\pm0.011$ & $4.96\pm0.14$ \\    
T13642 & 187.94181 & 12.32521 & 20.39 & $0.869\pm0.004$ & $6.32\pm0.02$ \\    
T13648 & 187.98298 & 12.32301 & 21.79 & $0.731\pm0.020$ & $3.99\pm0.21$ \\    
T13899 & 187.84033 & 12.42620 & 22.38 & $0.964\pm0.009$ & $7.53\pm0.31$ \\    
T13901 & 187.87673 & 12.42616 & 21.83 & $0.853\pm0.013$ & $5.56\pm0.20$ \\    
T14060 & 187.87089 & 12.38884 & 20.83 & $0.722\pm0.007$ & $4.39\pm0.06$ \\
T14108 & 187.83846 & 12.37504 & 19.82 & $0.908\pm0.002$ & $6.23\pm0.01$ \\
T14133 & 187.84645 & 12.37052 & 20.69 & $0.980\pm0.005$ & $6.28\pm0.06$ \\
T14149 & 187.87374 & 12.36463 & 21.84 & $0.773\pm0.013$ & $6.42\pm0.31$ \\    
T14228 & 187.84434 & 12.35020 & 20.44 & $0.785\pm0.004$ & $5.93\pm0.03$ \\
T15685 & 188.04605 & 12.36726 & 22.08 & $0.730\pm0.017$ & $5.37\pm0.31$ \\    
T15777 & 188.14226 & 12.45638 & 21.93 & $0.796\pm0.013$ & $5.41\pm0.19$ \\   
T15863 & 188.17885 & 12.36688 & 21.36 & $0.753\pm0.008$ & $5.64\pm0.11$ \\  
T15886 & 188.15205 & 12.34920 & 22.33 & $0.743\pm0.021$ & $4.99\pm0.29$ \\    
T15900 & 188.21484 & 12.33844 & 20.6 & \nodata\ & $4.20\pm0.06$ \\
T15908 & 188.15419 & 12.33126 & 21.19 & $0.781\pm0.007$ & $4.38\pm0.08$ \\  
T16080 & 188.27954 & 12.34479 & 21.0 & \nodata\ & $6.21\pm0.07$ \\  
T17211 & 188.33727 & 12.48077 & 21.0 & \nodata\ & $5.08\pm0.08$ \\    
VUCD6 & 187.86816 & 12.41766 & 18.63 & $0.896\pm0.001$ & $5.69\pm0.01$ \\
VUCD8 & 188.01813 & 12.34176 & 18.96 & $0.855\pm0.001$ & $6.06\pm0.01$ \\
\enddata
\tablecomments{Note that the $(g-i)_0$ color uncertainties are based on
shot-noise only, and are useful for relative comparisons of colors in this dataset.
The absolute uncertainties from calibration, reddening correction,
background estimation, etc., will be larger.  The equivalent iron metallicities
shown on the right-hand axis of Figure~\ref{fig:metals}(a) are derived
using the expression: [Fe/H]~$\simeq0.438\times$CaT$-3.641$.}
\end{deluxetable}

This test does not prove unequivocally that these colors are a perfect proxy for
metallicity, given various unresolved issues about CaT-based metallicities,
and the possibility of some residual age effects on the colors.
However, it does provide assurance that the current conventional wisdom about
GC color as a reliable metallicity proxy is probably sound in the halo of M87
(i.e., at radii of $\sim$~45--220~kpc). 

Previous spectroscopic work using bluer wavelengths
at smaller radii ($\sim$~3--40~kpc) also found M87 GC metallicity indices to
correlate strongly with $U-R$ color, and for the ages to be almost all
larger than 10~Gyr \citep{1998ApJ...496..808C}.
We revisit these data by comparing our $(g-i)_0$ colors to these authors' metallicity
estimates in Figure~\ref{fig:metals}(b), finding a tight correlation between the two quantities.
Again, this supports the color-metallicity link.

\begin{figure*}
\epsscale{0.56}
\plotone{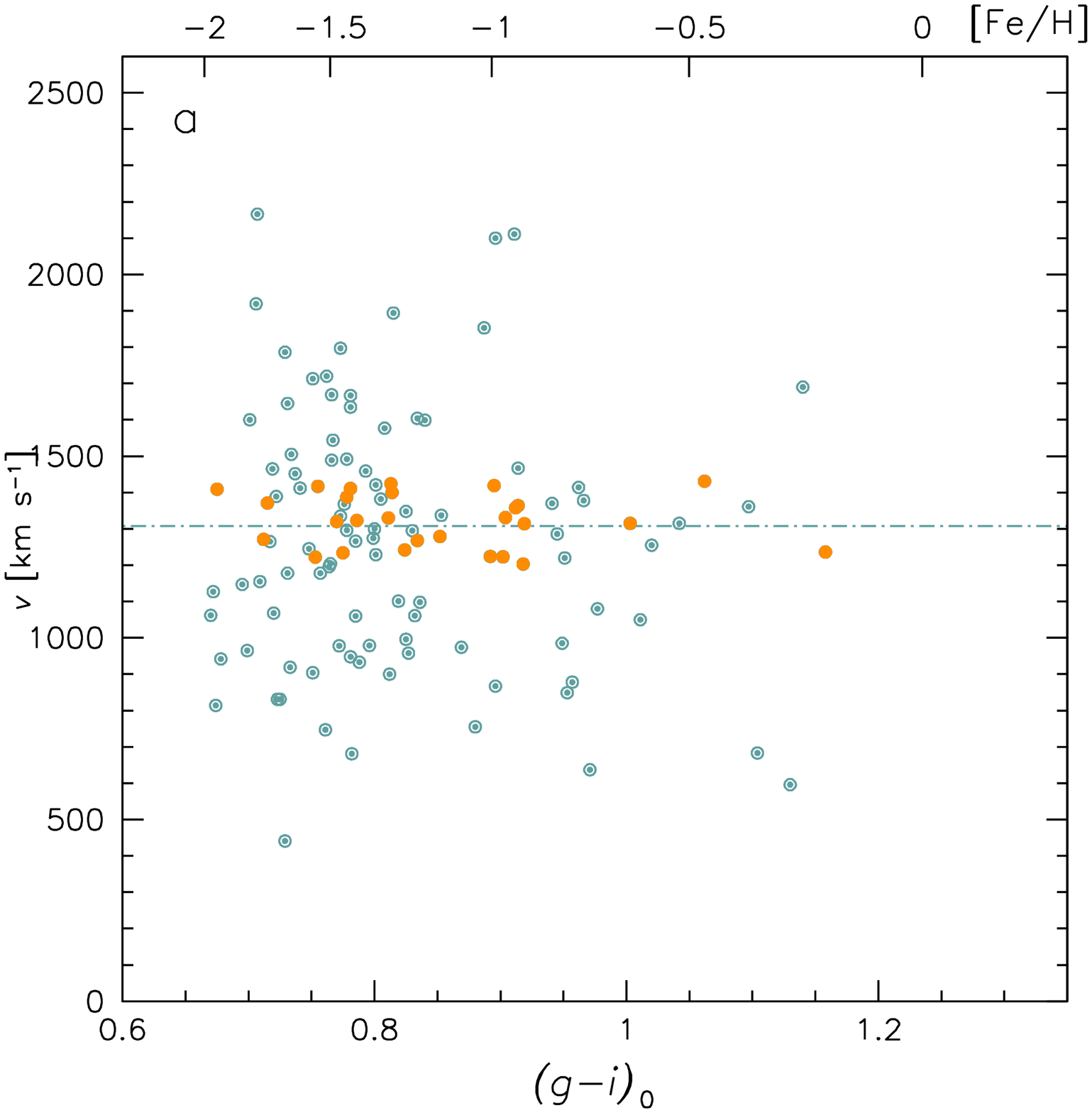}
\hspace{0.1cm}
\plotone{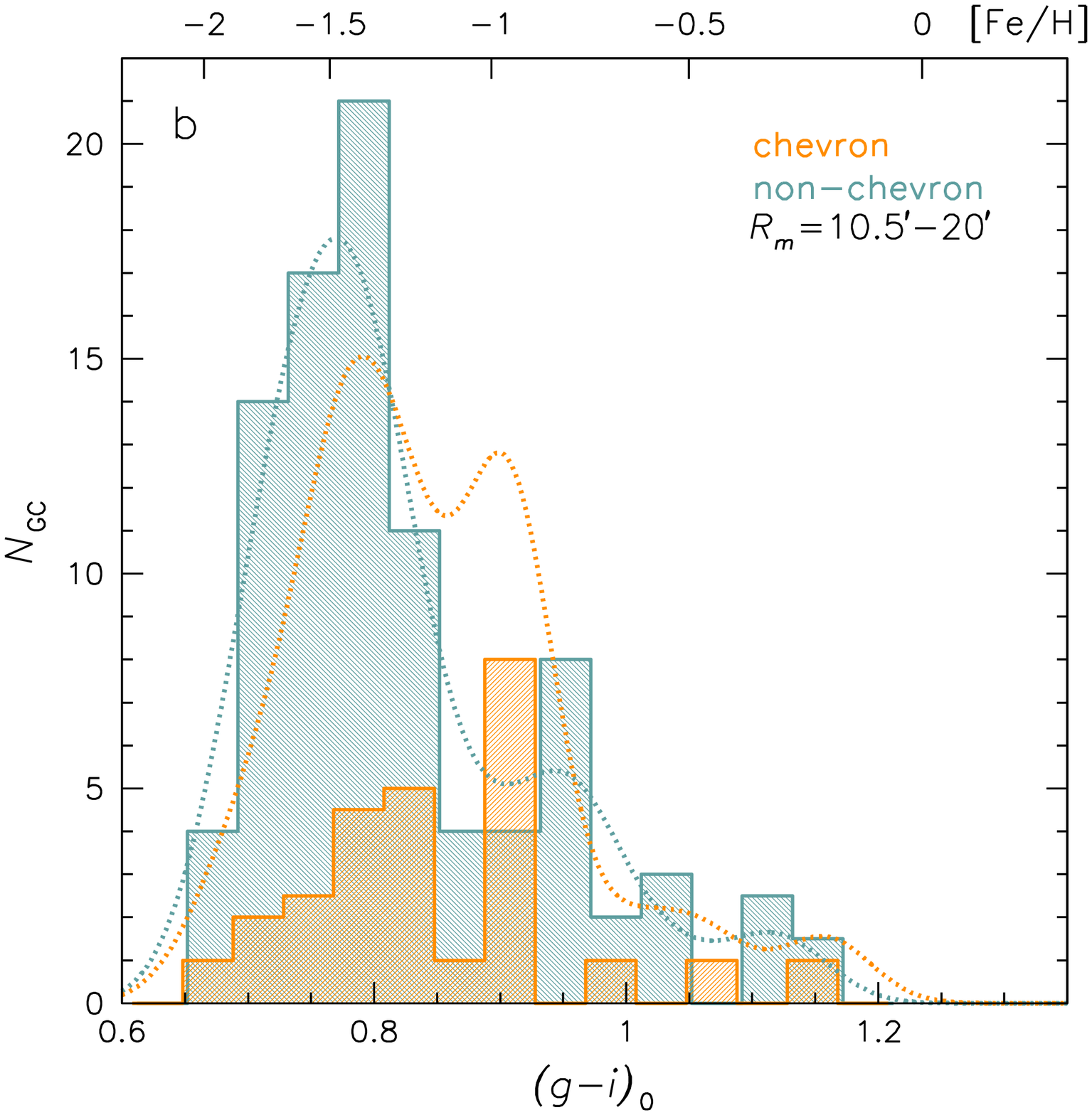}
\figcaption[colors]{\label{fig:colors}
Correlations between phase space and color properties of M87 GCs
in the $R_m=$~10.5--20~arcmin shell region.
(a): line-of-sight velocity versus color.
A ``cold'' velocity component appears at colors redder than $(g-i)_0 \sim 0.77$.
GCs assigned to the chevron are marked as filled orange circles.
(b): the chevron and non-chevron subsamples (with 27 and 92 objects, respectively).
Histograms show the raw counts in color bins of 0.04-mag width.
Alternative, unbinned color distributions are shown by the dotted curves, which are
produced by smoothing the discrete data with a 0.036-mag optimum Gaussian kernel
(a standard technique in studies of GC color distributions; \citealt{2006ApJ...639...95P}).
The amplitude of the smoothed chevron distribution has been rescaled to the same
integrated number as the non-chevron.
The chevron GCs appear to have a slightly redder ``blue peak'' than the non-chevron GCs,
as well as a strong intermediate-color peak.
}
\end{figure*}

Despite these arguments that color is a good proxy for metallicity in our data set, 
it is important to note that our analysis below is independent of this link.
Whatever the root cause of variations in color, we can still use it as an empirical
tagging factor.  Then for our current purposes the issue is more a matter of semantics:
whether we are analyzing chemo-dynamics or chromo-dynamics.

Our first basic goal is to use color as a completely independent dimension in the data,
to see if it supports the differences among GCs that emerged from the
position-velocity phase-space analyses.
We plot velocity versus color in the shell region in Figure~\ref{fig:colors}(a).
In M87, as in galaxies in general, the population of metal-poor GCs has
a more extended distribution from their host galaxy's center than the more
metal-rich GCs.
Given that all of these objects reside in the same gravitational potential, and assuming
that they have similar orbital anisotropies, their spatial differences are
expected to be related to their velocity dispersions.
The metal-rich GCs should generally have lower dispersions, and Figure~\ref{fig:colors}(a) is
expected to show a systematic decrease in dispersion with color, with perhaps a strong
transition between the two main subpopulations, at $(g-i)_0\sim$~0.9.

Figure~\ref{fig:colors}(a) does show a kinematic transition with color, but not in the way
just described.
There is an overall pattern of a cold component
near $v_{\rm sys}$ embedded in a hotter component (cf. Figure~\ref{fig:losvds}),
but the cold component does not appear bluewards of $(g-i)_0 \sim 0.77$.
This color is in the middle of the main blue peak for M87's GC system,
and implies that there is kinematical substructure within this peak.\footnote{This
is again a test that is insensitive to the radius transformation $R_m$.}
We have looked at the same information using the CaT-based metallicities,
with consistent results (colder material at CaT~$\gsim 5.3$~\AA\ or [Fe/H] $\gsim -1.3$~dex),
but the statistics are too poor to be conclusive (17 GCs in the shell region).

We next take a different approach, of separating GCs into probable
shell and non-shell objects, based on the phase-space
maximum likelihood models constructed previously.
We then examine the color distributions of the two subsamples.
For each of the models (tapered Gaussian, chevron, wedge), the
two color distributions are different, with the chevron case
shown in Figure~\ref{fig:colors}(b) as an example.\footnote{Among
our probable shell sample, there are two ultra-compact dwarf
candidates: VUCD1 and H59823.}
Here the shell appears to contain an excess population of
intermediate-color GCs, peaking at $(g-i)_0 \sim 0.90$.

Motivated by this result,
we also show the velocity distribution of all the intermediate color objects 
in the shell region, with $(g-i)_0=$~0.85--0.95, in Figure~\ref{fig:losvds2}(a).
These objects do show a double-peaked profile as expected in the chevron model,
except for an additional low-velocity peak that accounts for much of the candidate
outer shell.
We have also fitted the chevron model to the phase-space data for the blue,
intermediate-color, and red GCs separately.  The resulting parameters for
each of these fits are mutually consistent, and color does not discriminate sharply
between shell and non-shell objects,
but we do find that the shell fraction is highest in the intermediate-color
population, with $f_{\rm s} \sim 0.5$.

We evaluate the statistical significance of the differences between the
chevron and non-chevron color distributions using a Kolmogorov-Smirnov test.
We find a $D$ value of 0.25, with a 13\% probability of occurrence by chance.
According to the test, the largest difference between the distributions
comes at the far blue end, supporting the transition seen in Figure~\ref{fig:colors}(a).
The same is true for all three shell models, with the far-blue difference
being strongest for the tapered-Gaussian case, and the intermediate-color peak
for the chevron case.
The similarity of the three models also suggests that these conclusions
are insensitive to the choice for radius $R_m$, as is
also explicitly the case for the test in Figure~\ref{fig:colors}(a).

We have found that metallicity information (primarily using color as a proxy)
provides independent support for the presence of the M87 shell.
The colors do not yet differentiate clearly between the alternative phase-space
models for the shell, but even our unprecedented data set only scratches
the surface of what can now be accomplished.
It is technically feasible to increase the number of measured GC velocities by 
an order of magnitude, e.g., to $\sim$~1500 objects in the shell region alone.

\subsection{Progenitor inferences from GCs}\label{sec:prog}

Having established 
the existence of the shell, and some of its properties, we now attempt to
identify the nature of its progenitor, assuming that the accretion and disruption
of a single galaxy was responsible.
Unfortunately, we will see that it is difficult to derive a self-consistent
solution from the basic shell parameters, so we will carry out an inventory
of different types of constraints to see if an overall solution emerges
(with an estimate of the progenitor's $V$-band stellar luminosity, $L_V$, as a basic goal).

\begin{figure*}
\epsscale{0.75}
\plotone{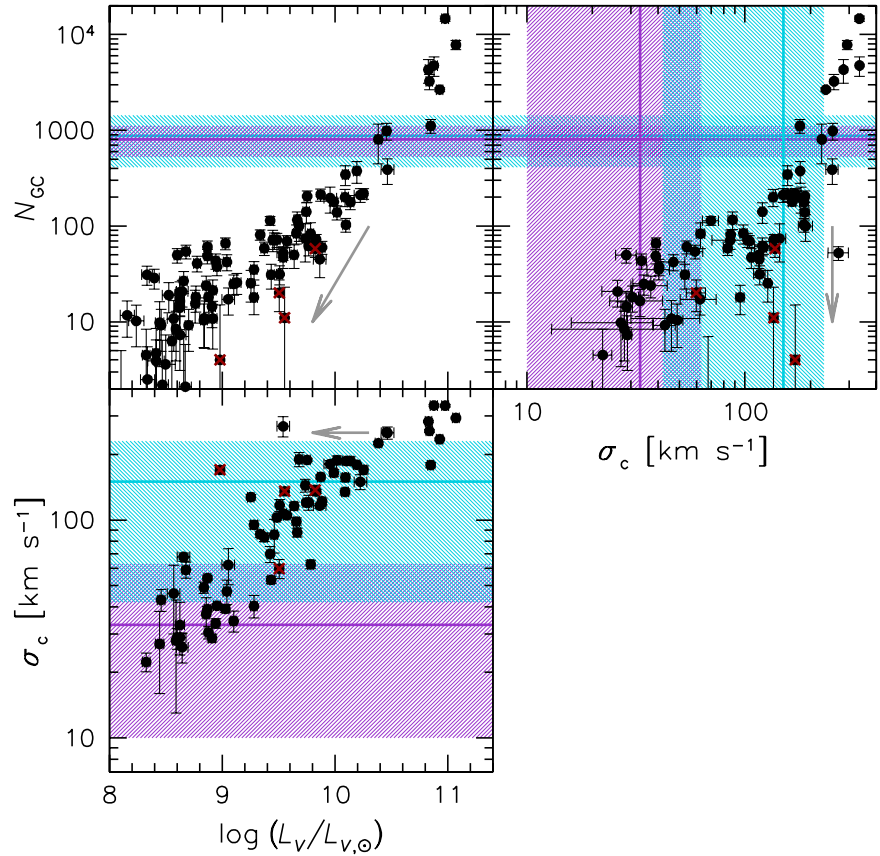}
\figcaption[VCC2]{\label{fig:grid}
Parameter-space of candidate progenitor galaxies for the M87 shell,
plotting $V$-band luminosity, central stellar velocity dispersion, and GC numbers.
The bands show constraints from modeling the shell GCs,
with the narrower purple bands stemming from our default chevron-based analysis,
and the wider blue bands showing a model that does not incorporate spatial
information (yielding a conservative upper limit on the dispersion).
The points with error bars show early-type galaxies in Virgo from the ACSVCS survey,
with $N_{\rm GC}$ and $L_V$ from \citet{2008ApJ...681..197P}.
The dispersions are mostly from HyperCat with some additions from
\citet{2009MNRAS.394.1229C} and \citet{2011A&A...526A.114T}.
Points with red $\times$ symbols superimposed show four low-luminosity ellipticals near
to M87, whose positions in these diagrams are suggestive of histories of stripping---as
illustrated schematically by the gray arrows.
The constraints from the chevron model do not match up well with
any realistic candidate progenitor galaxies, 
while the conservative model is more viable
(see text for details).
}
\end{figure*}

Since in the previous Section we have just discussed the color distribution of the shell GCs, it is natural to try using these colors to constrain the progenitor $L_V$.
Empirically, 
GC colors correlate with host galaxy luminosity 
(e.g., \citealt{2001AJ....121.2974L,2004AJ....127.3431S,2004ApJ...613..262L}),
which is conceptually related to the correlations between substructure metallicities
and their progenitor masses (e.g., \citealt{2008ApJ...689..936J}).
This framework was used to estimate the relative masses of accreted 
galaxies in the halo of 
a nearby S0 in \cite{2011ApJ...736L..26A}.

In the case of M87,
we can make use of the correlations found 
from the ``ACSVCS'' large survey of elliptical and lenticular galaxies,
both giant and dwarf, in the Virgo cluster \citep{2006ApJ...639...95P}.
Although the GC color distributions in these galaxies are in general complex,
they show a general tendency for {\it bimodality}, with distinct blue and red peaks.
The color-luminosity relation is therefore reported separately for the blue
and red GCs.
The underlying basis for these correlations is thought to be a mass-metallicity
relation, but again, our analysis does not rely on this conclusion.

We have modeled the color distributions of the chevron and non-chevron GCs
(Figure~\ref{fig:colors}(b)) using Kaye's
Mixture Models based on heteroscedastic Gaussian color subpopulations \citep{1994AJ....108.2348A}.
We find marginal significance ($p=0.325$) for bimodality among the chevron GCs,
and high significance ($p<0.001$) for the non-chevron GCs.
Beginning with the {\it non-chevron} GCs, we find that they
have a primary peak at
$(g-i)_0 \sim$~0.77 and a possible weak secondary peak at $(g-i)_0\sim$~0.94,
which correspond to $(g-z)_0 \sim$~0.92 and $\sim$~1.17, or
[Fe/H]~$\sim$~$-1.5$ and $-0.8$, respectively
(where we have made empirical color conversions 
using a large set of M87 GCs that overlap between ACS and CFHT imaging).

We next assume for simplicity that the GC color peaks in the surviving Virgo early-type 
galaxies are representative of the
disrupted progenitors that built up the M87 halo.\footnote{There
is evidence that galaxies accreted during earlier
epochs had systematically redder GC colors at a fixed mass \citep{2006ARA&A..44..193B},
which would mean that our luminosities for the typical M87 halo progenitors
will be overestimated.}
Considering first the peak locations of the non-chevron GCs and
comparing them to the non-parametric peaks determined for the ACSVCS galaxies \citep{2006ApJ...639...95P},
we find the former to be fairly atypical when considered jointly
(a puzzle that would be worsened if the chevron GCs were included).
This makes the progenitor inference unclear, perhaps owing to the superposition
of color peaks from multiple progenitors,
but our best guess would be a typical progenitor magnitude of
$M_V \sim -18$ [$\log\,(L_V/L_{V,\odot}) \sim 9.5$].

Approximating the $V$-band stellar mass-to-light ratios as roughly constant,
the implication would be that the accretion events which built up the bulk
of the M87 halo had characteristic stellar-mass merger ratios of $\sim$~1:30.
This may be consistent with current ideas of massive galaxies' stellar
envelopes being assembled through minor mergers (e.g., \citealt{2011arXiv1106.5490O}).

Turning now to the {\it chevron} GCs, they have a primary peak at
$(g-i)_0 \sim$~0.79 and a secondary peak at $(g-i)_0\sim$~0.90.
corresponding to $(g-z)_0 \sim$~0.95 and $\sim$~1.11,
[Fe/H]~$\sim$~$-1.4$ and $-1.0$.
This red peak location is very atypical for the ACSVCS sample, and may have been
inaccurately estimated through our procedure of statistically disentangling
the chevron and non-chevron GCs.  We therefore consider the color constraint on the
shell progenitor luminosity to be indeterminate.
%

The other two major clues about the shell origin are its {\it number} of GCs,
$\ngc$, and their mutual {\it velocity dispersion}, $\sgc$.
For both of these properties, we use the results of maximum likelihood
models from Section~\ref{sec:max}.
While our default constraint comes from the ``chevron'' model, we will also
account for the systematic uncertainty in the shell morphology (e.g., owing to the
radius conventions discussed in Section~\ref{sec:rad}) by considering
an alternative scenario of maximal ignorance of the morphology.
For this case we adopt the double-Gaussian analysis of Section~\ref{sec:max},
which leads to a much larger value of $\sgc$ (50--120~\kms\ rather than 10--30~\kms).


To make use of the $\ngc$ and $\sgc$ constraints, we must correct both observational
measurements to the relevant quantities for the progenitor galaxies.
In the case of $\ngc$, we extrapolate the fraction of shell
GCs in our spectroscopic sample to the much larger sample of unobserved GCs
in the same region.
We have constructed a smooth model surface density profile of the M87 GC system,
based on both the CFHT photometric data as well as spectroscopic data to confirm
the non-GC contamination rate (see S+11).
This model is initially based on our photometric limit of $i=22.5$
($z\simeq22.4$), but
to get total GC numbers we need to extrapolate to fainter luminosities,
using an evolved Schechter function fitted to the central GCs of M87 \citep{2007ApJS..171..101J}.
This provides a scaling factor of 2.76.\footnote{With a Gaussian luminosity function,
this factor would be 2.50, but for direct comparison with $\ngc$ values from
the ACSVCS, we adopt 2.76.}

Now integrating the revised surface density between 10 and 20 arcmin where the shell
signal is clear, we find a total of $3500\pm400$~GCs. 
Multiplying by the inferred $f_{\rm s}$
values for both the chevron and double-Gaussian models
(see Section~\ref{sec:max}), we estimate $\ngc=800^{+320}_{-270}$ and
$870^{+560}_{-460}$ total GCs in the shell, respectively (integrating over all magnitudes).
These numbers could be lower if we have detected a disproportionate fraction of
the shell objects by spatial bias in our spectroscopic selection that coincides
with a spatial clumping of the shell; or higher if the shell extends farther
in radius and velocity than in our simple model (see, e.g., Figure~\ref{fig:sims}(b)).
It is difficult to quantify such uncertainties, but we suspect that they could involve
an additional factor of $\sim$~2, which would accommodate a range of 
up to $\sim$~200--2900 GCs.
If $\ngc \sim 800$, then the shell accounts for
$\sim$~20\% of the GCs around M87 within the 50--100~kpc range,
and $\sim$~5\% of the total GC system.\footnote{To
derive the latter value, we have integrated the GC spatial density profile
out to a radius of 45~arcmin (220~kpc), which is where the planetary
nebula (PN) kinematics suggest
a transition between objects bound and not bound to M87 (\citealt{2009A&A...502..771D};
the kinematics data for the GCs themselves are less extensive at large radii and do not yet show
any sign of such a transition).
The total GC population is then $\sim$~16200.}

What kinds of galaxies are observed so many GCs?
To provide a visual overview of this constraint, 
we plot the number of GCs versus host luminosity for the ACSVCS sample
in the upper left panel of Figure~\ref{fig:grid},
finding that $\sim L^*$ luminosities of
of $\log\,(L_V/L_{V,\odot})\sim$~10.1--10.8 are implied.

Next we consider the shell's internal velocity dispersion, which as a 
first approximation, we assume to be similar
to the internal velocity dispersion of its progenitor,
as might be expected under the simplest tidal stripping models.
Here the comparison to the ACSVCS galaxies is less straightforward, since 
very few of these have direct measurements available
of their GC velocity dispersions.  Instead, we will adopt a conversion
between $\sgc$ and the widely-available central stellar velocity dispersion $\sigc$.
Based on the forthcoming kinematical analysis of a handful of low-luminosity E/S0s (Pota et al.,
in preparation), we find that the ratio $\sigc/\sgc$ varies from galaxy
to galaxy over a range $\sim$~1.3--2.3, with a typical value of $\sim$~1.8.
We therefore infer $\sigc \sim 35\pm25$~\kms\ and $\sim 150\pm100$~\kms\ 
based on the chevron and double-Gaussian models, respectively.

We plot this constraint again versus the ACSVCS sample in the bottom left
panel of Figure~\ref{fig:grid}.
Here we see that the chevron-based velocity dispersion
suggests a dwarf E/S0 with $\log\,(L_V/L_{V,\odot})\la$~9.5,\footnote{An
alternative mechanism for producing a low velocity dispersion is through tidal
tails stripped out from the cold disk of an infalling spiral galaxy
 (e.g., \citealt{1992ApJ...399L.117H,1995AJ....110..140H}).
However, we do not know of any galaxy with hundreds of GCs contained in a cold disk.
Alternatively, some GCs could have been {\it formed} from cold gas in
such a merger, and then deposited on closely associated orbits.
Because of the age-metallicity degeneracy, the shell GCs might in principle be
young and metal-rich rather than old and metal-poor.
Full analysis of the GC ages is beyond the scope of this paper, but our 
initial inspection of the Hectospec data indicates there are no objects as
young as $\sim$~1~Gyr.}
while the conservative upper-limit dispersion allows for luminosities
of up to $\log\,(L_V/L_{V,\odot})\sim$~10.6.

For the chevron model, the dispersion constraints on luminosity are thus inconsistent with 
the constraints from GC numbers.
To make this comparison more directly, we show dispersion versus GC numbers
in the upper right panel of Figure~\ref{fig:grid}.
The intersections of the vertical and horizontal bands mark out possible
solutions for the progenitor based on the shell analyses---but
these do not overlap with the properties of known Virgo galaxies.
%
{\it It does not seem plausible that either a dwarf galaxy hosted hundreds of
GCs, or that an $\sim L^*$ galaxy left extensive debris with a velocity dispersion
of $\lsim$~30~\kms.}

The conservative model fares much better in this respect, as
several ACSVCS galaxies are consistent with both its $\ngc$ and
$\sgc$ constraints. Some examples include NGC~4435 and NGC~4473, with
$\log\,(L_V/L_{V,\odot})\sim$~10.1--10.2.
This model seems promising, but
one should remember that is intended to representing a bracketing case
of maximal ignorance about the shell morphology, while there does seem
to be some degree of substructure tapering with radius in phase-space,
independently of the radius convention.
Therefore it is worthwhile reviewing any additional lines of evidence
about the accretion progenitor.

If the progenitor were very massive, then it should have left other clear
signatures of its passing, besides the phase-space shell.  One might expect
to see obvious disturbances and asymmetries in the light distribution of M87,
as well as strong merger-induced halo rotation, which is not observed (S+11).
If we assume that this rules out a mass ratio of $\sim$~1:3 or higher, it implies
$\log\,(L_V/L_{V,\odot}) \la 10.4$ for the progenitor, or perhaps somewhat
higher if dark matter is taken into account.

Besides the impact of the accretion event on M87, the incoming galaxy should
have deposited not only GCs but field stars, which would leave a large-scale
surface brightness fluctuation across the face of M87.
In the shell region where the typical surface brightness is $\mu_V\sim$~27~mag~arcsec$^{-2}$, 
we can rule out obvious fluctuations at the $\sim$~0.5~mag~arcsec$^{-2}$ level
(see Figure~\ref{fig:main}(a) and \citealt{2010ApJ...715..972J}).
We can therefore rule out $\log\,(L_V/L_{V,\odot}) \gsim$~9.8 of accreted
stars that are distributed uniformly over the shell region.\footnote{This 
constraint is relevant not only to our default single-progenitor scenario,
but also to a scenario involving multiple small galaxies
that together deposited a large number of GCs with cold kinematics:
e.g., from $\sim$~5 dwarf E/S0s with $\log\,(L_V/L_{V,\odot})\sim$~9.3 each.
Besides the low probability of accreting such galaxies on near-identical trajectories,
the associated stellar debris could violate the surface brightnesses constraint.
}

This luminosity limit is lower if the stars are confined to only one side of M87,
as Figure~\ref{fig:main}(a) suggests; and higher if the stars have become
spatially decoupled from the GCs and reside at smaller radii where they are
either hard to detect, or are still intact as one of the nearby low-luminosity
elliptical galaxies.  
Such spatial segregation is plausible at some level since the GCs
would be expected to inhabit preferentially the outer parts of the incoming galaxy,
and thus be stripped off before the stars.
Note that a fluctuation in the surface density of the GCs themselves would be
challenging to detect at the $\sim$~50\% level.


As mentioned previously, there are no major signatures of interactions previously
known for M87.  Some candidates for large-scale, low-surface brightness
stellar features have been identified
\citep{1971ApJ...163..195A,1997ApJ...490..664W,2010ApJ...720..569R}, 
but these are at spatially disparate locations from the GC shell, and in some cases
are likely to be manifestations of foreground cirrus.


\begin{figure}
\epsscale{1.16}
\plotone{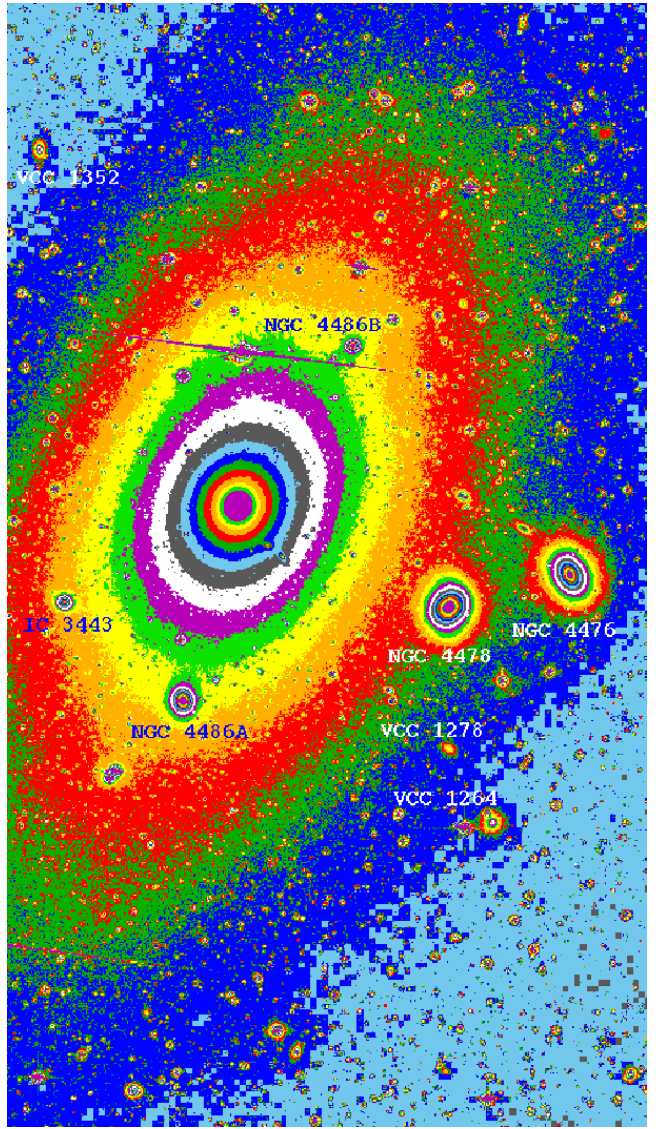}
\figcaption[compositedEc]{\label{fig:dEs}
Closer view of the M87 $V$-band image from Figure~\ref{fig:main}(a),
showing a 23.5\arcmin$\times$40\arcmin\ (110$\times$190~kpc) region.
Some small galaxies in the field are labeled.
The shading is logarithmically spaced by surface brightness,
and arbitrarily colored in order to enhance any isophote twisting
(e.g., in the case of NGC~4478).
}
\end{figure}

More intriguingly, as mentioned above, there are
five low-luminosity ellipticals within a
30--80~kpc projected radius of M87:
NGC~4476, NGC~4478, NGC~4486A, NGC~4486B, and IC~3443
(see Figures~\ref{fig:main} and \ref{fig:dEs}).
Some of these have previously been proposed as the stripped remnants of
larger galaxies, based on disturbances in their isophotes or unusual
stellar colors for their luminosities.
If any of these galaxies is being actively disrupted, some stars and
GCs could have already been stripped out and deposited within the halo of M87.

The first four of these galaxies are in the ACSVCS sample, so we consider
this scenario in more concrete terms by marking them in Figure~\ref{fig:grid} with
red $\times$ symbols.  Here we see from the top two panels
that all four galaxies appear depleted in
GCs relative to the average trend, which suggests stripping of their outer
regions.  The offsets are strongest for NGC~4486A and NGC~4486B, which are
statistically consistent with hosting {\it zero} GCs (see further discussion
in \citealt{2008ApJ...681..197P}). 
NGC~4486B is also a very unusual outlier in the $\sigc$-$L_V$ panel, 
and possibly in supermassive black hole correlations \citep{2009ApJ...698..198G},
implying that not only its GCs but also $\sim$~90\% of its stars have been stripped away.

The $\sigc$ values of NGC~4486A and NGC~4486B suggest
that they originally hosted $\sim$~70--200 and $\sim$~100--500 GCs, respectively.
Their phase-space positions in Figure~\ref{fig:main}(b),
where their velocities are $\sim$~750~\kms\ and $\sim$~1550~\kms, 
respectively, indicate that NGC~4486B could be associated with the shell.
NGC~4478 (at $\sim$~1400~\kms) is also consistent with the shell in phase-space;
it does not appear to be as severely stripped, but still could have lost up
to $\sim$~250~GCs.  

The colors of the NGC~4478 GCs also provide one of the closer
matches to the shell GC colors (this test is not possible for NGC~4486A and 
NGC~4486B since they have virtually no GCs remaining).
For both NGC~4478 and NGC~4486B,
the dynamical friction timescales at their projected radii are $\sim$~1--3~Gyr
or shorter, reinforcing the likelihood that these are
objects in the final stages of orbital decay toward the center of M87.
For both galaxies, there are also {\it outer} stellar velocity dispersion measurements
available, making the extrapolation to $\sgc$ less uncertain.
For NGC~4478 this is $\sim$~50--100~\kms\ \citep{2001MNRAS.326..473H},
and for NGC~4486B, it is $\sim$~100--120~\kms\ 
\citep{1997ApJ...482L.139K,2010MNRAS.408..254S}.

The present-day luminosity of NGC~4478, $\log\,(L_V/L_{V,\odot})=9.8$, 
suggests that it may have experienced very little stellar stripping 
(Figure~\ref{fig:grid}, lower-left), which would explain the lack of 
surface brightness variations around M87.
For NGC~4486B, the implied luminosity at infall of $\log\,(L_V/L_{V,\odot})\sim$~10.1
would require that much of the stellar debris is hidden at small galactocentric radii,
which is plausible given that this stripped galaxy is
currently at a radius of $\sim$~30~kpc
(note that the missing debris is an issue even without considering the GC shell).

Both of these galaxies appear to be possible shell progenitors,
as their properties are consistent with all of the constraints discussed above, 
within the uncertainties, except for the low velocity dispersion in
the chevron model (this dispersion could then be considered either as an
artifact of the adopted model, or as a product of stripping dynamics that
are not yet understood).
In fact, the problem could be turned around:  given that there are 
stripped galaxies around M87 which must have lost hundreds of accompanying GCs,
should we not expect to see some evidence for them in phase-space?

The multiplicity of candidate shell progenitors is also a reminder that our
initial single-progenitor assumption may be oversimplified.
It is indeed possible that the phase-space of GCs in the M87 halo contains multiple,
overlapping, prominent shells and streams.\footnote{There may be another, related
feature in M87: S+11 identified
a peculiar velocity offset in the metal-rich GCs inside $\sim$~10~kpc.}
More GC spectra are clearly needed to better determine the substructure properties,
and to make clearer progenitor inferences.

As a final plausibility check, we examine whether or not our preferred
solution, with an accreted luminosity of $\log\,(L_V/L_{V,\odot})\sim$~10,
(i.e., $\sim 0.5\,L^*$, or a $\sim$~1:10 merger),
represents a likely event in the context of other observational
constraints on merger rates in BCGs.
As discussed in Section~\ref{sec:intro}, the details of BCG assembly are not
well known, which motivates our phase-space study, but there are still
some plausible upper limits on merger rates.
For example, M87 cannot have accreted a galaxy with $\log\,(L_V/L_{V,\odot})\sim$~10
more often than once every $\sim$~1.5~Gyr, averaged over a Hubble time,
or else the total luminosity would be higher than is observed today.

The calculation is more complicated than this because the merger history will
be divided over a spectrum of mass ratios, and because the merger rate may change
with time. We therefore turn to the low-$z$ BCG survey 
carried out by \citet{2009MNRAS.396.2003L},
who found that major mergers (down to a $\sim$~1:4 mass ratio)
were visible in 1 case out of every 30.
If we assume for simplicity that mergers are detectable in phase-space for roughly
twice as long as in real-space (which is the only place that merger timescales
will enter into our calculations), 
then recent major merger signatures could be found in $\sim$~7\% of BCGs.

Next, we should like to extrapolate to lower-mass ratios.  The simplest assumption
we can make is that the accreted objects follow a standard Schechter luminosity
function, where 1:4 events correspond to $\sim L^*$ in the case of M87.
We then calculate that events down to 1:10 should be present in $\sim$~25\% of BCGs.
Thus it does not seem improbable that the shell in M87 traces a $\sim$~1:10 merger
event, while more massive progenitors would begin to stretch the limits
of plausibility.

\begin{figure*}
\epsscale{0.52}
\plotone{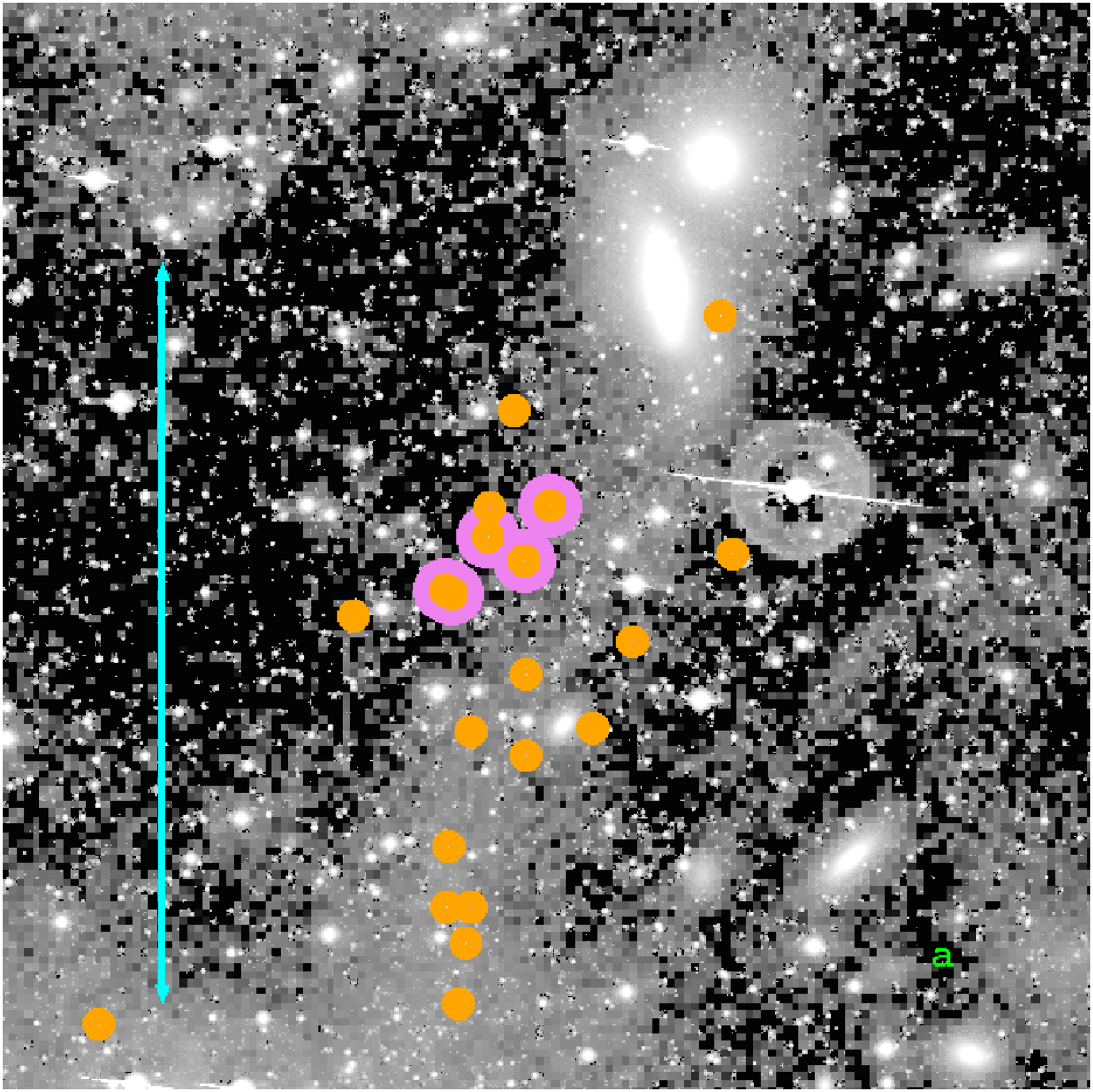}
\hspace{0.1cm}
\epsscale{0.56}
\plotone{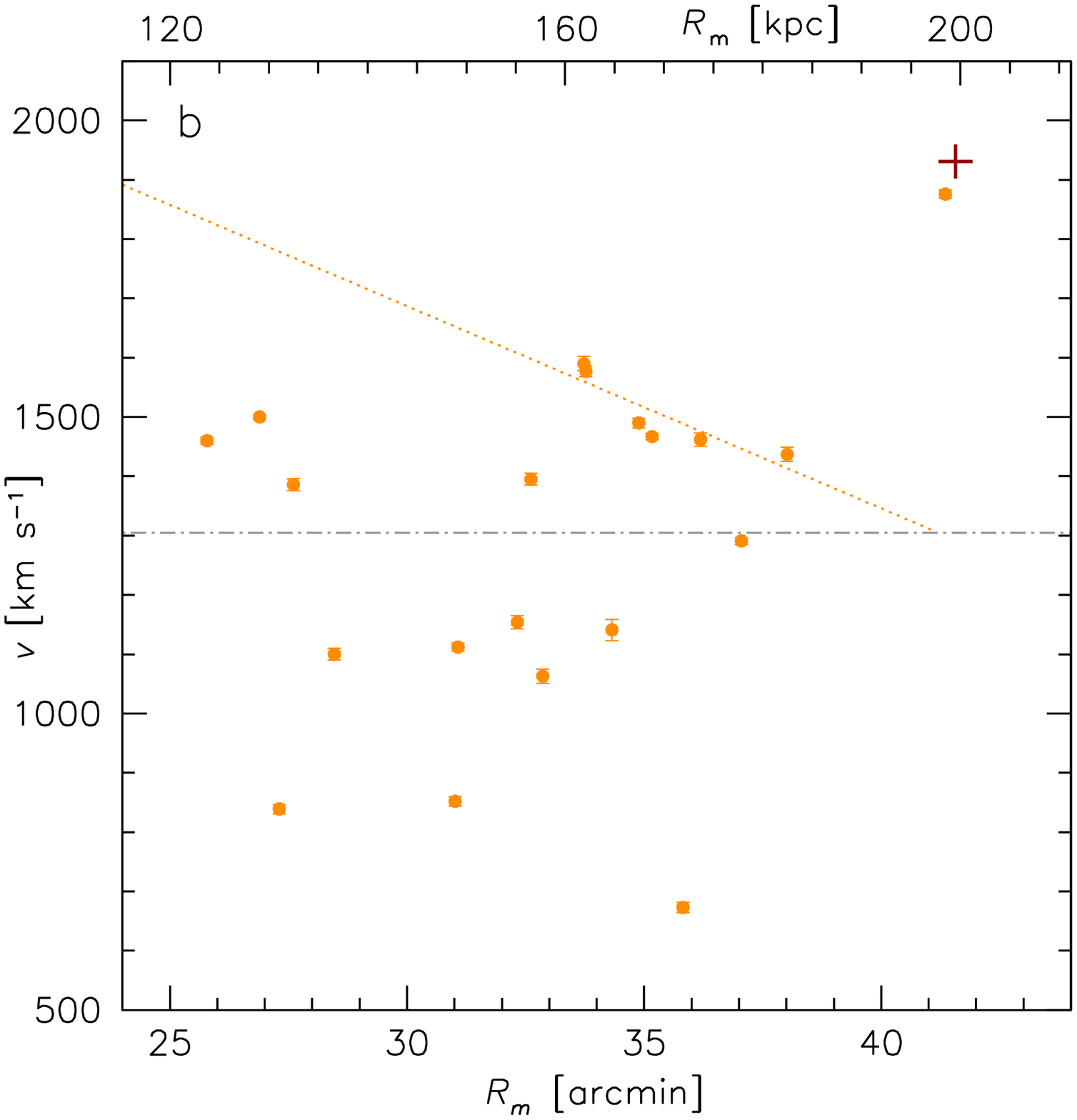}
\figcaption[stream]{\label{fig:stream}
Globular clusters near outer ``stream A''.
(a): positional space (a zoom-in of Figure~\ref{fig:main}(a)).
The five candidate stream GCs are outlined in purple
(the two easternmost ones are almost overlapping).
The bar with arrows again shows a 100~kpc scale.
(b): phase-space (using distance from the center of M87).
Velocity uncertainties are indicated by error bars.
One GC at upper right is probably bound to the low-luminosity S0 NGC~4461 (red cross).
The systemic velocity is shown by a horizontal line, and a diagonal
line illustrates our best-fit stream model.
}
\end{figure*}


\section{Outer stream analysis}\label{sec:outer}

We next analyze the GCs in the northwest area, around ``stream A''.
Figure~\ref{fig:stream} summarizes the positional space and phase space properties of this area.
The known stellar stream has a $V$-band luminosity of $(2.4\pm0.3)\times10^8 L_{V,\odot}$
over a $\sim$~15~$\times$~100~kpc region \citep{2010ApJ...715..972J}.

It is important to keep in mind that the spectroscopic observations were tiled
closely around the stellar stream, and much of the apparent narrowness in real
space of the overall GC distribution is an artifact of this bias.
The fractional return of bona fide GCs from the spectroscopic observations was indeed
higher ``on-stream'' than ``off-stream'', but the statistics here are relatively poor
and a more in-depth analysis will be required to consider this question further.

We focus instead on a possible cold phase-space substructure within the spectroscopic sample,
where the hotter background may be the stream or the general diffuse outer GC system
of M87.  This tentative substructure consists of 6 GCs (out of 19 in the sample apart
from one associated with a nearby galaxy) 
that are clustered in real-space and follow a cold diagonal path in phase space
(drawn schematically in Figure~\ref{fig:stream}(b)).
Two of these objects are a very close pair,
separated spatially by only 2.6~arcsec (0.2~kpc) and in
velocity by only 13~\kms, which is consistent with zero within the uncertainties.

To estimate the chances of this substructure being a chance clumping,
we pursue friends-of-friends group finding as previously carried out for the shell
(Section~\ref{sec:group}), but reverting
to a traditional three-dimensional ($x,y,v$) phase-space search.
In order to pick up the apparent group in the real data, we adopt
$w_x=w_y=$~50~arcsec (4.0~kpc), $w_v=$~30~\kms, $\lambda=0.29$.
We then simulate mock data sets using the same positions while drawing
velocities from a Gaussian LOSVD with dispersion 271~\kms,
and find groups with 6 or more members by chance only 9\% of the time.
Again, this analysis is conservative in that the mock data sets start from
spatial positions that may not be random, but not conservative in that
the group-finding parameters are optimized a posteriori.

We next carry out maximum-likelihood model fits to the data, using a 
Gaussian plus one-sided chevron model.
We estimate that $4^{+3}_{-2}$ of the GCs in this region belong to the stream,
which has a dispersion of $33^{+11}_{-25}$~\kms.
This model is intended as a characterization of the stream properties rather than
as a robust test of its reality,
since the fits are somewhat unstable to the starting conditions.
The model does not incorporate the full spatial information about the region,
and given that the stream objects are also clumped in the azimuthal direction,
we consider it likely that the 5 most closely associated objects seen in
Figure~\ref{fig:stream} comprise the stream.
Extrapolating to fainter magnitudes from our spectroscopic limit of $i_0\sim$~22.5,
we estimate a total of $\sim$~15~GCs in the stream clump.
However, our spectroscopic survey is incomplete, and there might easily be
$\sim$~20--30 stream GCs.

\begin{figure}
\epsscale{1.2}
\plotone{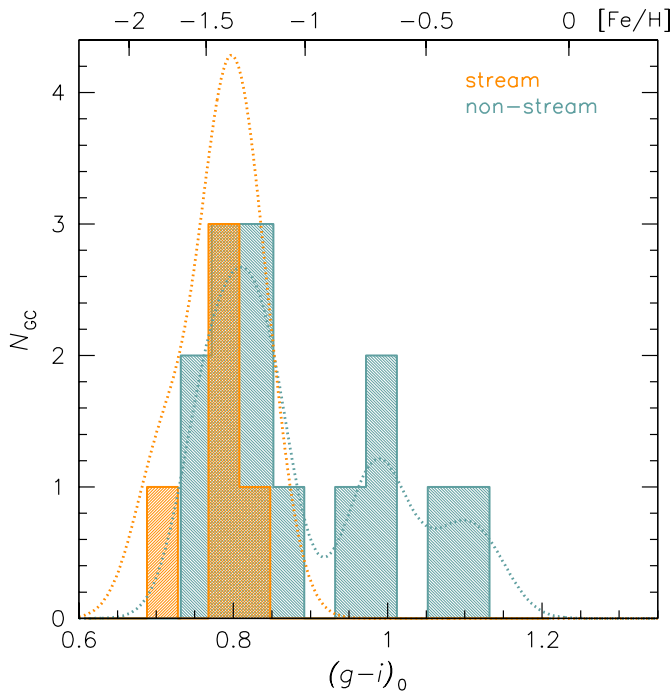}
\figcaption[streamcol]{\label{fig:streamcol}
Color distribution of GCs in the stream A region.
The two colored histograms show the stream and non-stream objects, per the legend,
with alternative smoothed distributions also shown (see Figure~\ref{fig:colors}(b)).
The stream GCs do not include any red objects, but the difference with the non-stream
GCs is not statistically significant.
}
\end{figure}

As with the shell analysis, we now consider the independent dimension of
metallicity (using color as a proxy).
Figure~\ref{fig:streamcol} shows that the stream GCs may have a different color distribution from
the rest of the GCs.  The former are narrowly confined to a blue color range of
$(g-i)_0=$~0.71--0.83 (a range in [Fe/H] of $\sim$~0.5~dex), while the latter
include a similar blue peak but also a second, much redder peak, with $(g-i)_0=$~0.97--1.13.
From the K-S test, this difference has a 22\% chance of occurring randomly.
As with the shell, the color information of the stream provides independent
confirmation of its distinctiveness.

As discussed for the shell region, we could in principle connect the peak color of the
stream GCs to a possible progenitor, but the statistics are currently too poor
for strong constraints.  Even so, we can still mention
one example of a potential analogue to the stream progenitor:
the Virgo dwarf elliptical IC~3328, with a stellar dispersion of $\sim$~35~\kms,
hosting $\sim$~40 GCs with a color peak near $(g-z)_0\sim$~0.92, i.e., $(g-i)_0\sim$~0.77.

Could this apparent GC stream from a dwarf galaxy be associated with the 
previously identified stellar stream?
The latter has a luminosity of $\log\,(L_V/L_{V,\odot})\sim$~8.4
that is consistent with this hypothesis
(compare the $L_V$-$\sigc$ trends for ACSVCS galaxies
in the  lower-left panel of Figure~\ref{fig:grid}).
Its color of $(B-V)\sim0.8$ also suggests a dwarf progenitor \citep{2010ApJ...720..569R}, although 
more massive progenitors are also possible within the uncertainties.

A puzzle is presented by the spatial offset of $\sim$~10~kpc between
the stellar and GC substructures (Figure~\ref{fig:stream}(b)).
One possibility that will require further modeling is
that the GCs and stars followed different trajectories owing to their different initial
binding energies.
The nucleus of the dwarf could also very well still be visible around M87,
although we have not identified an obvious candidate.

Another interesting possibility is that both the stream and the shell are part
of the same accretion event, since their peak GC colors and velocity dispersions
are similar.  Exploring this scenario will require more extensive 
simulations than we can attempt here.

\section{Theoretical analysis}\label{sec:sim}

We now attempt to model the dynamics of the observed substructures,
based mostly on numerical simulations but also making use of analytic
representations of orbit kinematics.

The narrow diagonal tracks of the stream and shell in phase space can be understood 
in simple terms as the near-radial infall of objects with similar initial potential
energies.  Analogous features are found in Local Group accretion
events \citep{2007ApJ...668..245G} and in simulations of minor mergers 
(e.g., \citealt{1984ApJ...279..596Q,1986A&A...166...53D,2007MNRAS.380...15F}).

\begin{table*}
\begin{center}
\caption{Parameters of Accretion Simulations}
\noindent{\smallskip}\\
\begin{tabular}{l c c c c c c c c c c c}
\hline
& Cold shell & Cold stream & Hot shell \\
\hline
satellite effective radius, $R_{\rm e}$ [kpc] &  1 & 1 & 4 \\
satellite velocity dispersion, $\sigma_{\rm p}(R_{\rm e})$ [\kms] & 35 & 35 & 200 \\
satellite mass, $M_{\rm sat}$ [$M_\odot$] & $1.8\times10^9$ & $1.8\times10^9$ & $3.2\times10^{11}$ \\
particle number, $N_{\rm p}$ & 65,536 & 65,536 & 65,536 \\
particle mass, $M_{\rm p}$ [$M_\odot$] & $2.75\times10^4$ & $2.75\times10^4$ & $4.9\times10^6$ \\
gravitational softening length, $r_{\rm l}$ [pc] & 20 & 20 & 22 \\
initial transverse velocity, $v_{\rm t}/v_{\rm c}$ & 0.05 & 0.1 & 0.1 \\
initial apocentric radius, $r_{\rm apo}$ [kpc] & 90 & 200 & 90 \\
initial pericentric radius, $r_{\rm peri}$ [kpc] & 2 & 12 & 4.7 \\
host dark halo virial mass, $M_{200}$ [$M_\odot$] & $4.2\times10^{14}$ & $4.2\times10^{14}$ & $4.2\times10^{14}$ \\
host dark halo virial radius, $r_{200}$ [Mpc] & 1.55 & 1.55 & 1.55 \\
host dark halo scale radius, $r_{\rm s}$ [Mpc] & 0.564 & 0.564 & 0.564 \\
host stellar effective radius, $R_{\rm e}$ [kpc] & 7 & 7 & 7 \\
host circular velocity at $R_{\rm e}$, $v_{\rm c}$ [\kms] & 475 & 475 & 475 \\
timestep, $t_{\rm s}$ [Myr] & 0.23 & 0.23 & 0.14 \\
\hline\label{tab:sims}
\end{tabular}
\end{center}
\end{table*}

The phase space structure of tidally stripped material can be influenced by many factors 
(e.g., \citealt{1988ApJ...331..682H,1989ApJ...342....1H,1998ApJ...495..297J,1999Natur.402...53H,2002ApJ...570..656J,2004ApJ...610L..97H,2009ApJ...699.1518R}),
which we summarize in broad terms as follows.
More massive satellites produce more diffuse streams owing to the higher
internal velocity dispersions and larger sizes of these systems.
In cuspy potentials, low angular momentum orbits can
bring the satellites close to the potential center and lead to broader shells
or fan-like features, while in orbits farther out, the weaker tidal shear
produces narrower loops and streams.
Finally, any dynamically cold substructure will tend to diffuse over time
due either to nonaxisymmetry in the host potential or to
perturbations from other satellites.

As a proof of principle, we carry out a basic set of simulations
inspired by M87 and its substructures.
These are not intended to explore the (very large)
orbital and progenitor parameter space for the infalling satellite galaxy,
or to provide a unique, robust match to the data,
but to illustrate qualitatively how
comparable features can emerge in phase-space
from the tidal stripping of infalling galaxies.

We conduct $N$-body
simulations of self-gravitating low mass galaxies orbiting in a fixed
spherical gravitational potential. For this cluster-mass potential,
we use a Navarro-Frenk-White model for the dark matter halo 
\citep{1996ApJ...462..563N,1999ApJ...512L...9M},
and embed a \citet{1990ApJ...356..359H} profile to represent the stellar mass of
the central galaxy.
The relevant parameters are reported in Table~\ref{tab:sims}.

For the inner shell, 
the main features that we wish to reproduce are the apparent chevron morphology
in phase space including a low velocity dispersion, with a high
degree of azimuthal mixing in real-space (see Figure~\ref{fig:main}).
For the outer stream, the goal is to find a thin structure in both
real and phase-space, which is not closely connected to the progenitor galaxy
or nucleus.

For both features, we adopt the same low mass galaxy model for the infalling satellite,
and consider a modest number of orbital configurations.
The infalling satellite galaxy follows a single-component Hernquist model,
with characteristic parameters summarized in Table~\ref{tab:sims}.
A real galaxy will consist of multiple components (stars, GCs, gas, and dark matter),
but this introduces additional degrees of freedom to the problem which we 
are for now keeping as conceptually simple as possible.

We place the satellite on elongated orbits around the potential center,\footnote{Our
adopted apocentric distances of $\sim$~100--200~kpc may not seem consistent with 
satellites that fall in from the large-scale environment, but our goal is to
set up an idealized model of accretion dynamics once an object has migrated
inwards by scattering or dynamical friction (e.g., \citealt{2005MNRAS.362..498F}).}
and run the simulations using an $N$-body treecode \citep{1987ApJS...64..715H}, which simulates
the fully self-gravitating response of the infalling satellite.
The simulations were evolved for several Gyr using a fixed timestep of 0.25
Myr. We create real-space and phase-space plots of all the satellite particles
for each time-step, using a small selection of viewing angles.
We then select by eye the best combination of timestep and viewing angles that
qualitatively resembles the data.

Next, we sample a subset of the particles to represent the GC observations,
plotting $10^4$ particles
to obtain a general sense of the substructure morphology, and an additional
subsample of 100 particles
for more direct comparison with the data.
Our default selection consists of an unbiased, random sample,
which can be considered as representing the limiting case where most of
the dark halo has been stripped away, with the remaining self-gravitating material 
traced well by the stars and GCs.
In almost every real galaxy, the metal-poor GCs occupy a more extended distribution
than the field stars, so we have tried the alternative limiting case where
the least bound particles are selected as ``GCs''.
The results from the two sampling schemes turn out to be qualitatively similar.

Clearly, our procedures are only the initial steps toward modeling the M87
halo kinematics.  Ideally, one would like to: use a live, triaxial potential that
includes the effects of dynamical friction and other substructures; 
explore a wide range of satellite and orbital parameters;
and use an objective, quantitative metric for optimizing the model fits.
However, our current approach is adequate for providing a broad sense of
some of the types of substructure signatures that are physically plausible, 
which one can see in related analyses of other systems
\citep{1988ApJ...331..682H,1998MNRAS.297.1292M,2001AJ....122.1397H,2005ApJ...635..931B,2007ApJ...668..245G,2007MNRAS.380...15F,2010MNRAS.408L..26P}.

The best qualitative matches that we identified for the shell and the stream (separately)
were shown in Figure~\ref{fig:sims}.  
The shell progenitor has made a close passage to the center of
M87 and was largely disrupted, leaving a classic chevron signature in phase space,
while also being spread out in positional space.
The stream progenitor has a larger pericenter and is disrupted more gradually,
maintaining integrity as a narrow stream in both phase space and positional space for
many orbits.  

In a real system, any triaxiality of the potential would cause the orbits to
phase-mix more rapidly, and the time-evolution of the mass distribution would
presumably smear out the cold substructures even further
(as well as inhibit any orbital resonances that might in principle
be an alternative mechanism for producing cold kinematical features).
Simulations of cluster-scale tidal streams
in a cosmological context \citep{2009ApJ...699.1518R} indicate that streams
are disrupted over timescales of $\sim$~2--4 times the local dynamical timescale,
$t_{\rm dyn} \sim r/v_{\rm c}$.
For the shell and stream, $t_{\rm dyn} \sim$~0.2 and 0.5~Gyr, respectively, so their
most likely lifetimes are $\sim$~0.5~Gyr and $\sim$~1.5~Gyr.

One important feature of coherent substructures is that they trace out a close family
of orbits over large scales in galaxy halos, and can thus be used to provide unique
constraints on the gravitational potential, including the radial distribution and shape
of the dark matter halo (e.g., 
\citealt{2009ApJ...703L..67L,2010ApJ...712..260K,2011MNRAS.417..198V}).
The case of a spherical system is conceptually simple:
the potential energy lost by particles as they fall in towards the center of the host
galaxy is reflected in a monotonic trend of increasing velocities with decreasing radius,
modulo projection effects \citep{1998MNRAS.297.1292M}.

We derive a new formulation of this method, assuming a spherical gravitational
field of a power-law form, where the total density, circular velocity, and potential are:
\begin{equation}
\rho_{\rm tot} \propto r^{-\alpha} , \\
\end{equation}
\begin{equation}
v_{\rm c} = v_0 \left(\frac{r}{r_0}\right)^{1-\frac{\alpha}{2}} ,
\end{equation}
\begin{equation}
\Phi(r) = \frac{v_0^2}{2-\alpha} \left(\frac{r}{r_0}\right)^{2-\frac{\alpha}{2}} ,
\end{equation}
where $\alpha < 2$, as is probably appropriate on $\sim$~10--100~kpc scales in M87
(cf. also section~2.2(f) of \citealt{2008gady.book.....B}).
Following equation (3) of \citet{1998MNRAS.297.1292M}, 
we consider a thin shell of radius $r_{\rm apo}$
where particles have zero velocity
(i.e., they are on radial orbits at apocenter).
By conservation of energy and by geometry, 
their line-of-sight velocities at 3-D radius $r$ and 2-D radius $R$ are then
given by
\begin{equation}
v_{\rm LOS}^2 = 2 \left(1-\frac{R^2}{r^2}\right) \left[\Phi(r_{\rm apo})-\Phi(r)\right] .
\end{equation}
The velocity ``edge'' of the shell is then determined by solving for the maximum
of $v_{\rm LOS}$ at a given radius $R$.
By fitting this simple model to a cold phase-space feature, we can then estimate
the value of $v_{\rm c}$ near $r_{\rm apo}$.
An even more simplified way to understand this constraint \citep{1998MNRAS.297.1292M} 
is that the slope of the shell velocity edge with radius in phase-space is approximately
equal to $v_{\rm c}/r_{\rm apo}$.
In practice, the dominant effect is the change of potential energy, so the determination
of $v_{\rm c}$ is fairly insensitive to the form adopted for $\Phi(r)$.

\begin{figure}
\epsscale{1.2}
\plotone{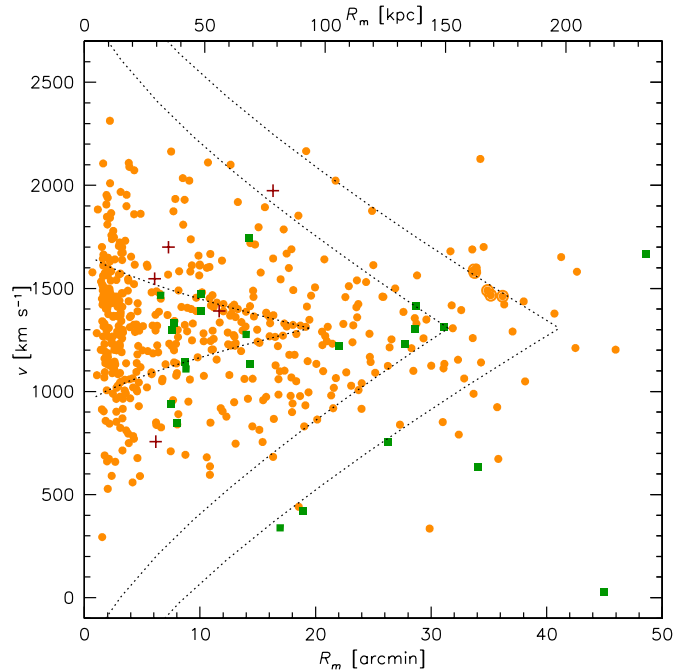}
\figcaption[potent]{\label{fig:potent}
Phase-space diagram of M87 as in Figure~\ref{fig:main}(b), with three monoenergetic model curves over-plotted
(see text for details).
GC and satellite galaxy data are shown by orange dots and red crosses as before, with
planetary nebulae \citep{2009A&A...502..771D} also now included as green squares.
The outer stream subgroup of GCs (at $R_m \sim$~35~arcmin) is highlighted with larger circles.
The inner shell and outer stream GCs can be represented well by model curves as shown,
and there is a hint of an intermediate-radius shell as marked, whose apex could explain
the low velocity dispersion found in a subgroup of the PNe.
\vskip 10pt
}
\end{figure}

\begin{figure*}
\epsscale{0.56}
\plotone{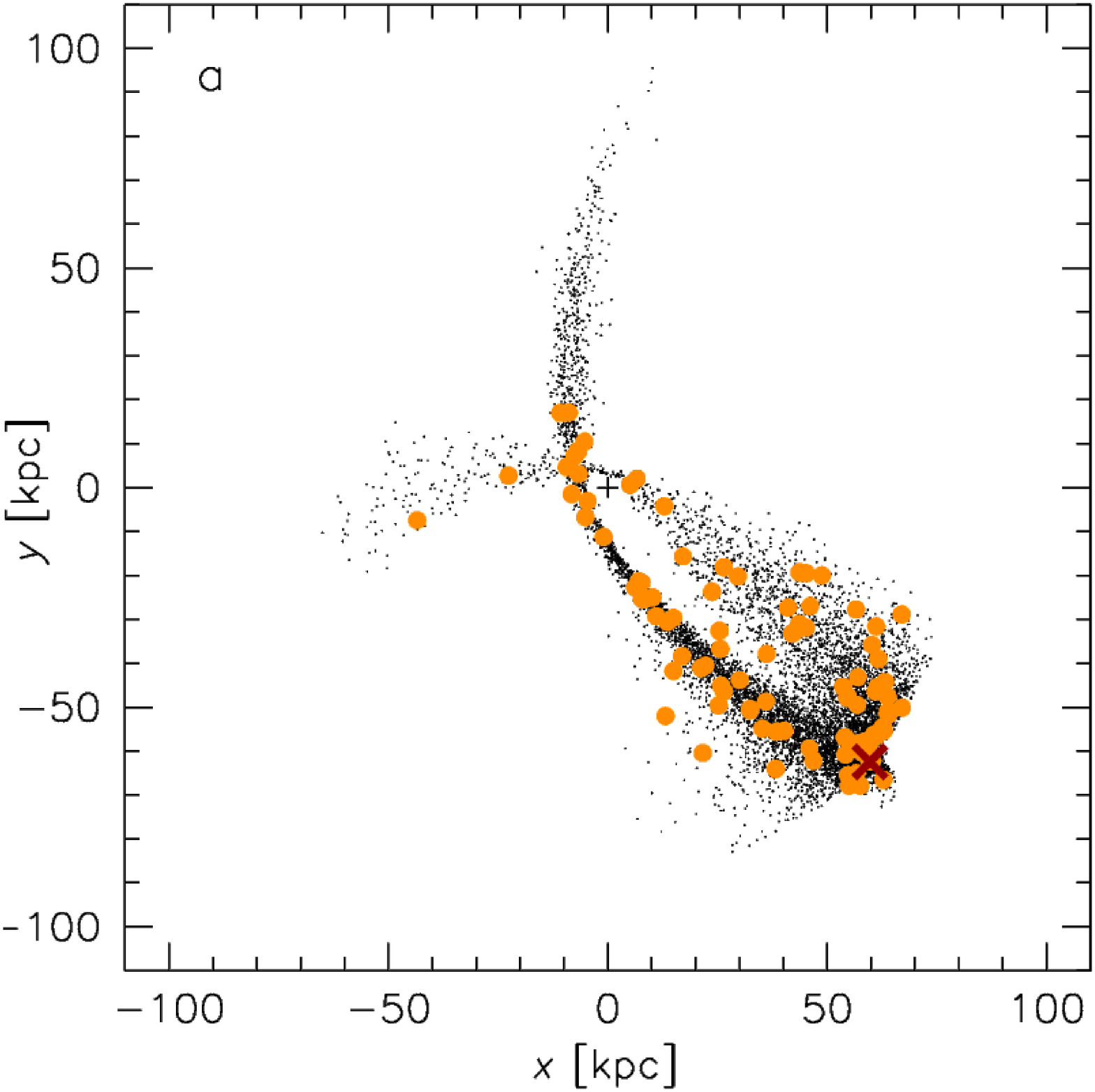}
\hspace{0.1cm}
\plotone{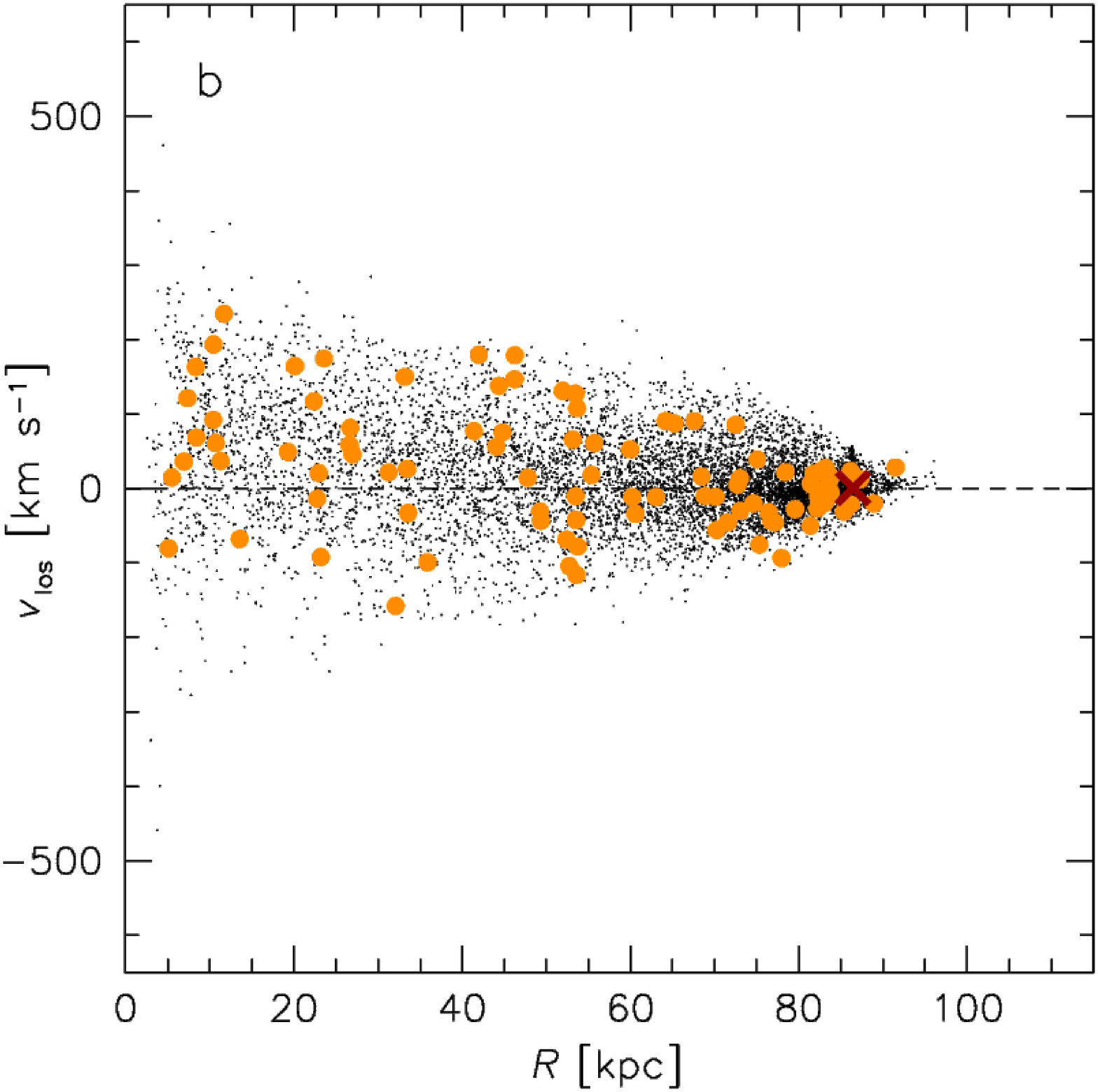}
\figcaption[shell9ishell9h]{\label{fig:cold}
Cold accretion model with higher initial angular momentum,
viewed near face-on (compare Figure~\ref{fig:sims};
note that both the vertical and horizontal scales have been
shrunk in tandem for a fair comparison of angles).
The snapshot corresponds to 2~Gyr after the initial apocenter.}
The opening angle of the chevron is smaller than in Figure~\ref{fig:sims}(b),
and in position-space the stream stays highly coherent for many orbits
(in the absence of perturbations in the potential).
\end{figure*}

In Figure~\ref{fig:potent} we show this model applied to the stream and to two possible shells in M87,
where we have taken the maximum-likelihood linear-slope fits from Sections~\ref{sec:max}
and \ref{sec:outer} as starting points for approximate fits in phase-space.
The model reproduces nicely the observed radial trends, 
and suggests some possible connections in phase-space between the shells and nearby 
low-luminosity ellipticals.
Also shown in the figure are halo PNe, which show hints of tracing
some of the same substructures seen in the GCs.
More precise orbital fits to the substructures and the potential progenitors will
require dedicated dynamical modeling (e.g., \citealt{2001ApJ...553..722R,2011ApJ...729..129M}).

There is a hitch with these model curves:
the implied circular velocities differ radically from typical estimates for the
M87 potential.
At $r \sim$~20, 32, 41~arcmin ($\sim$~90, 150, 200~kpc), we infer 
$v_{\rm c} \sim$~270, 1200, 1400~\kms, albeit with substantial uncertainties.
Previous estimates from dynamical and X-ray analyses yielded
$v_{\rm c} \sim$~650--900~\kms\ over this entire radial range \citep{2010MNRAS.409.1362D,2011ApJ...729..129M}, while our own equilibrium-based dynamical analysis
indicates $v_{\rm c} \sim$~400--550~\kms\ (S+11).\footnote{The local
escape velocity is {\it at least} $v_{\rm e}=2^{1/2}\,v_{\rm c}$.  We regard the
$v_{\rm c}\sim$~550~\kms\ model as plausibly describing a massive group-sized
dark matter halo around M87, in which case $v_{\rm e} \ga 800$~\kms.
This would indicate a bound-velocity range of $\sim$~500--2100~\kms, and probably
wider, suggesting that most or all of the objects in Figure~\ref{fig:potent}
are bound to M87 (modulo the unknown proper-motion velocities).}
These discrepancies should not be a major concern at this point because there are
various complications to consider including 
the radius convention of Section~\ref{sec:rad}, as well as
non-spherical effects \citep{2010ASPC..423..243J}.
In particular, a planar orbit viewed near face-on will show depressed line-of-sight
velocities, and a reduced opening angle of the chevron.

An illustration of the latter effect is provided in 
Figure~\ref{fig:cold}, which shows an alternative orbit for the shell that 
has higher angular momentum than our
fiducial case, and maintains better planar integrity.
Here the ``measured'' $v_{\rm c}$ would be $\sim$~250~\kms, despite the
true value in the model being 635~\kms.
Note that the simulation shown is also representative of a large family of orbits that are
not a good match to the shell observations because of the high degree of spatial coherence.

As discussed at length in Section~\ref{sec:prog}, there is an apparent conflict
between the shell's inferred kinematical coldness and the large number of GCs,
since these imply very different progenitor masses.
We consider this issue more explicitly by changing the simulated shell progenitor
to correspond to a more massive galaxy, 
with parameters summarized in Table~\ref{tab:sims}.
%
%

\begin{figure*}
\epsscale{0.56}
\plotone{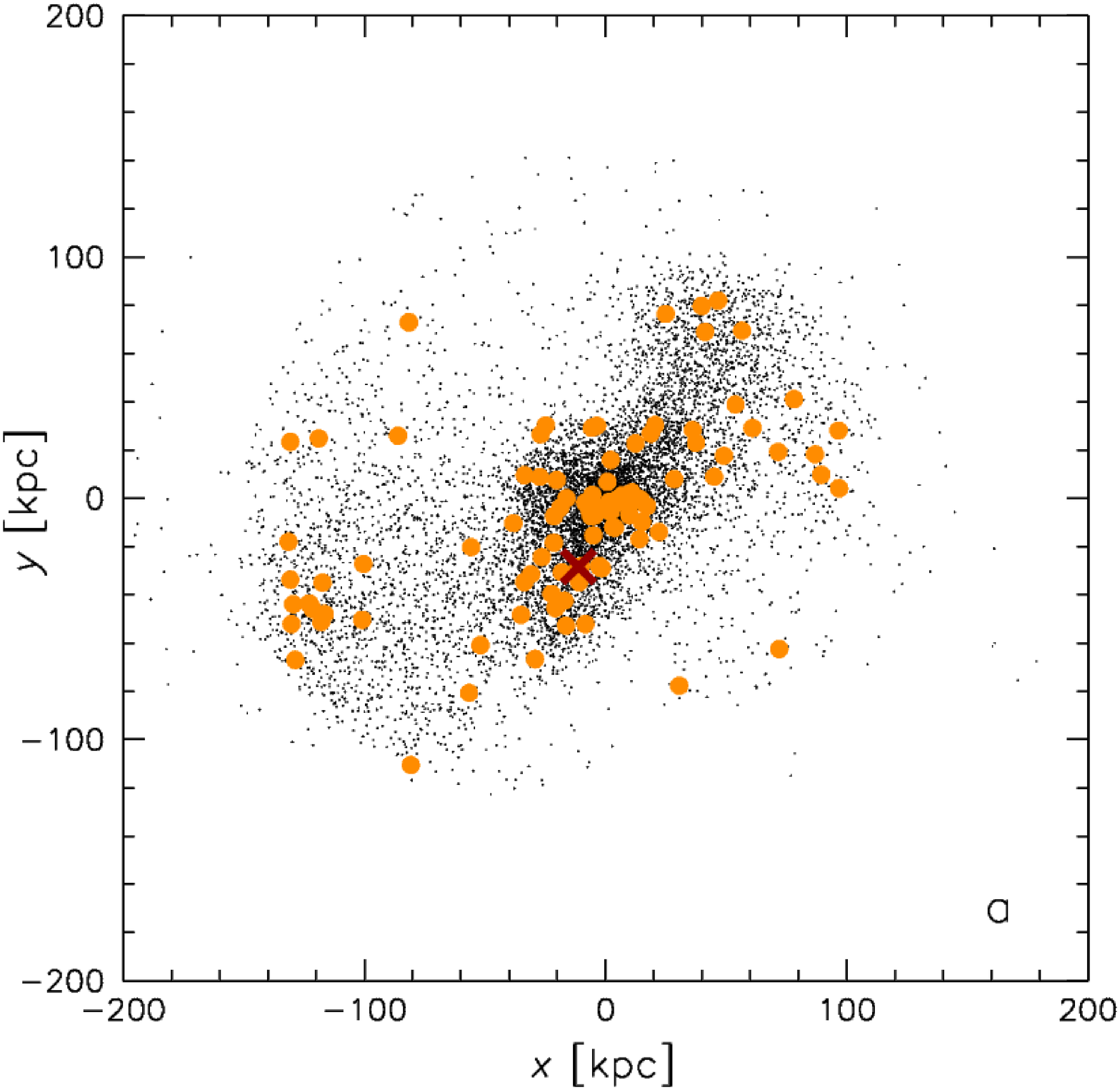}
\hspace{0.1cm}
\plotone{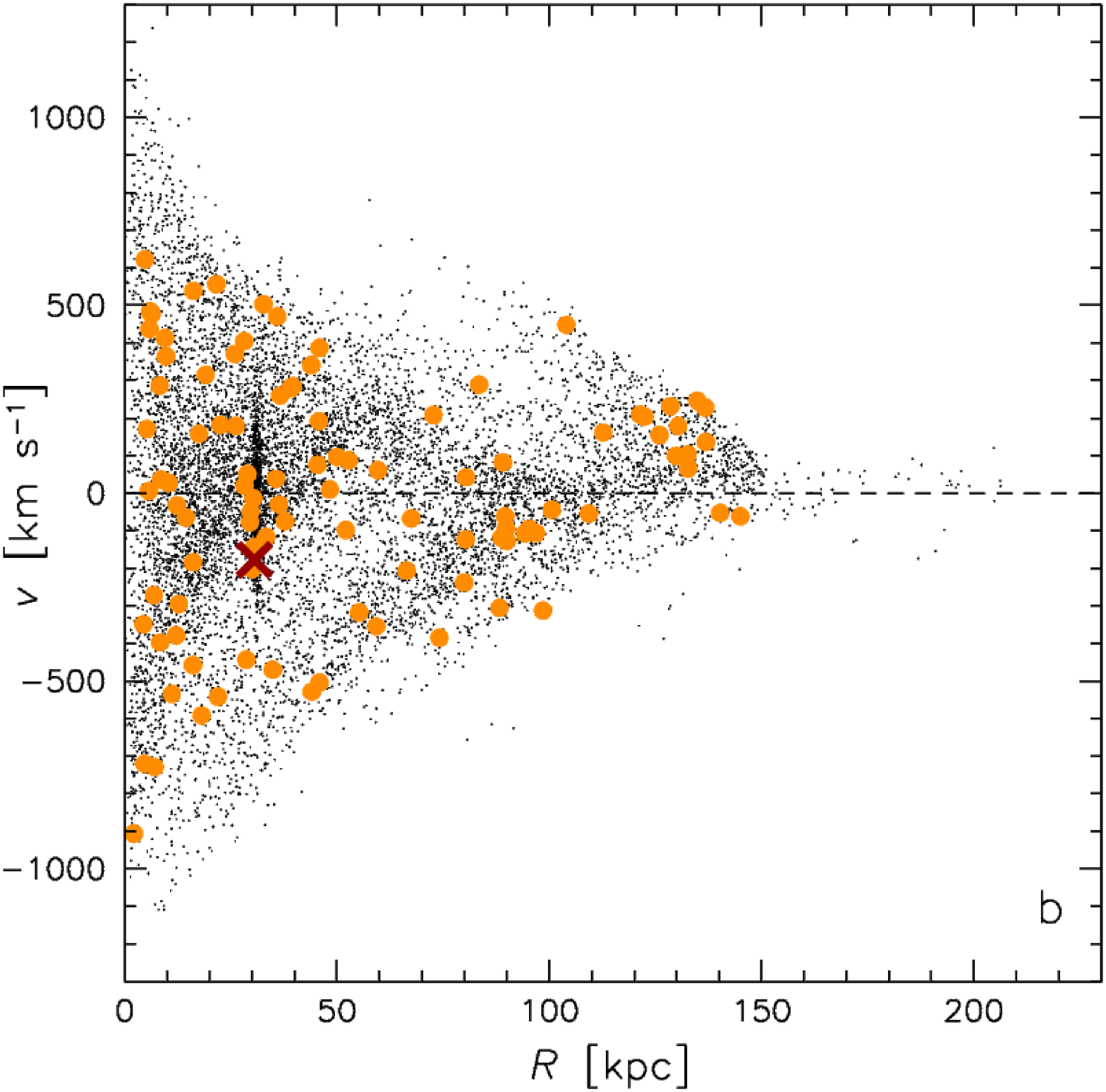}
\figcaption[shell6ushell6t]{\label{fig:sims2}
Hot accretion model, similar to the shell simulation in Figure~\ref{fig:sims} but
with a 100$\times$ more massive progenitor.
Here the snapshot corresponds to 1.5~Gyr after the initial apocenter.
Similar features are present as with the low-mass galaxy but are less distinct.
}
\end{figure*}

We show an example post-disruption snapshot in
Figure~\ref{fig:sims2}, finding
as expected that the resulting shell kinematics are hotter than in 
the dwarf-galaxy simulation, and do not compare as favorably to the observations
(compare Figures~\ref{fig:sims}(b) and \ref{fig:radii}(b)).
However, as previously discussed, this tension is alleviated if we modify our
default radius convention and chevron-based model
(compare Figure~\ref{fig:radii}(a)).

These simulations bracket the simple cases of very low- and high-mass progenitor
galaxies,
but a worthwhile endeavor for the future would be simulating
more detailed intermediate-mass cases, including bulge, disk, stellar halo,
and dark matter halo subcomponents.
An additional aspect that could be pursued is the modeling of
velocity gradients with azimuthal angle, which should be generically found in
tidal features (e.g., \citealt{2003ApJ...592L..25H}).
Our initial look at the M87 shell data shows no clear sign of such effects.

\section{Summary}\label{sec:concl}

We have used an unprecedentedly wide-field data set of high-precision globular cluster
velocities to search for phase-space substructure around the central cluster galaxy M87.
The large majority of the data are from S+11, with a few additional velocities obtained
from the literature, and from a new sample of Keck/DEIMOS spectroscopy presented here that
targets a known filament of stellar light in the outer halo.
Analysis of ($g-i$) colors and spectroscopic metallicities for a subset of our GC sample
confirms that the colors provide a good proxy for metallicity, allowing for
chemical tagging of substructures.

We identify two candidate features: one a small stream near to the outer-halo stellar filament,
and the other a previously unknown colossal ``shell'' in the inner halo.
We have carried out extensive statistical tests and characterizations of these substructures.

The outer GC stream appears to be a collection of $15\pm10$ objects (after extrapolating
to fainter magnitudes) residing in a $\sim$~5~$\times$~20~kpc clump
with an internal velocity dispersion of $\sim 30\pm20$~\kms.
We have found the presence of this stream to be significant both in a
group-finding analysis $(p=0.09)$, from a maximum likelihood model $(p=0.025)$,
and from its distinctive color distribution compared to the surrounding GCs $(p=0.22)$.

The stream dispersion, its GC colors and numbers counts, the color of the stellar filament,
and the filament's narrowness all suggest the accretion of a dwarf galaxy.
Low-mass kinematical substructures in elliptical galaxies have been
inferred in other studies \citep{2003ApJ...591..850C,2011AJ....141...27W},
but in this case the association with a visible counterpart
should allow for a much clearer determination of the accretion dynamics.

The shell of GCs has a chevron-like shape in phase-space that 
resembles classic expectations for a disrupted
infalling system \citep{1988ApJ...331..682H}, although
the interpretation in detail is more challenging.
The appearance of the shell is sharpened by
a change of coordinates from simple radius to elliptical, circular-equivalent radius.
The significance of the shell in phase-space is established using a group-finding algorithm ($p<0.01$),
a crude entropy metric ($p=0.31$), and a maximum-likelihood chevron-based model ($p<0.01$ compared to a pure Gaussian model).
The color distribution of the shell and non-shell GCs from the same region is different $(p=0.13)$, providing independent evidence for the existence of the shell.
The phase-space morphology resembles a simple chevron model more than a wedge or tapered Gaussian, but may very well have a complex, intermediate morphology;
however, such uncertainties do not compromise our basic finding of
a relatively massive and cold substructure.

We find from maximum-likelihood fitting that the shell 
contains 500--1100 GCs, which is $\sim$~5 times the entire Milky Way system,
and comprises around 20\% of M87's GCs in the 50--95~kpc region.
This number suggests a stellar debris field of (2--6)$\times10^{10} L_\odot$,
brought in by either a large group of dwarf galaxies, or by
a single giant elliptical or lenticular galaxy.  

The dramatic appearance of the shell in phase-space is due to its 
relatively cold kinematics,
with an estimated velocity dispersion of 10--30~\kms\ 
(or up to $\sim$~120~\kms\ if
we make generous allowances for uncertainties concerning the shell morphology).
The lower velocity dispersion estimates
are difficult to reconcile with the number of inferred GCs,
as they most naturally imply a dwarf galaxy hosting only a handful of GCs.
However, after allowing for systematic uncertainties in our analyses,
we suggest a solution involving an E/S0 progenitor with a luminosity of
$\sim 0.5\,L^*$, which could very well also be the parent galaxy for one
of the tidally-stripped galaxies now visible close in to M87.

There is a critical need for additional data from the shell region, including new velocities
of both GCs and PNe, in order to confirm the presence of the substructure and to 
refine our estimates of its characteristics.
Along these lines, we have recently obtained another cycle of GC spectroscopy from
MMT/Hectospec that includes $\sim$~35 velocity measurements of independent objects
in the shell region.
A preliminary analysis yields no obvious sign of substructure,
but the statistics from the new data are poor, and at least another $\sim$~100 velocities 
are needed.

We have carried out simplified $N$-body simulations of satellite galaxy 
accretion in a static cluster$+$BCG potential,
finding that these readily produce streams and chevrons in phase space.  
In one simulation, a dwarf galaxy has recently fallen in and is partially disrupted,
forming a classic, narrow tidal stream in both positional and phase space.
In a second case, the initial infall orbit had much lower angular momentum,
so that the satellite galaxy passed close to the center of M87 and was quickly disrupted.
Its remnants are strewn more haphazardly across
positional space and could be difficult to discern against the main body of M87,
while a distinct chevron persists in phase space, with the apex corresponding
to a shell turning point.

These simulations compare favorably to the M87 observations because the low
internal velocity dispersions of the dwarf progenitors result in cold kinematics
for the tidal debris.  Simulations of a more massive progenitor produce
hotter kinematics that do not as strongly resemble the data for the ``shell'', 
reiterating the tension between the different constraints on the shell progenitor mass,
which remains unresolved pending further observations and modeling.
It would be worth also investigating alternative explanations for the substructure,
such as a shell-like boundary between well-mixed GC subpopulations formed or accreted in
different epochs.

We have explored the use of the velocity-radius slopes of the substructures in phase-space 
estimate the underlying gravitational potential and dark matter distribution
around M87.
However, the interpretation of the data appears to be complicated by non-spherical effects.

Taken at face value, the shell GCs imply a substantial, recent accretion event in the halo of M87.
This is the first large substructure identified around a central cluster galaxy,
the first large stellar substructure with a clear kinematical detection 
in {\it any} type of galaxy beyond the Local Group, 
and by far the largest shell or stream discovered in the Local Universe.

M87 was a favored target for our survey as well as previous ones because of its
proximity, rich GC system, and {\it general lack of obvious dynamical disturbance}.
The large shell we have discovered is an example of the lurking accretion
signatures that could be found from detailed phase-space studies of
apparently placid galaxies.

There are however more indirect lines of evidence for a recent minor merger
in M87, such as an off-center supermassive black hole \citep{2010ApJ...717L...6B}, 
a lopsided central GC velocity distribution (S+11),
an accretion powered jet, and several surrounding
peculiar low-luminosity satellites that could be the nuclei of since-shredded
disk galaxies.
Further theoretical work is needed to see if these features could all be
explained by
a unified scenario involving a recent gas-rich minor merger.

The shell and stream in M87 have probable lifetimes of $\sim$~0.5~Gyr and $\sim$~1.5~Gyr, respectively.  Given our best guesses for the progenitors of these features, the total luminosity of M87 would be built up by $\sim$~10 or $\sim$~1000 events corresponding to the shell or the stream, respectively.
If these recent events are representative of the long-term evolution of M87, and by
extension of the class of brightest cluster galaxies in general,
then they support a picture of late-epoch growth by hierarchical assembly
\citep{2007MNRAS.375....2D}.

More generally, the M87 results are another demonstration of the potency of 
GCs for providing unique information about the stellar halos of galaxies.
We note the concluding prediction of \citet{2009ApJ...699L.178N} 
for their landmark two-phase galaxy formation scenario:
``...the outer parts of massive giant ellipticals will tend to be old,
blue, metal-poor, and relatively uniform from galaxy to galaxy
since they are all composed essentially of the debris from tidally
destroyed accreted small systems.''
Globular clusters have in fact long provided evidence for exactly this scenario.

\vskip 5pt

\acknowledgements

We thank Nelson Caldwell 
for assistance with observations; Jason X. Prochaska and
Kate Rubin for software help; Magda Arnaboldi for providing results in electronic form;
Lars Hernquist, Bill Mathews, and Mike Merrifield for helpful discussions;
and Eric Emsellem for a constructive review.\\
J.B. and A.J.R. acknowledge support from the NSF through grants AST-0808099, AST-0909237,
AST-1101733, and AST-1109878,
and by the UCSC University Affiliated Research Center's Aligned Research Program.
J.S. was supported by NASA through a Hubble Fellowship, administered by the Space Telescope Science Institute, which is operated by the Association of Universities for Research in Astronomy, Incorporated, under NASA contract NAS5-26555. 
J.C.M. has been supported by the NSF through grants AST-0607526 and AST-0707793.
We acknowledge financial support from the {\it Access to Major Research Facilities
Programme}, a component of the {\it International Science
Linkages Programme} established under the Australian
Government's innovation statement, {\it Backing Australia's Ability}.\\
Much of the data presented herein were obtained at the W.~M.~Keck Observatory, which is operated as a scientific partnership among the California Institute of Technology, the University of California and the National Aeronautics and Space Administration. The Observatory was made possible by the generous financial support of the W.~M.~Keck Foundation.
Observations reported here were obtained at the MMT Observatory, a joint facility of the Smithsonian Institution and the University of Arizona.
Based in part on data collected at Subaru Telescope (which is operated by the National Astronomical
Observatory of Japan),
via a Gemini Observatory time exchange (GN-2009A-C-204).
This research used the facilities of the Canadian Astronomy Data Centre operated by the National Research Council of Canada with the support of the Canadian Space Agency. 
We acknowledge the usage of the HyperLeda database (http://leda.univ-lyon1.fr).
This research has made use of the NASA/IPAC Extragalactic Database (NED) which is operated by the Jet Propulsion Laboratory, California Institute of Technology, under contract with the National Aeronautics and Space Administration. 
Based on observations made with the NASA/ESA Hubble Space Telescope, and obtained from the Hubble Legacy Archive, which is a collaboration between the Space Telescope Science Institute (STScI/NASA), the Space Telescope European Coordinating Facility (ST-ECF/ESA) and the Canadian Astronomy Data Centre (CADC/NRC/CSA).


\begin{thebibliography}{}

\bibitem[Arnold et al.(2011)]{2011ApJ...736L..26A} Arnold, J.~A., Romanowsky, A.~J., Brodie, J.~P., Chomiuk, L., Spitler, L.~R., Strader, J., Benson, A.~J., \& Forbes, D.~A.\ 2011, \apjl, 736, L26 
\bibitem[Arp \& Bertola(1971)]{1971ApJ...163..195A} Arp, H., \& Bertola, F.\ 1971, \apj, 163, 195 
\bibitem[Ascaso et al.(2011)]{2011ApJ...726...69A} Ascaso, B., Aguerri, J.~A.~L., Varela, J., Cava, A., Bettoni, D., Moles, M., \& D'Onofrio, M.\ 2011, \apj, 726, 69 
\bibitem[Ashman et al.(1994)]{1994AJ....108.2348A} Ashman, K.~M., Bird, C.~M., \& Zepf, S.~E.\ 1994, \aj, 108, 2348 
\bibitem[Batcheldor et al.(2010)]{2010ApJ...717L...6B} Batcheldor, D., Robinson, A., Axon, D.~J., Perlman, E.~S., \& Merritt, D.\ 2010, \apjl, 717, L6 
\bibitem[Bellazzini et al.(2003)]{2003AJ....125..188B} Bellazzini, M., Ferraro, F.~R., \& Ibata, R.\ 2003, \aj, 125, 188 
\bibitem[Belokurov et al.(2006)]{2006ApJ...642L.137B} Belokurov, V., et al.\ 2006, \apjl, 642, L137 
\bibitem[Bergond et al.(2006)]{2006A&A...448..155B} Bergond, G., Zepf, S.~E., Romanowsky, A.~J., Sharples, R.~M., \& Rhode, K.~L.\ 2006, \aap, 448, 155 
\bibitem[Bernardi(2009)]{2009MNRAS.395.1491B} Bernardi, M.\ 2009, \mnras, 395, 1491 
\bibitem[Binney \& Tremaine(2008)]{2008gady.book.....B} Binney, J., \& Tremaine, S.\ 2008, Galactic Dynamics: Second Edition, Princeton University Press
\bibitem[Brodie \& Strader(2006)]{2006ARA&A..44..193B} Brodie, J.~P., \& Strader, J.\ 2006, \araa, 44, 193
\bibitem[Brodie et al.(2011)]{2011AJ....142..199B} Brodie, J.~P., Romanowsky, A.~J., Strader, J., \& Forbes, D.~A.\ 2011, \aj, 142, 199
\bibitem[Brough et al.(2011)]{2011MNRAS.414L..80B} Brough, S., Tran, K.-V., Sharp, R.~G., von der Linden, A., \& Couch, W.~J.\ 2011, \mnras, 414, L80 
\bibitem[Buitrago et al.(2008)]{2008ApJ...687L..61B} Buitrago, F., Trujillo, I., Conselice, C.~J., Bouwens, R.~J., Dickinson, M., \& Yan, H.\ 2008, \apjl, 687, L61 
\bibitem[Bullock \& Johnston(2005)]{2005ApJ...635..931B} Bullock, J.~S., \& Johnston, K.~V.\ 2005, \apj, 635, 931 
\bibitem[Carollo et al.(2007)]{2007Natur.450.1020C} Carollo, D., et al.\ 2007, \nat, 450, 1020 
\bibitem[Cassata et al.(2011)]{2011arXiv1106.4308C} Cassata, P., et al.\ 2011, \apj, in press, arXiv:1106.4308 
\bibitem[Chilingarian(2009)]{2009MNRAS.394.1229C} Chilingarian, I.~V.\ 2009, \mnras, 394, 1229 
\bibitem[Coccato et al.(2009)]{2009MNRAS.394.1249C} Coccato, L., et al.\ 2009, \mnras, 394, 1249 
\bibitem[Coccato et al.(2010)]{2010MNRAS.407L..26C} Coccato, L., Gerhard, O., \& Arnaboldi, M.\ 2010, \mnras, 407, L26 
\bibitem[Cohen et al.(1998)]{1998ApJ...496..808C} Cohen, J.~G., Blakeslee, J.~P., \& Ryzhov, A.\ 1998, \apj, 496, 808 
\bibitem[Cohen \& Ryzhov(1997)]{1997ApJ...486..230C} Cohen, J.~G., \& Ryzhov, A.\ 1997, \apj, 486, 230 
\bibitem[Collins et al.(2009a)]{2009Natur.458..603C} Collins, C.~A., et al.\ 2009a, \nat, 458, 603 
\bibitem[Collins et al.(2009b)]{2009MNRAS.396.1619C} Collins, M.~L.~M., et al.\ 2009b, \mnras, 396, 1619
\bibitem[Cooper et al.(2011)]{2011arXiv1111.2864C} Cooper, A.~P., Martinez-Delgado, D., Helly, J., et al.\ 2011, ApJL, in press, arXiv:1111.2864 
\bibitem[Cortesi et al.(2011)]{2011MNRAS.414..642C} Cortesi, A., Merrifield, M.~R., Arnaboldi, M., et al.\ 2011, \mnras, 414, 642 
\bibitem[C{\^o}t{\'e} et al.(1998)]{1998ApJ...501..554C} C{\^o}t{\'e}, P., Marzke, R.~O., \& West, M.~J.\ 1998, \apj, 501, 554 
\bibitem[C{\^o}t{\'e} et al.(2003)]{2003ApJ...591..850C} C{\^o}t{\'e}, P., McLaughlin, D.~E., Cohen, J.~G., \& Blakeslee, J.~P.\ 2003, \apj, 591, 850 
\bibitem[Damjanov et al.(2009)]{2009ApJ...695..101D} Damjanov, I., et al.\ 2009, \apj, 695, 101 
\bibitem[Das et al.(2010)]{2010MNRAS.409.1362D} Das, P., Gerhard, O., Churazov, E., \& Zhuravleva, I.\ 2010, \mnras, 409, 1362
\bibitem[De Lucia \& Blaizot(2007)]{2007MNRAS.375....2D} De Lucia, G., \& Blaizot, J.\ 2007, \mnras, 375, 2 
\bibitem[Doherty et al.(2009)]{2009A&A...502..771D} Doherty, M., et al.\ 2009, \aap, 502, 771 
\bibitem[Dupraz \& Combes(1986)]{1986A&A...166...53D} Dupraz, C., \& Combes, F.\ 1986, \aap, 166, 53 
\bibitem[Durrell et al.(2003)]{2003ApJ...582..170D} Durrell, P.~R., Mihos, J.~C., Feldmeier, J.~J., Jacoby, G.~H., \& Ciardullo, R.\ 2003, \apj, 582, 170 
\bibitem[Evstigneeva et al.(2007)]{2007AJ....133.1722E} Evstigneeva, E.~A., Gregg, M.~D., Drinkwater, M.~J., \& Hilker, M.\ 2007, \aj, 133, 1722  
\bibitem[Fardal et al.(2007)]{2007MNRAS.380...15F} Fardal, M.~A., Guhathakurta, P., Babul, A., \& McConnachie, A.~W.\ 2007, \mnras, 380, 15 
\bibitem[Feldmeier et al.(2003)]{2003ApJS..145...65F} Feldmeier, J.~J., Ciardullo, R., Jacoby, G.~H., \& Durrell, P.~R.\ 2003, \apjs, 145, 65 
\bibitem[Forbes \& Bridges(2010)]{2010MNRAS.404.1203F} Forbes, D.~A., \& Bridges, T.\ 2010, \mnras, 404, 1203 
\bibitem[Forbes et al.(2011)]{2011MNRAS.413.2943F} Forbes, D.~A., Spitler, L.~R., Strader, J., Romanowsky, A.~J., Brodie, J.~P., \& Foster, C.\ 2011, \mnras, 413, 2943 
\bibitem[Forte et al.(1982)]{1982AJ.....87.1465F} Forte, J.~C., Martinez, R.~E., \& Muzzio, J.~C.\ 1982, \aj, 87, 1465 
\bibitem[Foster et al.(2010)]{2010AJ....139.1566F} Foster, C., Forbes, D.~A., Proctor, R.~N., Strader, J., Brodie, J.~P., \& Spitler, L.~R.\ 2010, \aj, 139, 1566 
\bibitem[Foster et al.(2011)]{2011MNRAS.415.3393F} Foster, C., Spitler, L.~R., Romanowsky, A.~J., et al.\ 2011, \mnras, 415, 3393 
\bibitem[Gao et al.(2007)]{2007ChJAA...7..111G} Gao, S., Jiang, B.-W., \& Zhao, Y.-H.\ 2007, CJAA, 7, 111 
\bibitem[Geisler et al.(2007)]{2007PASP..119..939G} Geisler, D., Wallerstein, G., Smith, V.~V., 
\& Casetti-Dinescu, D.~I.\ 2007, \pasp, 119, 939 
\bibitem[Gilbert et al.(2007)]{2007ApJ...668..245G} Gilbert, K.~M., et al.\ 2007, \apj, 668, 245
\bibitem[Gilbert et al.(2009)]{2009ApJ...705.1275G} Gilbert, K.~M., et al.\ 2009, \apj, 705, 1275 
\bibitem[G{\"u}ltekin et al.(2009)]{2009ApJ...698..198G} G{\"u}ltekin, K., Richstone, D.~O., Gebhardt, K., et al.\ 2009, \apj, 698, 198 
\bibitem[Halliday et al.(2001)]{2001MNRAS.326..473H} Halliday, C., Davies, R.~L., Kuntschner, H., et al.\ 2001, \mnras, 326, 473 
\bibitem[Hanes et al.(2001)]{2001ApJ...559..812H} Hanes, D.~A., C{\^o}t{\'e}, P., Bridges, T.~J., McLaughlin, D.~E., Geisler, D., Harris, G.~L.~H., Hesser, J.~E., \& Lee, M.~G.\ 2001, \apj, 559, 812
\bibitem[Harding et al.(2001)]{2001AJ....122.1397H} Harding, P., Morrison, H.~L., Olszewski, E.~W., Arabadjis, J., Mateo, M., Dohm-Palmer, R.~C., Freeman, K.~C., \& Norris, J.~E.\ 2001, \aj, 122, 1397 
\bibitem[Harris(2009)]{2009ApJ...703..939H} Harris, W.~E.\ 2009, \apj, 703, 939
\bibitem[Ha{\c s}egan et al.(2005)]{2005ApJ...627..203H} Ha{\c s}egan, M., et al.\ 2005, \apj, 627, 203
\bibitem[Ha{\c s}egan(2007)]{2007PhDT.........4H} Ha{\c s}egan, I.~M.\ 2007, Ph.D.~Thesis, Rutgers Univ.
\bibitem[Helmi et al.(1999)]{1999Natur.402...53H} Helmi, A., White, S.~D.~M., de Zeeuw, P.~T., \& Zhao, H.\ 1999, \nat, 402, 53 
\bibitem[Helmi et al.(2003)]{2003ApJ...592L..25H} Helmi, A., Navarro, J.~F., Meza, A., Steinmetz, M., \& Eke, V.~R.\ 2003, \apjl, 592, L25 
\bibitem[Helmi(2004)]{2004ApJ...610L..97H} Helmi, A.\ 2004, \apjl, 610, L97 
\bibitem[Helmi(2008)]{2008A&ARv..15..145H} Helmi, A.\ 2008, \aapr, 15, 145 
\bibitem[Hernquist(1987)]{1987ApJS...64..715H} Hernquist, L.\ 1987, \apjs, 64, 715 
\bibitem[Hernquist(1990)]{1990ApJ...356..359H} Hernquist, L.\ 1990, \apj, 356, 359
\bibitem[Hernquist \& Quinn(1988)]{1988ApJ...331..682H} Hernquist, L., \& Quinn, P.~J.\ 1988, \apj, 331, 682 
\bibitem[Hernquist \& Quinn(1989)]{1989ApJ...342....1H} Hernquist, L., \& Quinn, P.~J.\ 1989, \apj, 342, 1 
\bibitem[Hernquist \& Spergel(1992)]{1992ApJ...399L.117H} Hernquist, L., \& Spergel, D.~N.\ 1992, \apjl, 399, L117 
\bibitem[Hibbard \& Mihos(1995)]{1995AJ....110..140H} Hibbard, J.~E., \& Mihos, J.~C.\ 1995, \aj, 110, 140 
\bibitem[Hopkins et al.(2010)]{2010MNRAS.401.1099H} Hopkins, P.~F., Bundy, K., Hernquist, L., Wuyts, S., \& Cox, T.~J.\ 2010, \mnras, 401, 1099 
\bibitem[Huchra \& Brodie(1987)]{1987AJ.....93..779H} Huchra, J., \& Brodie, J.\ 1987, \aj, 93, 779 
\bibitem[Ibata et al.(2001)]{2001Natur.412...49I} Ibata, R., Irwin, M., Lewis, G., Ferguson, A.~M.~N., \& Tanvir, N.\ 2001, \nat, 412, 49 
\bibitem[Janowiecki et al.(2010)]{2010ApJ...715..972J} Janowiecki, S., Mihos, J.~C., Harding, P., Feldmeier, J.~J., Rudick, C., \& Morrison, H.\ 2010, \apj, 715, 972 
\bibitem[J{\'{\i}}lkov{\'a} et al.(2010)]{2010ASPC..423..243J} J{\'{\i}}lkov{\'a}, L., Jungwiert, B., Kr{\'{\i}}zek, M., Ebrov{\'a}, I., Stoklasov{\'a}, I., Bart{\'a}kov{\'a}, T., \& Bartoskov{\'a}, K.\ 2010, Galaxy Wars: Stellar Populations and Star Formation in Interacting Galaxies, 423, 243 
\bibitem[Johnston(1998)]{1998ApJ...495..297J} Johnston, K.~V.\ 1998, \apj, 495, 297 
\bibitem[Johnston et al.(2002)]{2002ApJ...570..656J} Johnston, K.~V., Spergel, D.~N., \& Haydn, C.\ 2002, \apj, 570, 656 
\bibitem[Johnston et al.(2008)]{2008ApJ...689..936J} Johnston, K.~V., Bullock, J.~S., Sharma, S., Font, A., Robertson, B.~E., \& Leitner, S.~N.\ 2008, \apj, 689, 936 
\bibitem[Jord{\'a}n et al.(2007)]{2007ApJS..171..101J} Jord{\'a}n, A., et al.\ 2007, \apjs, 171, 101 
\bibitem[Jord{\'a}n et al.(2009)]{2009ApJS..180...54J} Jord{\'a}n, A., et al.\ 2009, \apjs, 180, 54
\bibitem[Khochfar \& Silk(2006)]{2006ApJ...648L..21K} Khochfar, S., \& Silk, J.\ 2006, \apjl, 648, L21 
\bibitem[Kissler-Patig \& Gebhardt(1998)]{1998AJ....116.2237K} Kissler-Patig, M., \& Gebhardt, K.\ 1998, \aj, 116, 2237 
\bibitem[Koch et al.(2008)]{2008ApJ...689..958K} Koch, A., et al.\ 2008, \apj, 689, 958 
\bibitem[Koposov et al.(2010)]{2010ApJ...712..260K} Koposov, S.~E., Rix, H.-W., \& Hogg, D.~W.\ 2010, \apj, 712, 260 
\bibitem[Kormendy et al.(1997)]{1997ApJ...482L.139K} Kormendy, J., Bender, R., Magorrian, J., et al.\ 1997, \apjl, 482, L139 
\bibitem[Kormendy et al.(2009)]{2009ApJS..182..216K} Kormendy, J., Fisher, D.~B., Cornell, M.~E., \& Bender, R.\ 2009, \apjs, 182, 216
\bibitem[Krick et al.(2011)]{2011ApJ...735...76K} Krick, J.~E., Bridge, C., Desai, V., Mihos, J.~C., Murphy, E., Rudick, C., Surace, J., \& Neill, J.\ 2011, \apj, 735, 76 
\bibitem[Larsen et al.(2001)]{2001AJ....121.2974L} Larsen, S.~S., Brodie, J.~P., Huchra, J.~P., Forbes, D.~A., \& Grillmair, C.~J.\ 2001, \aj, 121, 2974 
\bibitem[Law et al.(2009)]{2009ApJ...703L..67L} Law, D.~R., Majewski, S.~R., \& Johnston, K.~V.\ 2009, \apjl, 703, L67 
\bibitem[Lee et al.(2007)]{2007ApJ...661L..49L} Lee, Y.-W., Gim, H.~B., \& Casetti-Dinescu, D.~I.\ 2007, \apjl, 661, L49 
\bibitem[Liu et al.(2009)]{2009MNRAS.396.2003L} Liu, F.~S., Mao, S., Deng, Z.~G., Xia, X.~Y., \& Wen, Z.~L.\ 2009, \mnras, 396, 2003 
\bibitem[Lotz et al.(2004)]{2004ApJ...613..262L} Lotz, J.~M., Miller, B.~W., \& Ferguson, H.~C.\ 2004, \apj, 613, 262 
\bibitem[Mackey et al.(2010)]{2010ApJ...717L..11M} Mackey, A.~D., et al.\ 2010, \apjl, 717, L11 
\bibitem[Mart{\'{\i}}nez-Delgado et al.(2010)]{2010AJ....140..962M} Mart{\'{\i}}nez-Delgado, D., et al.\ 2010, \aj, 140, 962 
\bibitem[Faltenbacher \& Mathews(2005)]{2005MNRAS.362..498F} Faltenbacher, A., \& Mathews, W.~G.\ 2005, \mnras, 362, 498 
\bibitem[McIntosh et al.(2008)]{2008MNRAS.388.1537M} McIntosh, D.~H., Guo, Y., Hertzberg, J., Katz, N., Mo, H.~J., van den Bosch, F.~C., \& Yang, X.\ 2008, \mnras, 388, 1537 
\bibitem[McLaughlin(1999)]{1999ApJ...512L...9M} McLaughlin, D.~E.\ 1999, \apjl, 512, L9 
\bibitem[McNeil et al.(2010)]{2010A&A...518A..44M} McNeil, E.~K., Arnaboldi, M., Freeman, K.~C., Gerhard, O.~E., Coccato, L., \& Das, P.\ 2010, \aap, 518, A44 
\bibitem[Merrett et al.(2003)]{2003MNRAS.346L..62M} Merrett, H.~R., et al.\ 2003, \mnras, 346, L62 
\bibitem[Merrifield \& Kuijken(1998)]{1998MNRAS.297.1292M} Merrifield, M.~R., \& Kuijken, K.\ 1998, \mnras, 297, 1292 
\bibitem[Mihos et al.(2005)]{2005ApJ...631L..41M} Mihos, J.~C., Harding, P., Feldmeier, J., \& Morrison, H.\ 2005, \apjl, 631, L41
\bibitem[Miyazaki et al.(2002)]{2002PASJ...54..833M} Miyazaki, S., et al.\ 2002, \pasj, 54, 833
\bibitem[Mouhcine et al.(2011)]{2011MNRAS.415..993M} Mouhcine, M., Ibata, R., \& Rejkuba, M.\ 2011, \mnras, 415, 993 
\bibitem[Murphy et al.(2011)]{2011ApJ...729..129M} Murphy, J.~D., Gebhardt, K., \& Adams, J.~J.\ 2011, \apj, 729, 129 
\bibitem[Naab et al.(2009)]{2009ApJ...699L.178N} Naab, T., Johansson, P.~H., \& Ostriker, J.~P.\ 2009, \apjl, 699, L178 
\bibitem[Navarro et al.(1996)]{1996ApJ...462..563N} Navarro, J.~F., Frenk, C.~S., \& White, S.~D.~M.\ 1996, \apj, 462, 563 
\bibitem[Nolthenius \& Ford(1986)]{1986ApJ...305..600N} Nolthenius, R., \& Ford, H.\ 1986, \apj, 305, 600 
\bibitem[Oser et al.(2010)]{2010ApJ...725.2312O} Oser, L., Ostriker, J.~P., Naab, T., Johansson, P.~H., \& Burkert, A.\ 2010, \apj, 725, 2312 
\bibitem[Oser et al.(2011)]{2011arXiv1106.5490O} Oser, L., Naab, T., Ostriker, J.~P., \& Johansson, P.~H.\ 2011, in press, arXiv:1106.5490 
\bibitem[Paturel et al.(2003)]{2003A&A...412...45P} Paturel, G., Petit, C., Prugniel, P., et al.\ 2003, \aap, 412, 45 
\bibitem[Peek \& Graves(2010)]{2010ApJ...719..415P} Peek, J.~E.~G., \& Graves, G.~J.\ 2010, \apj, 719, 415
\bibitem[Pe{\~n}arrubia et al.(2010)]{2010MNRAS.408L..26P} Pe{\~n}arrubia, J., Belokurov, V., Evans, N.~W., Mart{\'{\i}}nez-Delgado, D., Gilmore, G., Irwin, M., Niederste-Ostholt, M., \& Zucker, D.~B.\ 2010, \mnras, 408, L26 
\bibitem[Peng et al.(2006)]{2006ApJ...639...95P} Peng, E.~W., et al.\ 2006, \apj, 639, 95 
\bibitem[Peng et al.(2008)]{2008ApJ...681..197P} Peng, E.~W., et al.\ 2008, \apj, 681, 197 
\bibitem[Perrett et al.(2003)]{2003ApJ...589..790P} Perrett, K.~M., Stiff, D.~A., Hanes, D.~A., \& Bridges, T.~J.\ 2003, \apj, 589, 790 
\bibitem[Proctor et al.(2009)]{2009MNRAS.398...91P} Proctor, R.~N., Forbes, D.~A., Romanowsky, A.~J., Brodie, J.~P., Strader, J., Spolaor, M., Mendel, J.~T., \& Spitler, L.\ 2009, \mnras, 398, 91 
\bibitem[Prugniel et al.(2011)]{2011A&A...528A.128P} Prugniel, P., Zeilinger, W., Koleva, M., \& de Rijcke, S.\ 2011, \aap, 528, A128 
\bibitem[Quinn(1984)]{1984ApJ...279..596Q} Quinn, P.~J.\ 1984, \apj, 279, 596 
\bibitem[Richtler et al.(2011)]{2011A&A...531A.119R} Richtler, T., Salinas, R., Misgeld, I., Hilker, M., Hau, G.~K.~T., Romanowsky, A.~J., Schuberth, Y., \& Spolaor, M.\ 2011, \aap, 531, A119 
\bibitem[Rix et al.(1997)]{1997ApJ...488..702R} Rix, H.-W., de Zeeuw, P.~T., Cretton, N., van der Marel, R.~P., \& Carollo, C.~M.\ 1997, \apj, 488, 702 
\bibitem[Romanowsky \& Kochanek(2001)]{2001ApJ...553..722R} Romanowsky, A.~J., \& Kochanek, C.~S.\ 2001, \apj, 553, 722 
\bibitem[Romanowsky et al.(2009)]{2009AJ....137.4956R} Romanowsky, A.~J., Strader, J., Spitler, L.~R., Johnson, R., Brodie, J.~P., Forbes, D.~A., \& Ponman, T.\ 2009, \aj, 137, 4956 
\bibitem[Rudick et al.(2009)]{2009ApJ...699.1518R} Rudick, C.~S., Mihos, J.~C., Frey, L.~H., \& McBride, C.~K.\ 2009, \apj, 699, 1518 
\bibitem[Rudick et al.(2010)]{2010ApJ...720..569R} Rudick, C.~S., Mihos, J.~C., Harding, P., Feldmeier, J.~J., Janowiecki, S., \& Morrison, H.~L.\ 2010, \apj, 720, 569 
\bibitem[Ruszkowski \& Springel(2009)]{2009ApJ...696.1094R} Ruszkowski, M., \& Springel, V.\ 2009, \apj, 696, 1094 
\bibitem[Schuberth et al.(2010)]{2010A&A...513A..52S} Schuberth, Y., Richtler, T., Hilker, M., Dirsch, B., Bassino, L.~P., Romanowsky, A.~J., \& Infante, L.\ 2010, \aap, 513, A52 
\bibitem[Searle \& Zinn(1978)]{1978ApJ...225..357S} Searle, L., \& Zinn, R.\ 1978, \apj, 225, 357
\bibitem[Shih \& M{\'e}ndez(2010)]{2010ApJ...725L..97S} Shih, H.-Y., \& M{\'e}ndez, R.~H.\ 2010, \apjl, 725, L97 
\bibitem[Spolaor et al.(2010)]{2010MNRAS.408..254S} Spolaor, M., Hau, G.~K.~T., Forbes, D.~A., \& Couch, W.~J.\ 2010, \mnras, 408, 254 
\bibitem[Starkenburg et al.(2009)]{2009ApJ...698..567S} Starkenburg, E., et al.\ 2009, \apj, 698, 567 
\bibitem[Stott et al.(2011)]{2011MNRAS.414..445S} Stott, J.~P., Collins, C.~A., Burke, C., Hamilton-Morris, V., \& Smith, G.~P.\ 2011, \mnras, 414, 445 
\bibitem[Strader et al.(2004)]{2004AJ....127.3431S} Strader, J., Brodie, J.~P., \& Forbes, D.~A.\ 2004, \aj, 127, 3431 
\bibitem[Strader et al.(2011)]{2011arXiv1110.2778S} Strader, J., Romanowsky, A., Brodie, J., et al.\ 2011, \apjs, in press, arXiv:1110.2778 (S+11)
\bibitem[Tal et al.(2009)]{2009AJ....138.1417T} Tal, T., van Dokkum, P.~G., Nelan, J., \& Bezanson, R.\ 2009, \aj, 138, 1417 
\bibitem[Toloba et al.(2011)]{2011A&A...526A.114T} Toloba, E., Boselli, A., Cenarro, A.~J., et al.\ 2011, \aap, 526, A114 
\bibitem[Trujillo et al.(2006)]{2006ApJ...650...18T} Trujillo, I., et al.\ 2006, \apj, 650, 18 
\bibitem[Valentinuzzi et al.(2010)]{2010ApJ...721L..19V} Valentinuzzi, T., et al.\ 2010, \apjl, 721, L19 
\bibitem[van der Wel et al.(2008)]{2008ApJ...688...48V} van der Wel, A., Holden, B.~P., Zirm, A.~W., Franx, M., Rettura, A., Illingworth, G.~D., \& Ford, H.~C.\ 2008, \apj, 688, 48 
\bibitem[van Dokkum et al.(2009)]{2009Natur.460..717V} van Dokkum, P.~G., Kriek, M., \& Franx, M.\ 2009, \nat, 460, 717 
\bibitem[van Dokkum et al.(2010)]{2010ApJ...709.1018V} van Dokkum, P.~G., et al.\ 2010, \apj, 709, 1018 
\bibitem[Varghese et al.(2011)]{2011MNRAS.417..198V} Varghese, A., Ibata, R., \& Lewis, G.~F.\ 2011, \mnras, 417, 198 
\bibitem[Vazdekis et al.(2003)]{2003MNRAS.340.1317V} Vazdekis, A., Cenarro, A.~J., Gorgas, J., Cardiel, N., \& Peletier, R.~F.\ 2003, \mnras, 340, 1317 
\bibitem[Vitvitska et al.(2002)]{2002ApJ...581..799V} Vitvitska, M., Klypin, A.~A., Kravtsov, A.~V., Wechsler, R.~H., Primack, J.~R., \& Bullock, J.~S.\ 2002, \apj, 581, 799 
\bibitem[Weil, Bland-Hawthorn, \& Malin(1997)]{1997ApJ...490..664W} Weil, M.~L., Bland-Hawthorn, J., \& Malin, D.~F.\ 1997, \apj, 490, 664 
\bibitem[Woodley et al.(2010)]{2010AJ....139.1871W} Woodley, K.~A., G{\'o}mez, M., Harris, W.~E., Geisler, D., \& Harris, G.~L.~H.\ 2010, \aj, 139, 1871
\bibitem[Woodley \& Harris(2011)]{2011AJ....141...27W} Woodley, K.~A., \& Harris, W.~E.\ 2011, \aj, 141, 27 
\bibitem[Whiley et al.(2008)]{2008MNRAS.387.1253W} Whiley, I.~M., et al.\ 2008, \mnras, 387, 1253 
\bibitem[Xue et al.(2011)]{2011ApJ...738...79X} Xue, X.-X., Rix, H.-W., Yanny, B., et al.\ 2011, \apj, 738, 79 
\bibitem[Zemp et al.(2009)]{2009MNRAS.394..641Z} Zemp, M., Diemand, J., Kuhlen, M., Madau, P., Moore, B., Potter, D., Stadel, J., \& Widrow, L.\ 2009, \mnras, 394, 641 
\end{thebibliography}
\end{document}